\documentclass[useAMS]{mn2e}
\usepackage{array}
\usepackage{epsfig}
\usepackage{lscape}
\usepackage{footnote}
\usepackage{graphics}
\usepackage{etoolbox}
\usepackage{leftidx}
\usepackage{footnote}
\usepackage{color}
\usepackage{cleveref}
\usepackage{caption}
\usepackage{subfig}
\usepackage{amssymb}
\usepackage{pdflscape}
\usepackage{threeparttable}
\usepackage{url}
\usepackage{amsmath}
\usepackage{letltxmacro}

\title[Extended Ly$\alpha$ emission around quasars]{
Extended Ly$\alpha$ emission around quasars with eclipsing damped
Ly$\alpha$ systems
\thanks{Based  on  data  obtained  with  MagE  at  the
Clay  telescope  of the  Las  Campanas  Observatory (CNTAC  Prgm. ID
CN2012B-51 and CN2013A-121) and with XSHOOTER at the ESO/VLT in
Chile as a part of ESO program 091.A-0299(A).}}
\author[H. Fathivavsari et al.]
  {H.~Fathivavsari$^{1}$,
  P.~Petitjean$^{1}$, P.~Noterdaeme$^{1}$,
  I.~P\^aris$^{2}$, H.~Finley$^{3,4}$,
  \newauthor % starts a new line in the
   S.~L{\'o}pez$^{5}$,
   R.~Srianand$^{6}$\\%             % author environment
  $^1$Institut d'Astrophysique de Paris, Universit\'e Paris 6-CNRS, UMR7095, 98bis Boulevard Arago, 75014 Paris, France\\
  $^2$Osservatorio Astronomico di Trieste, via G. B. Tiepolo 11, 34131 Trieste, Italy\\
  $^3$CNRS/IRAP, 14 Avenue E. Belin, F-31400 Toulouse, France\\
  $^4$University Paul Sabatier of Toulouse/ UPS-OMP/ IRAP, F-31400 Toulouse, France\\
  $^5$Departamento de Astronom\' ia, Universidad de Chile, Casilla 36-D, Santiago, Chile\\
  $^6$Inter-University Centre for Astronomy and Astrophysics, Post Bag 4, Ganeshkhind, 411 007, Pune, India\\
}

\begin{document}

\date{Accepted .......   Received ....... }

\pagerange{\pageref{firstpage}--\pageref{lastpage}} \pubyear{2016}

\maketitle

\label{firstpage}

\begin{abstract}

We present spectroscopic observations of six high redshift ($z_{\rm
em}$~$>$~2) quasars, which have been selected for their
Lyman\,$\alpha$ (Ly$\alpha$) emission region being only partially
covered by a strong proximate ($z_{\rm abs}$~$\sim$~$z_{\rm em}$)
coronagraphic damped Ly$\alpha$ system (DLA). We detected spatially
extended Ly$\alpha$ emission envelopes surrounding these six
quasars, with projected spatial extent in the range 26~$\le$~$d_{\rm
Ly\alpha}$~$\le$~51~kpc. No correlation is found between the quasar
ionizing luminosity and the Ly$\alpha$ luminosity of their extended
envelopes. This could be related to the limited covering factor of
the extended gas and/or due to the AGN being obscured in other
directions than towards the observer. Indeed, we find a strong
correlation between the luminosity of the envelope and its spatial
extent, which suggests that the envelopes are probably ionized by
the AGN. The metallicity of the coronagraphic DLAs is low and varies
in the range $-$1.75~$<$~[Si/H]~$<$~$-$0.63. Highly ionized gas is
observed to be associated with most of these DLAs, probably
indicating ionization by the central AGN. One of these DLAs has the
highest Al\,{\sc iii}/Si\,{\sc ii} ratio ever reported for any
intervening and/or proximate DLA. Most of these DLAs are redshifted
with respect to the quasar, implying that they might represent
infalling gas probably accreted onto the quasar host galaxies
through filaments.

\end{abstract}


\section{Introduction}

The study of the gas associated with quasars, either in the disk or
halo of the host galaxy is important in understanding the interplay
between the AGN, its host-galaxy and the galactic environment.
Linking the quasar host galaxy to the intergalactic medium (IGM),
the circum-galactic medium (CGM) may provide fuel for star formation
activities in the host galaxy. Indeed, gas from the IGM is being
accreted along filaments through cold streams onto quasar host
galaxies (Cantalupo et al. 2014; Martin et al. 2014). In addition,
the enrichment of the IGM by strong starburst- and/or AGN-driven
outflows also occurs through the CGM gas (Adelberger et al. 2005;
D'Odorico et al. 2013; Shull et al. 2015, Bordoloi et al. 2016). If
strong enough, these outflows can expel a large fraction of gas from
the CGM into the IGM and consequently suppress further star
formation in the galaxy (Silk $\&$ Rees 1998; Maiolino et al. 2012).
Observing the CGM around high redshift quasars will hence offer
important insights into the so-called 'AGN feedback' in galaxy
formation and evolution.

For decades, the preferred observational technique for studying the
CGM has been the analysis of absorption lines detected in the
spectra of background quasars (Bahcall \& Spitzer 1969; Petitjean et
al. 1996b; Croft et al. 2002; D'Odorico et al. 2002; Bergeron et al.
2004; Hennawi et al. 2006; Hennawi \& Prochaska 2007; Steidel et al.
2010; Churchill et al. 2013; Landoni et al. 2016, P\'eroux et al.
2016, Quiret et al. 2016). This observational technique revealed the
presence of infalling and outflowing absorbing material around high
redshift galaxies (Steidel et al. 2000; Simcoe, Sargent \& Rauch
2004; Bouch\'e et al. 2007; Ryan-Weber et al. 2009; Rubin et al.
2012; Shull, Danforth \& Tilton 2014). While most of the studies
used quasars as background sources, the method has been extended to
using galaxies (e.g. Steidel et al. 2010) as well as GRBs (e.g.
Petitjean \& Vergani 2011) as background sources. However, the one
dimensional nature of this method does not allow one to map the
spatial distribution of these material around an individual galaxy.

Alternatively, detecting the CGM in emission (e.g. Ly$\alpha$)
around high redshift quasars could in principle provide information
on the spatial distribution of the gas surrounding the quasar host
galaxies (see Cantalupo et al. 2014; Hennawi et al. 2015). Indeed,
extended Ly$\alpha$ emission have been found around high redshift
($z$~$>$~2) quasars, extending up to distances as large as
100$-$200~kpc with typical luminosity of 10$^{43}$\,erg\,s$^{-1}$
(Hu \& Cowie 1987; Petitjean et al. 1996a; Lehnert $\&$ Becker 1998;
Heckman et al. 1991a,b; Christensen et al. 2006, hereafter CJ06;
Courbin et al. 2008; North et al. 2012, hereafter NC12; Borisova et
al. 2016). In an effort to study extended Ly$\alpha$ emission around
high redshift quasars, Arrigoni Battaia et al. (2016) obtained 15
deep ($\sim$~2.0~$\times$~10$^{-18}$
erg~s$^{-1}$\,cm$^{-2}$\,arcsec$^{-2}$) narrow band images of
$z_{\rm em}$~$\sim$~2.25 quasars, using GMOS on Gemini-South
telescope. They detect Ly$\alpha$ emission around half of their
quasars, extending up to 32~kpc with a surface brightness in excess
of 10$^{-17}$ erg~s$^{-1}$\,cm$^{-2}$\,arcsec$^{-2}$. Several
mechanisms have been proposed to power this emission:
photo-ionization by the photons emitted by an AGN and/or a starburst
region (Weidinger, M\o ller \& Fynbo 2004; Weidinger et al. 2005),
ionization induced by the shock waves from radio jets (e.g. Heckman
et al. 1991a,b), and cooling radiation after the gravitational
collapse (a.k.a cold accretion, Haiman et al. 2000; Yang et al.
2006).

In the 'cold accretion' scenario, a spatially extended distribution
of infalling cold gas collapses into the potential wells of dark
matter and gets heated. The heat is then dissipated through the
emission of Ly$\alpha$ photons. The resulting line radiation could
potentially be observed as extended, low surface brightness
Ly$\alpha$ 'fuzz' around high redshift quasars (Haiman et al. 2000;
Fardal et al. 2000). Based on hydrodynamical simulations of galaxy
formation, Dijkstra $\&$ Loeb (2009) showed that this infall of
primordial gas into the centers of dark matter haloes could explain
the observed Ly$\alpha$ emission if only $\gtrsim$~10\% of the
gravitational potential energy is converted into the Ly$\alpha$
radiation. Here, the emission would come from the filamentary cold
streams of metal-poor intergalactic gas. The gas is optically thick
and self-shielded against external ionizing radiation sources and
therefore the emission would not depend on the presence of a central
AGN or a starburst region. Haiman $\&$ Rees (2001), on the other
hand, estimated that if a central quasar is switched on inside a
collapsing primordial gas, this could significantly enhance the
Ly$\alpha$ emission of the cloud. They predict angular diameter of a
few arcseconds and surface brightness $\sim$~10$^{-18}$ to
10$^{-16}$\,erg~s$^{-1}$\,cm$^{-2}$\,arcsec$^{-2}$ for the
Ly$\alpha$ fuzz. This surface brightness would be detectable with
current telescopes through spectroscopy or deep narrow-band imaging
(Steidel et al. 1991; Fried 1998; Bergeron et al. 1999; Fynbo et al.
2000; Bunker et al. 2003). Confirming the prediction by Haiman \&
Rees (2001), Weidinger et al. (2004) found low metallicity
primordial material falling onto the AGN through the ionizing cone
of the quasar, causing the gas to glow in Ly$\alpha$ due to
recombination. Moreover, Jiang et al. (2016) found a strong narrow
Ly$\alpha$ emission line in the trough of a proximate-DLA (PDLA)
towards the quasar SDSS~J0952+0114; and they argued that this
Ly$\alpha$ emission is most likely originated from some large scale
outflowing material driven by the central AGN.

It may well be that the spatially extended Ly$\alpha$ fuzz seen
around a high redshift galaxy (hosting a quasar) is powered by more
than one powering mechanism at the same time. It is also possible
that both infall and outflow of material are occurring
simultaneously in the CGM of a single galaxy. Using integral field
spectroscopy to study the giant Ly$\alpha$ nebula around a high
redshift radio galaxy at $z$~=~4.11, Swinbank et al. (2015) found
two blueshifted (with respect to the systemic redshift of the
galaxy) H\,{\sc i} absorbers with covering fraction $\sim$~1 against
the Ly$\alpha$ emission. They argued that the more blue-shifted
absorber, showing dynamics similar to that of the emitting gas, is
part of an expanding shell of gas that surrounds the Ly$\alpha$
emission nebula. They then suggested that the emission nebula itself
is probably powered by the cooling of primordial gas inside a dark
matter halo. In principle, combining the information from both the
1D and 2D spectra of an Ly$\alpha$ emission envelope would allow one
to determine which powering mechanism is at play (Weidinger et al.
2005).

With the aim of accessing spatially extended Ly$\alpha$ emission
regions around high redshift quasars, we recently searched the
SDSS-III Baryon Oscillation Spectroscopic Survey (BOSS; Dawson et
al. 2013) for strong damped Ly$\alpha$ absorption systems (DLAs;
log\,$N_{\rm HI}$~$>$~21.30) coincident in redshift with the quasar
(Finley et al. 2013). Each of these  'eclipsing DLAs', plays the
role of a natural coronagraph, completely extinguishing the broad
Ly$\alpha$ emission from the quasar, revealing some narrow
Ly$\alpha$ emission line in its trough (Hennawi et al. 2009; Finley
et al. 2013; Fathivavsari et al. 2015, hereafter paper~I; Jiang et
al. 2016). This narrow emission line originates from a source
located at approximately the same redshift as the quasar and
possibly physically associated with it. Ly$\alpha$ emission from the
quasar narrow line region (NLR), or from star-forming regions in the
quasar host galaxy, and Ly$\alpha$ emitting haloes around the quasar
could be responsible for this narrow Ly$\alpha$ emission line seen
in the DLA trough.

In Finley et al. (2013), we found 26 quasars with $z$~$>$~2, showing
narrow Ly$\alpha$ emission line in the trough of their eclipsing
DLAs. We then performed follow-up observations of six of these
systems using the Magellan-MagE echelle spectrograph in order to
gain insight into the origin of their extended emission. One of
these six objects (i.e. J0823+0529) was already studied in detail in
paper~I and the remaining objects are covered in this paper.

Observations and data reduction are described in section~2. The
properties of individual objects are discussed in section~3. Results
are given in section~4, and section~5 presents the conclusions. In
this work, we use a standard flat $\Lambda$CDM cosmology with
$\Omega_{\Lambda}$~=~0.73, $\Omega_{m}$~=~0.27, and
H$_0$~=70~km\,s$^{-1}$~Mpc$^{-1}$ (Komatsu et al. 2011).

\section{Observations and data reduction}

In the course of a survey for DLA systems having redshifts
coincident with the redshift of the quasar (eclipsing DLAs), we
found 26 quasars that exhibit Ly$\alpha$ emission at the bottom of
their eclipsing DLAs (Finley et al. 2013). We then performed
follow-up echelle spectroscopy of six of these quasars using the
Magellan Echellete spectrograph (MagE; Marshall et al. 2008) on the
6.5~m Clay telescope located at Las Campanas observatory in Chile.
The spectrograph is equipped with 10~arcsec long slits of various
widths and has a medium resolution (R~$\sim$~4000), covering the
full visible spectral range (3000 to 10\,000\,\textup{\AA}). For one
of the quasars (i.e. J0953+0349), we also have an XSHOOTER spectrum
(as a part of ESO program 091.A-0299(A)). The XSHOOTER spectrograph
(Vernet et al. 2011), mounted on the Very Large Telescope (VLT),
covers the full spectral region from 3000~\textup{\AA} to 2.5~$\mu$m
and has a resolution of $R$~$\sim$~5300 in the near-infrared (NIR)
arm when the slit width is 0.9~arcsec.

The MagE observations were performed by keeping the slit along two
perpendicular position angles, namely along north-south and
east-west directions. The observations were done in the 'derotator'
mode in order to fix the position angles during each exposure time.
The journal of the observations is shown in Table~\ref{journal}. For
each quasar, the spectrum of a standard star was also obtained,
immediately after the science exposure, to allow for a precise flux
calibration of the quasar spectra. The spectra were reduced and
extracted using the Mage\_Reduce pipeline written and developed by
George Becker\footnote{
\url{ftp://ftp.ast.cam.ac.uk/pub/gdb/mage_reduce/}}. Since the
spectral orders of the observed spectra are curved with respect to
the (x,y)-coordinate system defined by the CCD columns and rows
(Kelson 2003), we followed the method described in paper~I to
rectify the curvature of the orders. The final spectra are then
binned with a 3~$\times$~3 box, resulting in
FWHM~=~$\sim$~80~km\,s$^{-1}$. Each pixel has a width of
$\sim$~27~km\,s$^{-1}$, and each spectral resolution element is
sampled by 3 pixels. Each pixel along the slit covers
$\sim$\,0.3~arcsec. The pipeline uses optimum sky subtraction
techniques (Kelson 2003) to subtract the background emission from
the 2D spectra. We also checked that the residual flux outside the
quasar trace at $\sim$\,100~\textup{\AA} away on either side of the
extended Ly$\alpha$ emission is consistent with the background noise
levels. Therefore, the background subtraction has minimum effect on
the extent of the envelopes in our sample.

%************************ Table 1 *********************
 \begin{table}
 \centering
\caption{Journal of MagE observations.}
 \setlength{\tabcolsep}{2.2pt}
\renewcommand{\arraystretch}{1.25}
\begin{tabular}{c c c c c}

\hline
 QSO   & Date & Exp. time(s) & Seeing(") & Slit width(") \\
       &      &   (EW/NS)    & (EW/NS)& (EW/NS) \\
\hline
J0112$-$0048 & 2012-12-13 & 3600/3600 & 1.1/1.0 & 1.0/1.0 \\
J0953$+$0349 & 2013-02-13 & 3600/3600 & 1.2/1.1 & 0.85/0.85 \\
J1058$+$0315 & 2013-02-13 & 3600/3600 & 1.2/1.0 & 1.0/0.85 \\
J1154$-$0215 & 2012-01-22 & 3017/3600 & 0.9/0.9 & 1.0/1.0 \\
J1253$+$1007 & 2013-02-13 & 2700/3600 & 1.0/1.2 & 1.0/1.0 \\

\hline
\end{tabular}
 \label{journal}
\end{table}
%********************************************************

%************************ J0112_1D *********************
\begin{figure*}
\centering
\begin{tabular}{c}
\includegraphics[bb=70 471 554 738,clip=,width=0.65\hsize]{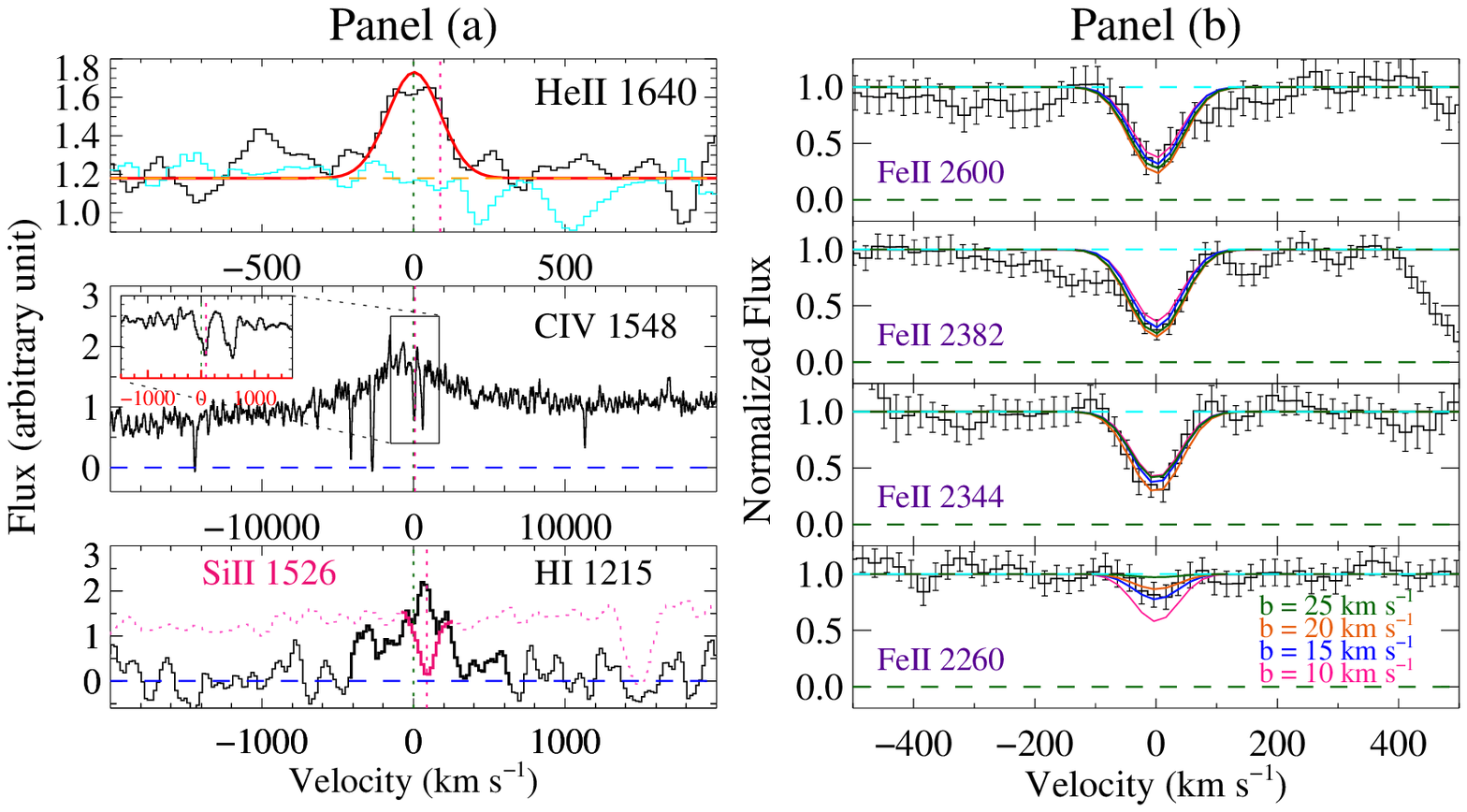}\\
\includegraphics[bb=64 351 542 729,clip=,width=0.65\hsize]{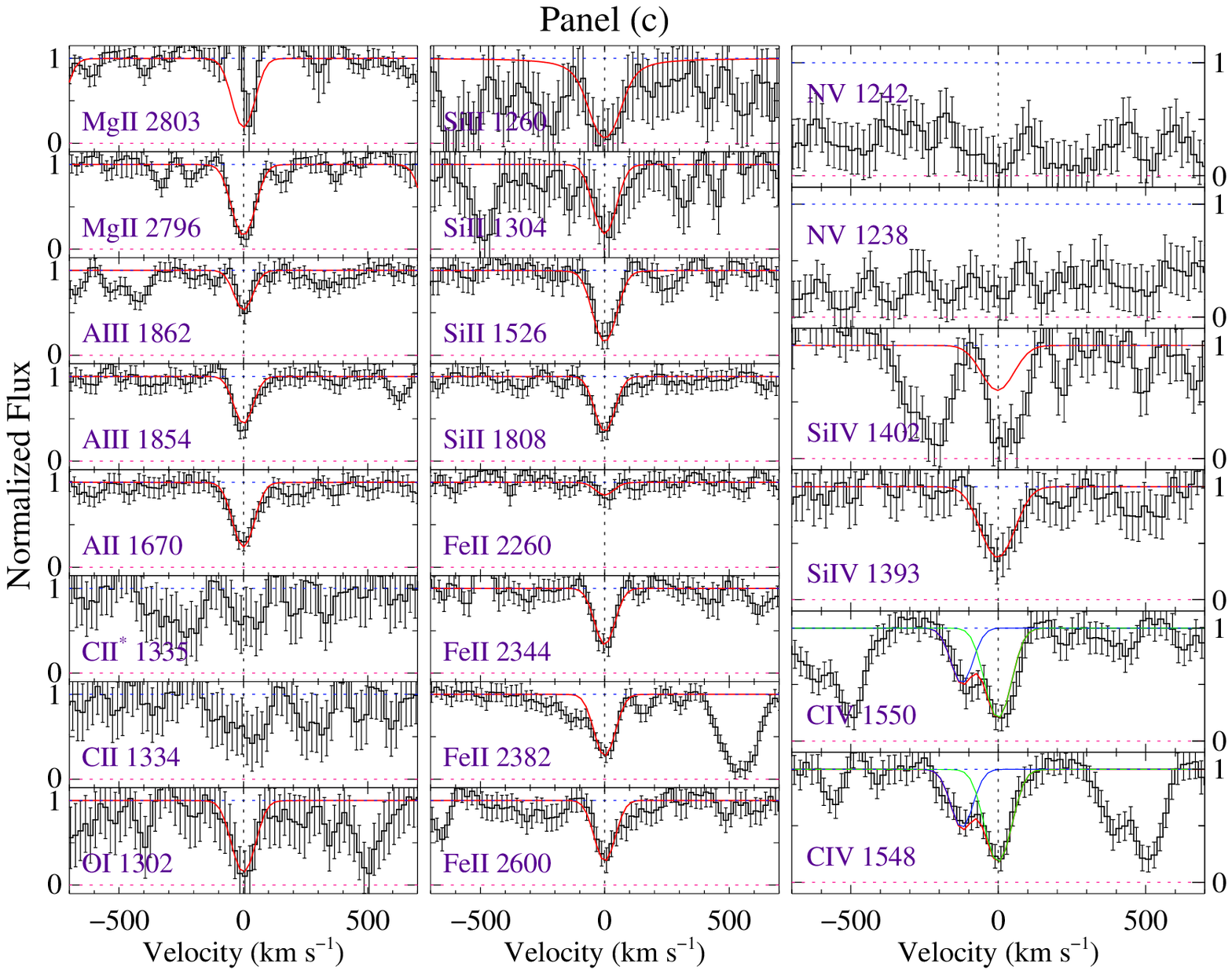}\\
\includegraphics[bb=58 435 533 594,clip=,width=0.65\hsize]{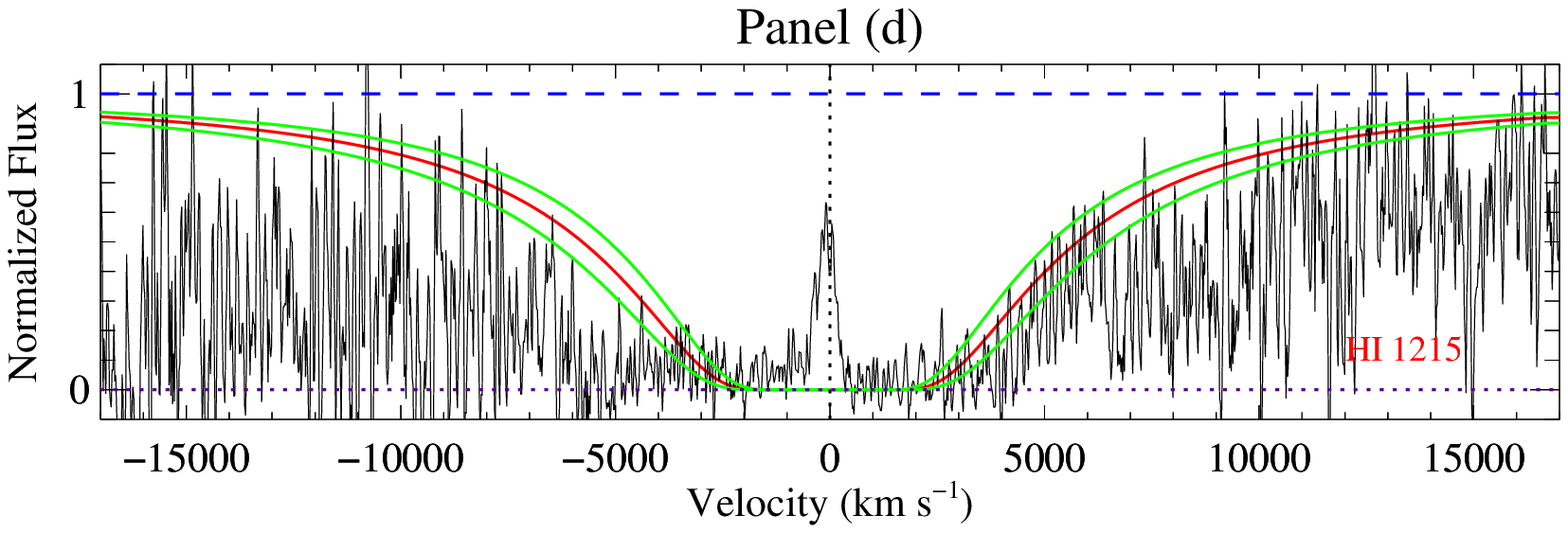}\\
\end{tabular}
\caption{{\sl Panel\,(a)}: Velocity profiles of some emission lines
detected in the spectrum of the quasar J0112$-$0048. The adopted
systemic redshift of the quasar ($z_{\rm em}$\,=\,2.1485) is used as
the origin of the velocity scale. The sky spectrum is overplotted as
a cyan curve in the top sub-panel. The pink vertical dotted lines
mark the redshift of the low-ion species. Note that in the H\,{\sc
i} emission line sub-panel, we overplot the Si~{\sc ii}$\lambda$1526
absorption as a pink curve. The inset in the middle sub-panel shows
a close-up view of the C\,{\sc iv} emission line region. {\sl
Panel\,(b)}: The optical depths of several Fe~{\sc ii} transitions
are used to constrain the Doppler parameter. The best fits for
different Doppler parameters are indicated by different colors. {\sl
Panel\,(c)}: Velocity plots and {\sc vpfit} solutions (where
applicable) of the species detected in the DLA at $z_{\rm
abs}$\,=\,2.1493. As shown in the C\,{\sc iv} doublet panel, the
C\,{\sc iv} absorption has an additional weaker component at
$v$~=~$-$117~km\,s$^{-1}$. For the low-ion species, we assume
$b$~=~17.4~km\,s$^{-1}$ as found from the analysis shown in
Panel~(b). {\sl Panel\,(d)}: the red curve is the single component
fit to the damped Ly$\alpha$ absorption profile with the two green
curves showing the $\pm$\,1\,$\sigma$ uncertainty.}
 \label{J0112_1D}
\end{figure*}
%*******************************************************

%************************ J0112_2D *********************
\begin{figure*}
\centering
\begin{tabular}{c}
\includegraphics[bb=73 390 546 675,clip=,width=0.65\hsize]{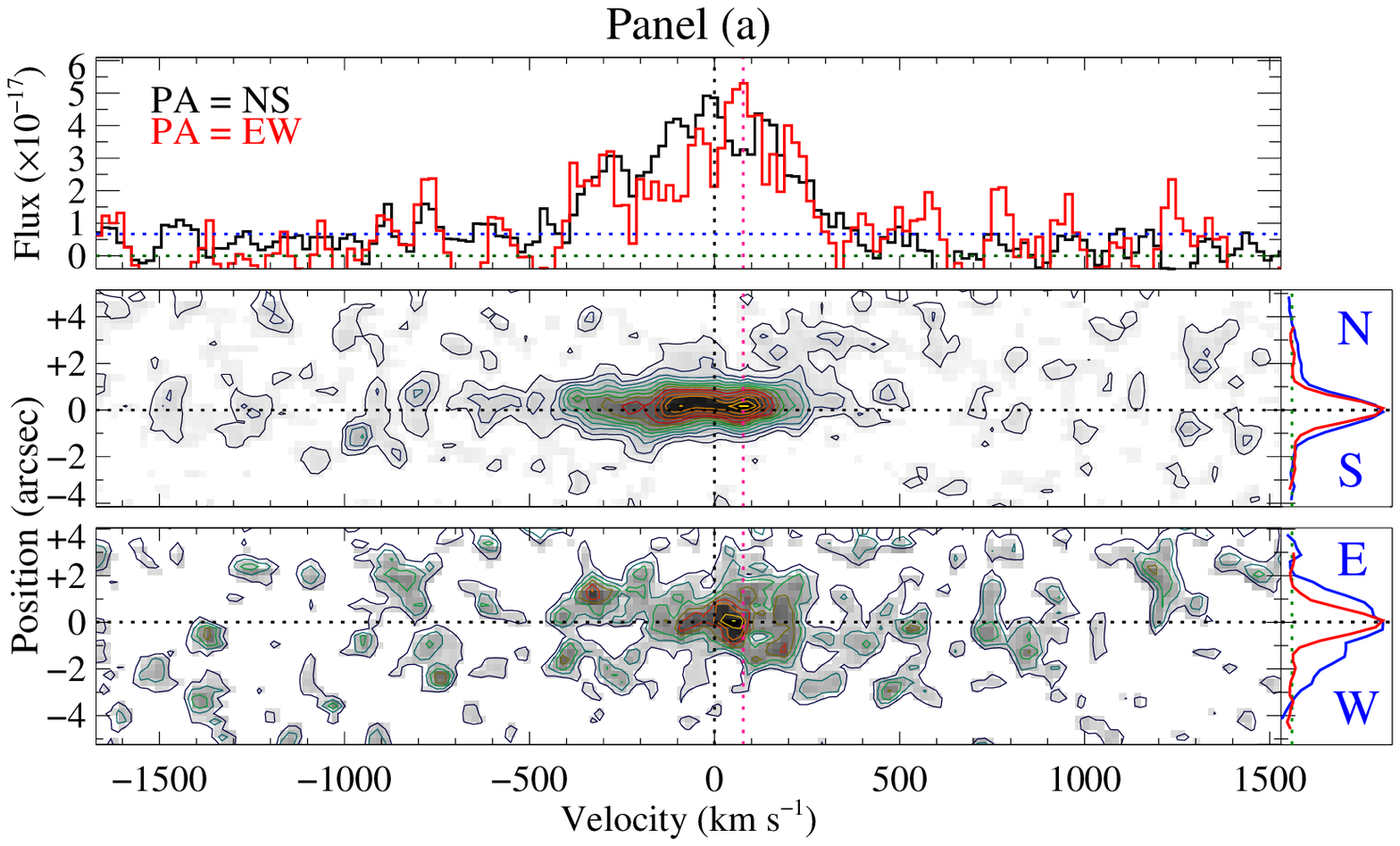}\\
\includegraphics[bb=43 399 566 629,clip=,width=0.65\hsize]{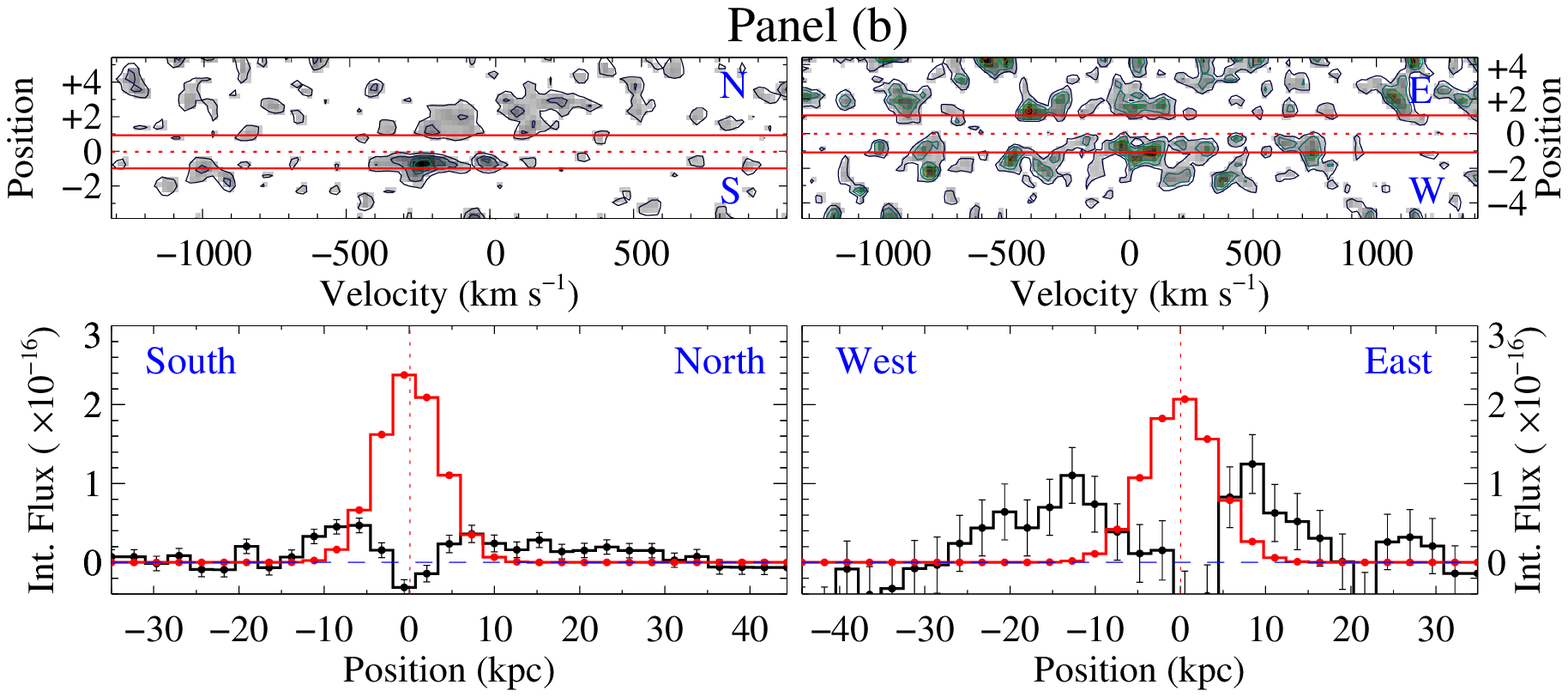}\\
\includegraphics[bb=49 363 566 575,clip=,width=0.65\hsize]{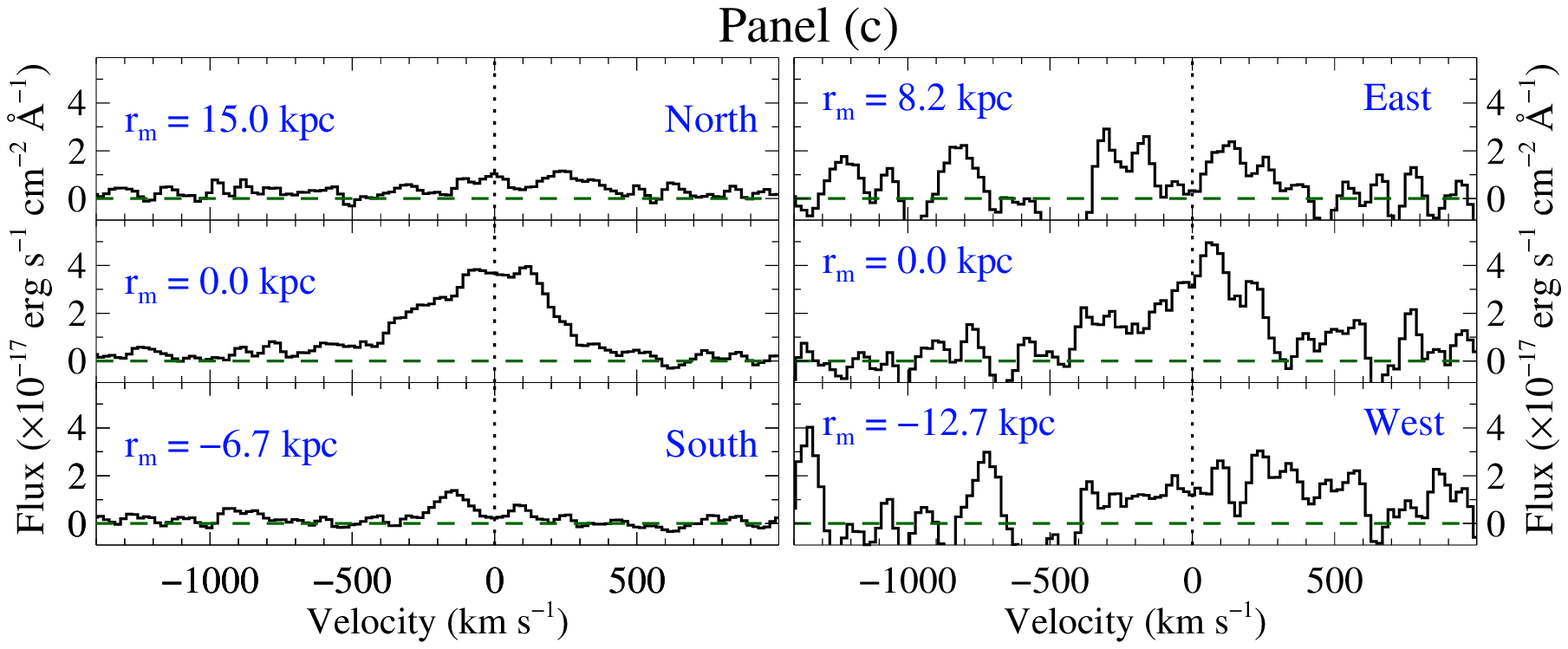}\\
\end{tabular}
\caption{1D and 2D spectra around the Ly$\alpha$ emission of quasar
J0112$-$0048. {\sl Panel\,(a)}: this panel comprises of three
sub-panels. The zero velocity is at $z_{\rm em}$\,=\,2.1485 and the
position of the DLA (at $z_{\rm abs}$\,=\,2.1493) is shown as a
vertical pink dotted line in the three sub-panels. The upper
sub-panel shows the 1D spectra of the Ly$\alpha$ emission seen at
the bottom of the DLA trough for the two PAs. The green and blue
horizontal dotted lines show the zero flux level and the 1\,$\sigma$
uncertainty on the flux, respectively. Here, the y-axis is in the
unit of erg~s$^{-1}$\,cm$^{-2}$\textup{\AA}$^{-1}$. The middle and
lower sub-panels show the 2D spectra of the Ly$\alpha$ emission seen
in the DLA trough for the two PAs. The outermost contour along PA=NS
(resp. PA=EW) corresponds to a flux density of
1.83\,$\times$\,10$^{-19}$ (resp. 9.13\,$\times$\,10$^{-20}$)
erg~s$^{-1}$\,cm$^{-2}$\textup{\AA}$^{-1}$ and each contour is
separated by 1.83\,$\times$\,10$^{-19}$ (resp.
9.13\,$\times$\,10$^{-20}$)
erg~s$^{-1}$\,cm$^{-2}$\textup{\AA}$^{-1}$ from its neighboring
contour. The black horizontal dotted lines mark the center of the
quasar trace. Panels on the right show the spatial profiles
integrated over the whole spectral width of the Ly$\alpha$ emission
(blue curves). The red curves show the profile of the quasar
continuum PSF. {\sl Panel\,(b)}: this panel comprises of two upper
sub-panels and two lower sub-panels. The upper sub-panels show the
2D spectra of the Ly$\alpha$ emission after subtracting the quasar
trace for the two PAs. The contour levels are the same as in panel
(a). The center of the trace is marked by the red horizontal dotted
line and the full width of the trace (i.e. 2\,$\times$\,FWHM) is
shown by the red solid horizontal lines. The y-axis here is position
in arcsecond. The lower sub-panels show the spatial profiles of the
extended Ly$\alpha$ emission (black histogram) and that of the
quasar trace (red histogram). Here, the y-axis is in
erg~s$^{-1}$\,cm$^{-2}$\,arcsec$^{-2}$. Panel\,(c): velocity
profiles of the spatially extended Ly$\alpha$ emission at several
distances from the AGN. The parameter $r_{\rm m}$ indicates the mean
distance to the central AGN.
%The zero velocity is at $z_{\rm em}$\,=\,2.1485
}
 \label{J0112_2D}
\end{figure*}
%*******************************************************

\section{Notes on individual objects}

For each quasar, we determine its systemic redshift, $z_{\rm em}$
(using quasar's intrinsic emission lines like He\,{\sc ii},
H$\beta$, C\,{\sc iv}, ..., etc), the redshift of the eclipsing DLA,
$z_{\rm abs}$ (using simultaneous fit of some low ion species), and
the redshift of the Ly$\alpha$ emission ($z_{\rm Ly\alpha}$) seen in
the trough of the eclipsing DLA. The parameters are listed in
Table~\ref{redshifts}. In this table, columns 8 and 9 are the
velocity widths of the Ly$\alpha$ emission ($\Delta$V$_{\rm
Ly\alpha}$) and of the C\,{\sc iv} absorption features
($\Delta$V$_{\rm CIV}$), respectively. The Ly$\alpha$ emission
(resp. C\,{\sc iv} absorption) velocity width is defined as the
wavelength region over which the flux level (resp. optical depth) is
larger than the 1\,$\sigma$ detection limit. Using {\sc
vpfit}\footnote{ \url{http://www.ast.cam.ac.uk/~rfc/vpfit.html}}, we
fit the low and high ion species associated with the eclipsing DLAs
and then determine the metallicity of the absorbers when possible.
To fit the absorption lines we used the same techniques described in
Fathivavsari et al. (2013). The plots showing the 1D and 2D emission
spectra along with the fits to the absorption lines for the quasar
J0112$-$0048 are presented in
Figs.~\ref{J0112_1D}\,\&\,\ref{J0112_2D} while the corresponding
plots of the remaining quasars are shown in Appendix~A. The
parameters of the fits are summarized in
Tables~\ref{Table_lowion}~\&~\ref{Table_hiion}. In
Table~\ref{Table_lowion}, the values in brackets are uncertain
because either the corresponding absorption lines are affected by
noise or all lines are strongly saturated.The metallicity and
relative abundances given in the last two rows of
Table~\ref{Table_lowion} are obtained using solar photospheric
abundances from Asplund et al. (2009).

In all quasar spectra, the Ly$\alpha$ emission detected in the
trough of the DLA is observed to be spatially extended. One can
think of this Ly$\alpha$ emission as a combination of two
components: the spatially unresolved narrow Ly$\alpha$ emission
(located on the quasar trace) and the second component which extends
beyond the quasar trace. The quasar trace is defined by the point
spread function (PSF) of the quasar continuum close to the
wavelength range where the Ly$\alpha$ emission is detected. The FWHM
of the PSF ($\sim$~1~arcsec) is approximately 8~kpc at the relevant
redshift. To determine the flux of the extended emission, we first
need to subtract the contribution from the unresolved component of
the emission which is seen on the quasar trace. To do so, we follow
the method described in section~5 of paper~I. Throughout this paper,
by envelope luminosity we mean the luminosity of the Ly$\alpha$
emission after subtracting that along the quasar trace. The
Ly$\alpha$ flux and luminosity of the extended envelopes, their
spatial extent, and their mean surface brightness are listed in
Tables~\ref{flux_extent}~\&~\ref{flux_luminosity}.

We emphasize that the extent of an envelope along the north, south,
east, or west direction is defined as the distance (in arcsecond or
kpc) from the center of the quasar trace up to the spatial pixel at
which the flux density per square arcsec reaches below the
1\,$\sigma$ detection limit. The 1\,$\sigma$ uncertainty in the flux
for each PA is reported in the sixth column of
Table~\ref{flux_extent}.

\subsection{SDSS J011226.76$-$004855.8}

For this quasar we detect Ly$\alpha$, C\,{\sc iv}, and He\,{\sc ii}
emission lines. To estimate the quasar redshift, we have fitted a
Gaussian function to the He\,{\sc ii} emission line which resulted
in $z_{\rm em}$~=~2.1485 (see panel~(a) of Fig.~\ref{J0112_1D}). We
adopt this value as the systemic redshift of the quasar.

Associated to the eclipsing DLA along this line of sight, we detect
absorption from low and high ionization species like, O\,{\sc i},
C\,{\sc ii}, Al\,{\sc ii}, Al\,{\sc iii}, Mg\,{\sc ii}, Fe\,{\sc
ii}, Si\,{\sc ii}, Si\,{\sc iv}, and C\,{\sc iv}. As shown in
panel~(c) of Fig.~\ref{J0112_1D}, the low-ion absorption profiles
are dominated by a single component at $z_{\rm abs}$~=~2.1493. The
DLA is located at $v$~$\sim$~\,+\,76~km\,s$^{-1}$ relative to the
quasar. We used several Fe\,{\sc ii} transitions (i.e. Fe\,{\sc
ii}$\lambda$2600, 2382, 2344, and 2260) to constrain the Doppler
parameter of this component. In panel~(b) of Fig.~\ref{J0112_1D},
the {\sc vpfit} solutions for different Doppler parameters are shown
by different colors.

%************************ Table 2 *********************
%\begin{landscape}
 \begin{table*}
  \begin{threeparttable}[b]
 \centering
\caption{The redshifts of the quasar, the Ly$\alpha$ emission, and
the DLA are given in columns 2 to 4 and the corresponding velocity
separations are in columns 5 to 7. The values in columns 5 and 6 are
with respect to the quasar's adopted systemic redshift while the
values in column 7 are relative to the DLA redshift, i.e. $z_{\rm
abs}$. Columns 8 and 9 give the velocity widths of the extended
Ly$\alpha$ emission and the C\,{\sc iv} absorption profile where the
signal is more than the 1$\sigma$ detection limit. Note that all
velocities are in km\,s$^{-1}$. The last column gives the upper
limit (in pc) on the size of the H\,{\sc i} cloud derived by
assuming that the hydrogen number density in our eclipsing DLAs is
$n_{\rm H}$~$>$~10~cm$^{-3}$ (see section 4.6.2).}
 \setlength{\tabcolsep}{8.35pt}
\renewcommand{\arraystretch}{1.3}
\begin{tabular}{c c c c c c c c c c c c c}

\hline

 ID   & $z_{\rm em}$ & $z_{\rm Ly\alpha}$ & $z_{\rm abs}$ & $\delta V_{\rm DLA-QSO}$ & $\delta V_{\rm Ly\alpha-QSO}$ & $\delta V_{\rm Ly\alpha-DLA}$ & $\Delta V_{\rm Ly\alpha}$ & $\Delta V_{\rm CIV}$ & $l$(H\,{\sc i}) \\

\hline

J0112$-$0048 &  2.1485 & 2.1482 & 2.1493 &  $+$76  &  $-$29     & $-$105  & 800  & 320  & $<$~289 \\
J0823$+$0529 &  3.1875 & 3.1925 & 3.1910 &  $+$250 &  $+$358    & $-$107  & 1400 & 754  & $<$~10\tnote{a} \\
J0953$+$0349 &  2.5940 & 2.5929 & 2.5961 &  $+$175 &  $-$92     & $-$267  & 1222 & 837  & $<$~289 \\
J1058$+$0315 &  2.3021 & 2.3029 & 2.2932 &  $-$809 &  $+$73     & $+$882  & 520  & 240  & $<$~204 \\
J1154$-$0215 &  2.1810 & 2.1842 & 2.1853 &  $+$405 &  $+$301    & $-$103  & 1157 & 1064 & $<$~182 \\
J1253$+$1007 &  3.0150 & 3.0308 & 3.0312 &  $+$1207&  $+$1177   & $-$30   & 1560 & 1030 & $<$~65 \\

\hline

\end{tabular}
 \label{redshifts}
 \begin{tablenotes}
            \item [a] See paper~I.
        \end{tablenotes}
        \end{threeparttable}
 \renewcommand{\footnoterule}{}
\end{table*}
%\end{landscape}

%************************ Table 3 *********************
%\begin{landscape}
 \begin{table*}
 \begin{threeparttable}[b]
 \centering
\caption{ Column densities and metallicities of the low-ionization
species detected in the eclipsing DLAs. The values in brackets are
uncertain because either the corresponding absorption lines are
affected by noise or all lines are strongly saturated. Column
densities are given in logarithmic units.}
 \setlength{\tabcolsep}{13.2pt}
\renewcommand{\arraystretch}{1.3}
\begin{tabular}{c c c c c c c}
\hline
            & J0112$-$0048   & J0823$+$0529 & J0953$+$0349 & J1058$+$0315 & J1154$-$0215 & J1253$+$1007 \\
\hline
%  $z_{abs}$ & 2.1493 & 3.1910 & 2.5961 & 2.2932 &  2.1853 & 3.0312 \\
  $b$\tnote{a} & 17.4                 & 20.0 & 20.0 & 15.0 &  20.0 & 37.0 \\
  N(O\,{\sc i})   & [17.47\,$\pm$\,0.24] &[17.08\,$\pm$\,0.50]& [17.11\,$\pm$\,0.20] & .....               &  ..... & 16.04\,$\pm$\,0.10 \\
  N(Si\,{\sc ii}) & 16.83\,$\pm$\,0.14   & 16.42\,$\pm$\,0.10 & 15.71\,$\pm$\,0.10   & 15.90\,$\pm$\,0.20  &  15.55\,$\pm$\,0.20   & 15.36\,$\pm$\,0.10 \\
  N(Fe\,{\sc ii}) & 15.05\,$\pm$\,0.20   & 15.33\,$\pm$\,0.20 & 15.43\,$\pm$\,0.20   & 15.74\,$\pm$\,0.15  &  15.35\,$\pm$\,0.20   & 14.27\,$\pm$\,0.10 \\
  N(C\,{\sc ii})  & .....                & .....              & [17.11\,$\pm$\,0.20] & .....               &  .....                & [16.12\,$\pm$\,0.20] \\
  N(C\,{\sc ii}$^{*}$)& .....            & .....              & [15.70\,$\pm$\,0.20] & .....               &  .....                & [14.80\,$\pm$\,0.20] \\
  N(Al\,{\sc ii}) & [14.77\,$\pm$\,0.20] & .....              & [14.09\,$\pm$\,0.15] & [15.56\,$\pm$\,0.16]&  [15.30\,$\pm$\,0.30] & [13.69\,$\pm$\,0.10] \\
  N(Al\,{\sc iii}) & 13.91\,$\pm$\,0.10  & 14.80\,$\pm$\,0.10 & 13.65\,$\pm$\,0.10   & 13.36\,$\pm$\,0.10  &  15.20\,$\pm$\,0.30   & 13.14\,$\pm$\,0.10 \\
  N(Mg\,{\sc ii}) & [15.90\,$\pm$\,0.25] & .....              & .....                & 16.96\,$\pm$\,0.20  &  [15.30\,$\pm$\,0.30] & ..... \\
  N(H\,{\sc i})   & 21.95\,$\pm$\,0.10   & 21.70\,$\pm$\,0.10 & 21.95\,$\pm$\,0.10   & 21.80\,$\pm$\,0.10  &  21.75\,$\pm$\,0.10   & 21.30\,$\pm$\,0.10 \\
  $[$Si/H$]$     &        $-$0.63     & $-$0.79 &       $-$1.75        &       $-$1.41        &          $-$1.71      &        $-$1.45       \\
  $[$Si/Fe$]$    &        1.77        & 1.08 &       0.27        &       0.15        &           0.19      &         1.08       \\
\hline
\end{tabular}
 \label{Table_lowion}
\begin{tablenotes}
            \item [a] Doppler parameter (km\,s$^{-1}$), see the text.
        \end{tablenotes}
        \end{threeparttable}
 \renewcommand{\footnoterule}{}
\end{table*}
%\end{landscape}

The fact that Fe\,{\sc ii}$\lambda$2260 is detected and is not
saturated allows us to constrain the Doppler parameter. As shown in
panel~(b) of Fig.~\ref{J0112_1D}, the best fit of the four Fe\,{\sc
ii} transitions results in $b$~=~17.4 km\,s$^{-1}$. To determine the
H\,{\sc i} column density of the DLA, a single component fit was
conducted to the DLA trough with the redshift fixed to the value
derived from the low-ion species, i.e. $z_{\rm abs}$~=~2.1493, and
we obtained log\,$N_{\rm HI}$~=~21.95~$\pm$~0.10. The fit is shown
in panel~(d) of Fig.~\ref{J0112_1D} as a red curve overplotted on
the observed spectrum (black curve), with the two green curves
indicating the corresponding $\pm$1\,$\sigma$ uncertainty. The fit
to other low-ion species was performed by fixing their corresponding
Doppler parameter and redshift to those obtained for Fe\,{\sc ii}.
The C\,{\sc iv} and Si\,{\sc iv} doublets were fitted independently
with all parameters allowed to vary during the fitting process. The
parameters of the fit are listed in
Tables~\ref{Table_lowion}~\&~\ref{Table_hiion}. As shown in
panel~(c) of Fig.~\ref{J0112_1D}, the expected position of the
N\,{\sc v} doublet falls on the red wing of the strong DLA profile
and is probably lost in the noise. The Si\,{\sc iv}$\lambda$1402
absorption profile is also blended with some unidentified features.
The C\,{\sc iv} doublet has two absorption components located at
$v\sim$\,$+$5 and $-$117\,km\,s$^{-1}$ relative to $z_{\rm abs}$.
While the component at $v\sim$\,$+$5 is also seen in Si\,{\sc iv},
the second component is only detected in C\,{\sc iv}, and no low-ion
absorption is associated with this second component.

As shown in Fig.~\ref{J0112_2D}, the Ly$\alpha$ emission at the
center of the DLA trough is spatially extended along both PAs. Along
the north-south (NS) direction, the emission is more extended to the
north up to $\sim$\,3.5 arcsec (30\,kpc) while the extension to the
south reaches up to $\sim$\,1.6 arcsec (13.5\,kpc) away from the
quasar trace. Along the east-west direction, the emission seems to
be more extended to the west ($\sim$\,3 arcsec or 25.5\,kpc) than to
the east ($\sim$\,2 arcsec or 16.5\,kpc). The Ly$\alpha$ flux and
luminosity of the envelope, its angular size, and the mean surface
brightness are listed in
Tables~\ref{flux_extent}~\&~\ref{flux_luminosity}.

From panel~(c) of Fig.~\ref{J0112_2D}, one can see that the
Ly$\alpha$ emission on the quasar trace is asymmetric, with a
steeper drop of the flux in its red wing and a shallower rise in the
blue wing. This asymmetry is less prominent along the east-west (EW)
direction due to the high level of noise. Since the emission seen on
the trace is not resolved, one would expect the same spectrum for
the quasar trace, regardless of the slit orientation. As illustrated
in panel~(c) of Fig.~\ref{J0112_2D}, the spectra on the trace for
the two PAs are consistent within the errors.

The Ly$\alpha$ emitting (resp. absorbing) gas is
$\sim$\,29~km\,s$^{-1}$ (resp. $\sim$\,76~km\,s$^{-1}$) blueshifted
(resp. redshifted) relative to the systemic redshift of the quasar.
These low velocity differences hint at the possibility  that the
emitting and/or absorbing cloud may originate from somewhere in the
quasar host galaxy or its close vicinity.

\subsection{SDSS J095307.13+034933.8}

This is the only quasar for which we also have an XSHOOTER spectrum.
Unfortunately, the SNR of this spectrum in the regions where we
expect to see redshifted Mg\,{\sc ii}, [O\,{\sc ii}] and [O\,{\sc
iii}] emission lines is not very high because the lines are either
located in bad portions of the spectrum or spoiled by strong sky
lines. However, we could still use the H$\beta$ emission line to get
an emission redshift of $z_{\rm em}$~=~2.5940 (see
Fig.~\ref{J0953_1D}). We adopt this redshift as the systemic
redshift of the quasar. As shown in Fig.~\ref{J0953_1D}, we also
detect O\,{\sc i}$\lambda$1304, C\,{\sc iii}$]$$\lambda$1909,
Si\,{\sc iii}$]$$\lambda$1892, and C\,{\sc iv}$\lambda$1548 emission
lines in the MagE spectrum of this quasar. Fitting Gaussians to
these emission lines yields $z$~=~2.5930, 2.5883, 2.5915, and
2.5880, respectively. The emission redshift from O\,{\sc
i}$\lambda$1304 and Si\,{\sc iii}$]$$\lambda$1892 are consistent
with that obtained from H$\beta$ within 200~km\,s$^{-1}$. But
C\,{\sc iii}$]$ and C\,{\sc iv} emission lines show, respectively,
475 and 500 km\,s$^{-1}$ blue-shift with respect to the adopted
systemic redshift of the quasar.

%************************ J1253_Ly_beta *********************
\begin{figure}
\centering
\begin{tabular}{c}
\includegraphics[bb=123 435 458 613,clip=,width=0.95\hsize]{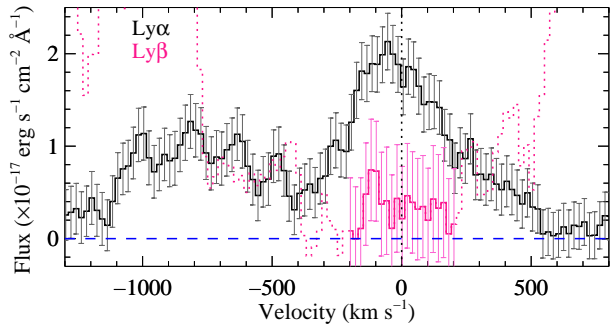}\\
\end{tabular}
\caption{Ly$\alpha$ (black curve) and Ly$\beta$ (pink curve)
emission lines detected in the trough of the damped Ly$\alpha$ and
Ly$\beta$ absorption lines along the line of sight towards the
quasar J1253+1007. }
 \label{Ly_beta}
\end{figure}
%*******************************************************

We fit a single Gaussian to the low ionization lines to get the
redshift $z_{\rm abs}$~=~2.5961 of the eclipsing DLA along this line
of sight which is 175~km\,s$^{-1}$ redshifted with respect to the
quasar. We fit the Ly$\alpha$ trough with a Voigt profile and
derived log\,$N_{\rm HI}$~=~21.95~$\pm$~0.10. We detect absorption
from O\,{\sc i}, C\,{\sc ii}, Si\,{\sc ii}, Al\,{\sc ii}, Fe\,{\sc
ii} in one component (see Fig.~\ref{J0953_1D}). To constrain the
corresponding Doppler parameter, we follow the same procedure
described in the previous section. As demonstrated in panel~(b) of
Fig.~\ref{J0953_1D}, Doppler parameters less than 20~km\,s$^{-1}$
(pink and blue curves) are clearly excluded while larger Doppler
parameters (i.e. $b$~$>$~25~km\,s$^{-1}$) are consistent with the
observations. Therefore, our reported column densities and
metallicity should be considered as upper limits. Parameters of the
fit are reported in Tables~\ref{Table_lowion}~\&~\ref{Table_hiion},
and the metallicity of the system is
[Si/H]\,$<$\,$-$1.75~$\pm$~0.20.

It is interesting to note that strong low ionization components
(O\,{\sc i} and C\,{\sc ii}, C\,{\sc ii}$^{*}$) are also seen at
$z$~=~2.5882 (660~km\,s$^{-1}$ blueshifted with respect to the DLA).
The C\,{\sc iv} and Si\,{\sc iv} absorption associated with this
component is strong and spread over $\sim$~250~km\,s$^{-1}$. N\,{\sc
v} may also be present but it seems to be strongly blended with some
intervening absorption.

As shown in Fig.~\ref{J0953_2D}, the Ly$\alpha$ emission at the
bottom of the DLA absorption line is extended along both PAs. From
this figure, we can see that the Ly$\alpha$ envelope is spatially
asymmetric. The emission extends out to a maximum distance of
$\sim$\,2.4~arcsec (20~kpc) and $\sim$\,1.9~arcsec (16~kpc) in the
south and east directions, respectively. Note that the emission
towards the north is so weak that we take it as a tentative
detection.

The spectra of the quasar trace for the two PAs are consistent with
each other within the errors (see panel~(c) of Fig.~\ref{J0953_2D}).
The DLA is redshifted by $\sim$~175~km\,s$^{-1}$ relative to the
systemic redshift of the quasar while the Ly$\alpha$ envelope and
the second low-ion system (located at $z_{\rm abs}$\,=\,2.5882) are
both blueshifted compared to $z_{\rm em}$ by 92 and
484~km\,s$^{-1}$, respectively. It seems possible that the low
metallicity DLA ([Si/H]\,$<$\,$-$1.75) is infalling towards the
central AGN and that the second low-ion system is part of some
outflowing material. It is also possible that both the absorption
and emission originate from the same structure. Note that the
transverse extensions of both of the two absorbing clouds should be
small compared to the NLR.

%************************ Table 4 *********************
%\begin{landscape}
 \begin{table}
 \centering
\caption{ Column densities of high ionization species. Note that the
second column gives the velocity separation of that component with
respect to the strongest component in the low-ion species (see
Table~2).}
 \setlength{\tabcolsep}{4.0pt}
\renewcommand{\arraystretch}{1.25}
\begin{tabular}{c c c c c}

\hline

 $z$   & $\delta$$V$(km\,s$^{-1}$) & Ion(X) & $b$(km\,s$^{-1}$) & logN(X)\\

\hline
\multicolumn{5}{c}{SDSS J0112$-$0048} \\
2.148066 & $-$117.5 &  C\,{\sc iv} & 10.0\,$\pm$\,2.0 & 15.67\,$\pm$\,0.30 \\
2.149356 & $+$5.3   &  C\,{\sc iv} & 16.7\,$\pm$\,5.0 & 16.70\,$\pm$\,0.30 \\
2.149338 & $+$3.6   &  Si\,{\sc iv}& 56.1\,$\pm$\,4.0 & 13.86\,$\pm$\,0.20 \\

\multicolumn{5}{c}{SDSS J0953+0349} \\
2.585898 & $-$851.7   &  C\,{\sc iv}& 22.2\,$\pm$\,9.0 & 14.24\,$\pm$\,0.24 \\
2.588245 & $-$665.6   &  C\,{\sc iv}& 68.8\,$\pm$\,7.7 & 15.06\,$\pm$\,0.04 \\
2.590421 & $-$473.8   &  C\,{\sc iv}& 62.5\,$\pm$\,24.0 & 15.04\,$\pm$\,0.12 \\
2.591778 & $-$360.5   &  C\,{\sc iv}& 10.0\,$\pm$\,5.0 & 15.72\,$\pm$\,0.30 \\
2.593449 & $-$221.1   &  C\,{\sc iv}& 40.0\,$\pm$\,8.7 & 14.00\,$\pm$\,0.04 \\
2.595836 & $-$22.0   &  C\,{\sc iv} & 41.9\,$\pm$\,5.4 & 14.65\,$\pm$\,0.05 \\
2.588187 & $-$660.4   &  Si\,{\sc iv}& 26.4\,$\pm$\,7.0 & 14.92\,$\pm$\,0.30 \\
2.589003 & $-$592.2   &  Si\,{\sc iv}& 13.0\,$\pm$\,5.0 & 13.30\,$\pm$\,0.30 \\
2.590254 & $-$487.8   &  Si\,{\sc iv}& 34.3\,$\pm$\,14.0 & 13.90\,$\pm$\,0.20 \\
2.595963 & $-$11.4   &  Si\,{\sc iv}& 29.0\,$\pm$\,7.8 & 14.36\,$\pm$\,0.21 \\

\multicolumn{5}{c}{SDSS J1058+0315} \\
2.293476 & $+$34.2   &  C\,{\sc iv}& 37.4\,$\pm$\,5.0 & 14.93\,$\pm$\,0.20 \\
2.293406 & $+$27.8   &  Si\,{\sc iv}& 24.4\,$\pm$\,2.3 & 15.50\,$\pm$\,0.30 \\

\multicolumn{5}{c}{SDSS J1154$-$0215} \\
2.177275 & $-$756.3   &  C\,{\sc iv}& 84.8\,$\pm$\,5.3 & 14.83\,$\pm$\,0.10 \\
2.178558 & $-$635.2   &  C\,{\sc iv}& 74.6\,$\pm$\,14.1 & 14.71\,$\pm$\,0.20 \\
2.185181 & $-$11.2   &  C\,{\sc iv}& 46.4\,$\pm$\,3.3 & 14.98\,$\pm$\,0.20 \\

\multicolumn{5}{c}{SDSS J1253+1007} \\
3.022167 & $-$672.5   &  C\,{\sc iv}& 60.4\,$\pm$\,2.2 & 14.95\,$\pm$\,0.05 \\
3.025153 & $-$450.0   &  C\,{\sc iv}& 15.0\,$\pm$\,2.0 & 13.75\,$\pm$\,0.11 \\
3.027819 & $-$251.5   &  C\,{\sc iv}& 40.0\,$\pm$\,5.0 & 16.20\,$\pm$\,0.13 \\
3.029663 & $-$114.3   &  C\,{\sc iv}& 35.0\,$\pm$\,5.0 & 14.10\,$\pm$\,0.15 \\
3.031250 & $+$3.7     &  C\,{\sc iv}& 33.0\,$\pm$\,2.0 & 15.70\,$\pm$\,0.10 \\
3.032776 & $+$117.2   &  C\,{\sc iv}& 30.0\,$\pm$\,3.0 & 13.30\,$\pm$\,0.10 \\
3.021977 & $-$686.7   &  Si\,{\sc iv}& 93.6\,$\pm$\,12.0 & 13.30\,$\pm$\,0.10 \\
3.027811 & $-$252.1   &  Si\,{\sc iv}& 45.8\,$\pm$\,2.0 & 14.55\,$\pm$\,0.10 \\
3.031253 & $+$4.0     &  Si\,{\sc iv}& 53.5\,$\pm$\,2.0 & 14.54\,$\pm$\,0.10 \\
3.021785 & $-$701.0   &  N\,{\sc v}& 40.2\,$\pm$\,4.0 & 15.23\,$\pm$\,0.12 \\
3.023130 & $-$600.7   &  N\,{\sc v}& 11.9\,$\pm$\,4.5 & 14.73\,$\pm$\,0.20 \\
3.027967 & $-$240.5   &  N\,{\sc v}& 39.4\,$\pm$\,2.5 & 14.80\,$\pm$\,0.10 \\

%0.000000 & $+$0.0   &  XX\,{\sc iv}& 00.0\,$\pm$\,0.0 & 00.00\,$\pm$\,0.00 \\
\hline

\end{tabular}
 \label{Table_hiion}
 \renewcommand{\footnoterule}{}
\end{table}
%\end{landscape}

\subsection{SDSS J105823.73+031524.4}

In addition to Ly$\alpha$, we detect emission from C\,{\sc iii]},
C\,{\sc iv}, and He\,{\sc ii} in the spectrum of this quasar (see
Fig.~\ref{J1058_1D}). The C\,{\sc iii]}\,$\lambda$1909 emission line
falls on top of a strong sky emission line at 6300\,$\textup{\AA}$
and hence cannot yield a reliable redshift estimate. On the other
hand, the blue wing of the He\,{\sc ii}\,$\lambda$1640 emission line
is affected by noise. However, we could still use the red wing of
the profile to carefully constrain the fit and obtain $z_{\rm
em}$~=~2.3021 which was adopted as the systemic redshift of the
quasar. Interestingly, the C\,{\sc iv} emission line indicates a
similar redshift.

The broad Ly$\alpha$ emission line of this quasar is extinguished by
the presence of a strong DLA located at $z_{\rm abs}$~=~2.2932
(derived through simultaneous fit of the low ionization species)
which is $\sim$~810~km\,s$^{-1}$ blue-shifted relative to the quasar
systemic redshift. The H\,{\sc i} column density of log\,$N_{\rm
HI}$~=~21.80~$\pm$~0.10 is derived from a single component fit to
the DLA profile (see panel~(d) of Fig.~\ref{J1058_1D}). This DLA is
associated with a number of low and high ionization species
including O\,{\sc i}, C\,{\sc ii}, Si\,{\sc ii}, Fe\,{\sc ii},
Al\,{\sc ii}, Al\,{\sc iii}, Mg\,{\sc ii}, Si\,{\sc iv} and C\,{\sc
iv} (see Fig.\,\ref{J1058_1D}). The somewhat low SNR of the spectrum
does not allow us to uniquely constrain the Doppler parameter using
several transitions of the same species. However, we still used four
Fe\,{\sc ii} transitions to derive a lower limit on this parameter,
$b>15$~km\,s$^{-1}$ (see panel~(b) of Fig.~\ref{J1058_1D}). We
derive iron and silicon abundances of [Fe/H]~$<$~$-$1.56~$\pm$~0.20
and [Si/H]~$<$~$-$1.41~$\pm$~0.20.

Independent fits to the C\,{\sc iv} and Si\,{\sc iv} absorption
lines were conducted with one main component (see
Fig.~\ref{J1058_1D}). A Gaussian fit to the Ly$\alpha$ emission line
gives $z_{\rm Ly\alpha}$~=~2.3029 which is only
$\sim$~73~km\,s$^{-1}$ redshifted with respect to the systemic
redshift of the quasar. The striking feature in Fig.~\ref{J1058_1D}
is that the DLA is $\sim$~880~km\,s$^{-1}$ blueshifted with respect
to the Ly$\alpha$ emission. This is the largest velocity separation
seen between the DLA and the narrow Ly$\alpha$ emission line in our
sample. This is also the largest separation between the DLA and the
quasar. This large velocity offset suggests that the Ly$\alpha$
emission may be associated with the quasar and not the DLA.

Although SNR is not very high, we note that the flux on the trace
derived by our decomposition is smaller along PA~=~NS compared to
PA~=~EW (see panel~(c) of Fig.~\ref{J1058_2D}). Actually, the slit
width for the two PAs are not the same: 0.8 and 1.0 arcsec for
PA~=~NS and PA~=~EW, respectively. This is because the seeing was
better for PA~=~NS. We also note that a point-like source is present
on the 2D spectrum at PA~=~EW at the velocity of
$\sim$\,50~km\,s$^{-1}$ relative to the quasar (see panel~(a) of
Fig.~\ref{J1058_2D}). We believe that this point-source is
contributing to the flux on the trace. This source is not seen along
PA~=~NS because of the better seeing and smaller slit-width.

If we assume that this point-like source of Ly$\alpha$ emission
arises from a star-forming region inside a foreground (or the host)
galaxy, we can estimate the star formation rate (SFR) implied by its
Ly$\alpha$ luminosity. Integrating over the region where this
compact emission is detected, we derive a flux of
$f$~=~3.26\,$\times$\,10$^{-17}$ erg~s$^{-1}$\,cm$^{-2}$, with
corresponding Ly$\alpha$ luminosity of
$L$~=~1.40\,$\times$\,10$^{42}$ erg~s$^{-1}$. Assuming the standard
case B recombination theory and using the relation proposed by
Kennicutt (1998), we get SFR~=~1.3~M$_{\odot}$\,yr$^{-1}$. This is a
lower limit to the SFR as, in addition to the slit loss, dust
extinction and/or absorption by foreground neutral hydrogen could
diminish the actual Ly$\alpha$ flux radiated away from the galaxy.
Moreover, Erb et al. (2006) showed that, in Lyman break galaxies
(LBGs), I(Ly$\alpha$)/I(H$\alpha$)~=~0.6. Therefore, if our
point-like source is an LBG, this implies a higher SFR of
18.6~M$_{\odot}$\,yr$^{-1}$. The velocity separation of
$\sim$~54~km\,s$^{-1}$ of this point-like source from the quasar,
along with its small impact parameter ($\sim$~0.5~arcsec or
$\sim$~4~kpc) suggests that it might be inside the zone-of-influence
of the quasar. Therefore, the observed star formation activity might
have been triggered by the strong feedback from the AGN.
Alternatively, this compact Ly$\alpha$ emission could have been
induced by the quasar through the Ly$\alpha$ fluorescence emission
(Srianand et al. 2016).

The 2D spectra of this quasar reveal extended Ly$\alpha$ emission
along both PAs (Fig.~\ref{J1058_2D}). Along PA~=~NS, the emission is
equally extended up to $\sim$\,1.8 arcsec (15.5~kpc) to the north
and south directions. A notable feature along PA~=~EW is that the
velocity spread of the emission ($\sim$~500~km\,s$^{-1}$) is similar
along the trace and at $\sim$\,1.3 arcsec (11~kpc) away from the
trace to the west. The emission is also spatially more extended to
the west, reaching out to $\sim$\,2.4 arcsec (20~kpc). The presence
of the point-like source of Ly$\alpha$ emission could in part be
responsible for the high velocity extent of the envelope
($>$\,500~km\,s$^{-1}$) towards the west.

The eclipsing DLA along this line of sight is
$\sim$~810~km\,s$^{-1}$ blueshifted with respect to $z_{\rm em}$.
The lack of N\,{\sc v} and O\,{\sc vi} absorption lines and the low
metallicity of the absorber argue against a scenario in which the
DLA is associated with outflowing interstellar medium from the
quasar host. Alternatively, the absorber may arise from a foreground
galaxy unrelated to the quasar.

\subsection{SDSS J115432.67$-$021537.9}

For this quasar, we detect emission lines from C\,{\sc
iv}\,$\lambda$1548, He\,{\sc ii}\,$\lambda$1640, Si\,{\sc
iii}]\,$\lambda$1892, and C\,{\sc iii}]\,$\lambda$1909. To estimate
the systemic redshift of the quasar, a single Gaussian is fit to the
He\,{\sc ii} emission line with $z_{\rm em}$~=~2.1810. A similar fit
to the C\,{\sc iii}] emission line gives $z$~=~2.1830. We adopt the
former as the systemic redshift of the quasar because the He\,{\sc
ii} line is a more reliable redshift indicator. Panel~(a) of
Fig.~\ref{J1154_1D} shows the emission lines and their Gaussian
fits.

Associated with the eclipsing DLA along this line of sight, we
detect metal absorption lines from Si\,{\sc ii}, Fe\,{\sc ii},
Al\,{\sc ii}, Al\,{\sc iii}, Mg\,{\sc i}, Mg\,{\sc ii}, C\,{\sc iv},
Si\,{\sc iv}, and possibly N\,{\sc v}. The spectral regions
corresponding to the O\,{\sc i}\,$\lambda$1302, C\,{\sc
ii}\,$\lambda$1334, C\,{\sc ii}$^{*}$\,$\lambda$1335, and N\,{\sc
v}\,$\lambda\lambda$1339,1402 absorption lines are strongly affected
by noise (see panel~(c) of Fig.~\ref{J1154_1D}). We use four
Fe\,{\sc ii} absorption lines to derive the redshift of the damped
Ly$\alpha$ system ($z_{\rm abs}$~=~2.1853). We also tried to
constrain the Doppler parameter using some Fe\,{\sc ii} transitions
(see panel~(b) of Fig.\,\ref{J1154_1D}). Following the same approach
as in previous sections, one can see that the best $b$-value to fit
the four Fe\,{\sc ii} transitions is $b$~$\sim$~20~km\,s$^{-1}$. We
adopted this $b$-value and subsequently fit the other absorption
lines. Panel~(c) of Fig.~\ref{J1154_1D} illustrates the observed
velocity profiles and {\sc vpfit} solutions (where applicable) of
different species. The parameters are given in
Tables~\ref{Table_lowion}~\&~\ref{Table_hiion}. A damped profile
fitted to the Ly$\alpha$ absorption line resulted in log\,$N_{\rm
HI}$~=~21.75~$\pm$~0.10. The fit is shown as a red curve in
panel~(d) of Fig.\,\ref{J1154_1D} with the two green curves showing
the $\pm$1\,$\sigma$ uncertainty in the H\,{\sc i} column density.
This DLA is 103 and 405~km\,s$^{-1}$ blueshifted compared to the
Ly$\alpha$ emission and systemic redshift of the quasar,
respectively (see panel~(a) of Fig.~\ref{J1154_1D} and
Table~\ref{redshifts}). The derived metallicity of this absorber is
[Si/H]~=~$-$1.71~$\pm$~0.20.

The extended Ly$\alpha$ emission is detected along both PAs, with
maximum extension reaching $\sim$~3.2~arcsec (27~kpc) similarly to
the east and south of the quasar as shown in Fig.~\ref{J1154_2D}. We
checked that the emission seen on the trace in the velocity range
$-$500\,$<$\,$v$\,$<$\,0~km\,s$^{-1}$ along PA~=~NS is real. This
emission, however, is barely detected along PA~=~EW, both because
the emission is intrinsically faint and the SNR of the spectrum is
lower along this PA. Overall, the spectra of the quasar trace for
the two PAs are consistent within the errors (see panel~(c) of
Fig.~\ref{J1154_2D}).

As shown in panel~(c) of Fig.~\ref{J1154_2D}, the Ly$\alpha$ profile
on the quasar trace is clearly asymmetric. Here, the Ly$\alpha$ flux
sharply rises on the blue side and slowly decreases on the red side
of the profile. Note that the blue edge of the profile (marked by a
pink arrow) is close to the systemic redshift of the quasar
($\delta$V~$\sim$~60~km\,s$^{-1}$). Remarkably, we detect a small
peak on the blue side of the profile at
$v$~$\sim$~$-$300~km\,s$^{-1}$. The presence of this peak could be
understood in the context of resonant scattering of Ly$\alpha$
photons. Simulations have shown that Ly$\alpha$ radiation should
generally escape the galaxy with a double-peaked profile (Verhamme
et al. 2006; Laursen et al. 2009; Noterdaeme et al. 2012) but
observations reveal that the blue peak is usually diminished or in
some cases completely suppressed (Fynbo et al. 2010). The presence
of strong galactic winds, absorption and dust extinction could be
responsible for the suppression of the blue peak (Verhamme et al.
2008).

On the other hand, one could also fit a Gaussian function to the
Ly$\alpha$ emission seen on the quasar trace. This Gaussian function
is shown as a green dotted curve in panel~(c) of
Fig.~\ref{J1154_2D}. One could now reproduce the observed asymmetric
Ly$\alpha$ emission profile by putting some Ly$\alpha$ absorption
(blue dotted profile in panel~(c) of Fig.~\ref{J1154_2D}) on top of
the green dotted curve. Note that the blue dotted profile, located
at $v$~$\sim$~$-$40~km\,s$^{-1}$ relative to the quasar, is
$\sim$~150~km\,s$^{-1}$ blueshifted with respect to the green dotted
curve. This is similar to what is seen in Lyman break galaxies which
typically exhibit redshifted emission and blueshifted absorption in
Ly$\alpha$ (Shapley et al. 2003). No low-ion is associated with this
Ly$\alpha$ absorption while the C\,{\sc iv}, Si\,{\sc iv} and
possibly N\,{\sc v} absorption that are seen at
$v$~$\sim$~$-$635~km\,s$^{-1}$ could probably be partly associated
with it.

\subsection{SDSS J125302.00+100742.3}

The quasar has the best SNR spectrum in our sample. Unfortunately,
the O\,{\sc i}\,$\lambda$1304 and He\,{\sc ii}\,$\lambda$1640
emission lines are not detected and the C\,{\sc iii}]\,$\lambda$1909
emission line region is heavily affected by sky emission and
telluric absorption lines. However, we could still fit Gaussian
functions to the C\,{\sc iv} and Si\,{\sc iv} emission lines to get
$z$~=~3.0150 and 3.0170, respectively. In the analysis of this
system, we adopt $z_{\rm em}$~=~3.0150 as the systemic redshift of
the quasar. The emission lines and their Gaussian fits are shown in
panel~(a) of Fig.~\ref{J1253_1D}.

Metal absorption lines from O\,{\sc i}, C\,{\sc ii}, C\,{\sc
ii}$^{*}$, Fe\,{\sc ii}, Si\,{\sc ii}, Si\,{\sc ii}$^{*}$, Al\,{\sc
ii}, Al\,{\sc iii}, C\,{\sc iv}, and Si\,{\sc iv} are detected in
the eclipsing DLA along this line of sight (see
Fig.~\ref{J1253_1D}). A single component fit with $z_{\rm
abs}$~=~3.0312 and $b$~=~37.0~km\,s$^{-1}$ was conducted on the
low-ion species. The redshift and $b$-value were determined through
a simultaneous fit to the five Si\,{\sc ii} transitions (i.e.
Si\,{\sc ii}\,$\lambda$1190, $\lambda$1193, $\lambda$1304,
$\lambda$1526, and $\lambda$1808). It can be seen from panel~(b) of
Fig.~\ref{J1253_1D} that Doppler parameters less than
37~km\,s$^{-1}$ overpredict the observed Si\,{\sc ii}\,$\lambda$1808
optical depth. The parameters of the fit to the low and high ion
species are reported in
Tables~\ref{Table_lowion}~\&~\ref{Table_hiion}. The H\,{\sc i}
column density of log\,$N_{\rm HI}$~=~21.30~$\pm$~0.10 was
determined by fitting damping profiles on the Ly$\alpha$ and
Ly$\beta$ absorption lines, with the redshift fixed to that of the
low-ions (see panel~(d) of Fig.~\ref{J1253_1D}). By fitting the five
detected Si\,{\sc ii} transitions we derive:
[Si/H]~=~$-$1.45~$\pm$~0.20. We also tentatively derive
[C/H]~=~$-$1.61~$\pm$~0.20 and [O/H]~$>$~$-$1.95. From the Fe\,{\sc
ii}\,$\lambda$\,1608 absorption we derive
[Fe/H]~=~$-$2.53~$\pm$~0.20 indicating significant depletion of this
element on to dust.

While low-ion absorption profiles are dominated by a single
component at $v$~=~0~km\,s$^{-1}$ ($z_{\rm abs}$~=~3.0312), high-ion
species are mainly seen in three distinct components, namely at
$v$~=~0, $-$200, and $-620$~km\,s$^{-1}$ (see panel~(c) of
Fig.~\ref{J1253_1D}). Out of these three components, N\,{\sc v} and
O\,{\sc vi} are only seen at $v$~=~$-$620 and $-$200~km\,s$^{-1}$,
with no N\,{\sc v} associated with the DLA at $v$~=~0 km\,s$^{-1}$.
Note that N\,{\sc v} absorption is the strongest at
$v$~=~$-$620~km\,s$^{-1}$; this is also the case for C\,{\sc iv} and
O\,{\sc vi} but not Si\,{\sc iv} as it has the weakest absorption at
this velocity. This clearly shows that the gas is progressively
getting more ionized as we go to higher velocities relative to the
DLA. This gas must be located close to the AGN as indicated as well
by the presence of Si\,{\sc ii}$^{*}$ absorption (see Paper I). It
is interesting to note that Ly$\alpha$ emission on the trace is
detected over a large velocity range ($\sim$~1500~km\,s$^{-1}$)
which covers all these absorption components.

In addition to the narrow Ly$\alpha$ emission line that is detected
in the DLA trough, it seems that its corresponding narrow Ly$\beta$
emission line is also detected in the trough of the damped Ly$\beta$
absorption profile with the flux ratio of
$I$(Ly$\alpha$)/$I$(Ly$\beta$)~$<$~9.2, and the integrated flux of
$f_{\rm Ly
\beta}$~=~(1.8~$\pm$~0.4)~$\times$~10$^{-17}$~erg~s$^{-1}$\,cm$^{-2}$
(see Fig.~\ref{Ly_beta}). This ratio is for the strongest component
of the Ly$\alpha$ emission seen in the range
$-$400~$<$~$v$~$<$~400~km\,s$^{-1}$. For the whole Ly$\alpha$
emission, extending from $-$1100 to 400~km\,s$^{-1}$, this ratio is
$<$~14.6.  Note that for the temperature range of 100~$\le$~$T_{\rm
e}$~$\le$20\,000 this ratio should be in the range
9.5~$\le$~$I$(Ly$\alpha$)/$I$(Ly$\beta$)~$\le$~4.0 (Martin 1988).
Jiang et al. (2016) also reported the possible detection of
Ly$\beta$ emission in the Ly$\beta$ absorption trough of a PDLA
found towards the quasar SDSS~J0952+0114.

As shown in panels~(a) and (b) of Fig.~\ref{J1253_2D}, the
Ly$\alpha$ emission is spatially extended along both PAs. Maximum
extension is found to the north reaching out to $\sim$~3.7~arcsec
(30~kpc) from the nucleus emission. The velocity spread of the
Ly$\alpha$ emission on the quasar trace is $\sim$~1560~km\,s$^{-1}$
which is the largest in our quasar sample (see Table~\ref{redshifts}
column 8). The kinematic of the extended envelope towards the east
and up to a distance of 1.5~arcsec (12~kpc) is remarkably similar to
that of the trace (see panel~(b) of Fig.~\ref{J1253_2D}). The
velocity spread of the emission towards the north is
$\sim$~500~km\,s$^{-1}$ while it is $\lesssim$~300~km\,s$^{-1}$ to
the south and west direction. Note that the spectrum along the
east-west direction has a lower SNR because its exposure time is
$\sim$~25\,\% shorter than that of the other PA. However, the
emission on the quasar trace for the two PAs are still consistent
within the errors (see panel~(c) of Fig.~\ref{J1253_2D}).

The emission on the quasar trace has two main components at
100~$\lesssim$~$v$~$\lesssim$~800 and
900~$\lesssim$~$v$~$\lesssim$~1600~km\,s$^{-1}$ (see panel~(a) of
Fig.~\ref{J1253_2D}). The velocity difference between the center of
the two components is $\sim$~700~km\,s$^{-1}$. We note that the
velocity separation between the DLA and the stronger component of
the Ly$\alpha$ emission is $\sim$~30~km\,s$^{-1}$, possibly
suggesting they originate in the same region. Note that the velocity
separation between the Ly$\alpha$ emission and the quasar itself is
not robust as the quasar redshift is derived from the C\,{\sc iv}
emission line.

\section{Results and discussion}

The Ly$\alpha$ emission seen in the trough of the eclipsing DLAs in
our sample are all strong and spatially extended along the two
observed PAs. The observed properties of the Ly$\alpha$ emission are
summarized in Tables~\ref{flux_extent} and \ref{flux_luminosity}.
The values in Table~\ref{flux_luminosity} are corrected for the
Galactic extinction (using the dust map of Schlegel et al. 1998) and
the slit loss effect. The slit loss effect was corrected by assuming
that the Ly$\alpha$ nebulae are uniform face-on ellipses with major
and minor diameters equal to their extent along the two PAs. We then
estimated the amount of flux missed by the slit and consequently
corrected the observed luminosities.

%************************ z_J0112 *********************
\begin{figure*}
\centering
\begin{tabular}{c c}
\includegraphics[bb=38 360 580 611,clip=,width=0.70\hsize]{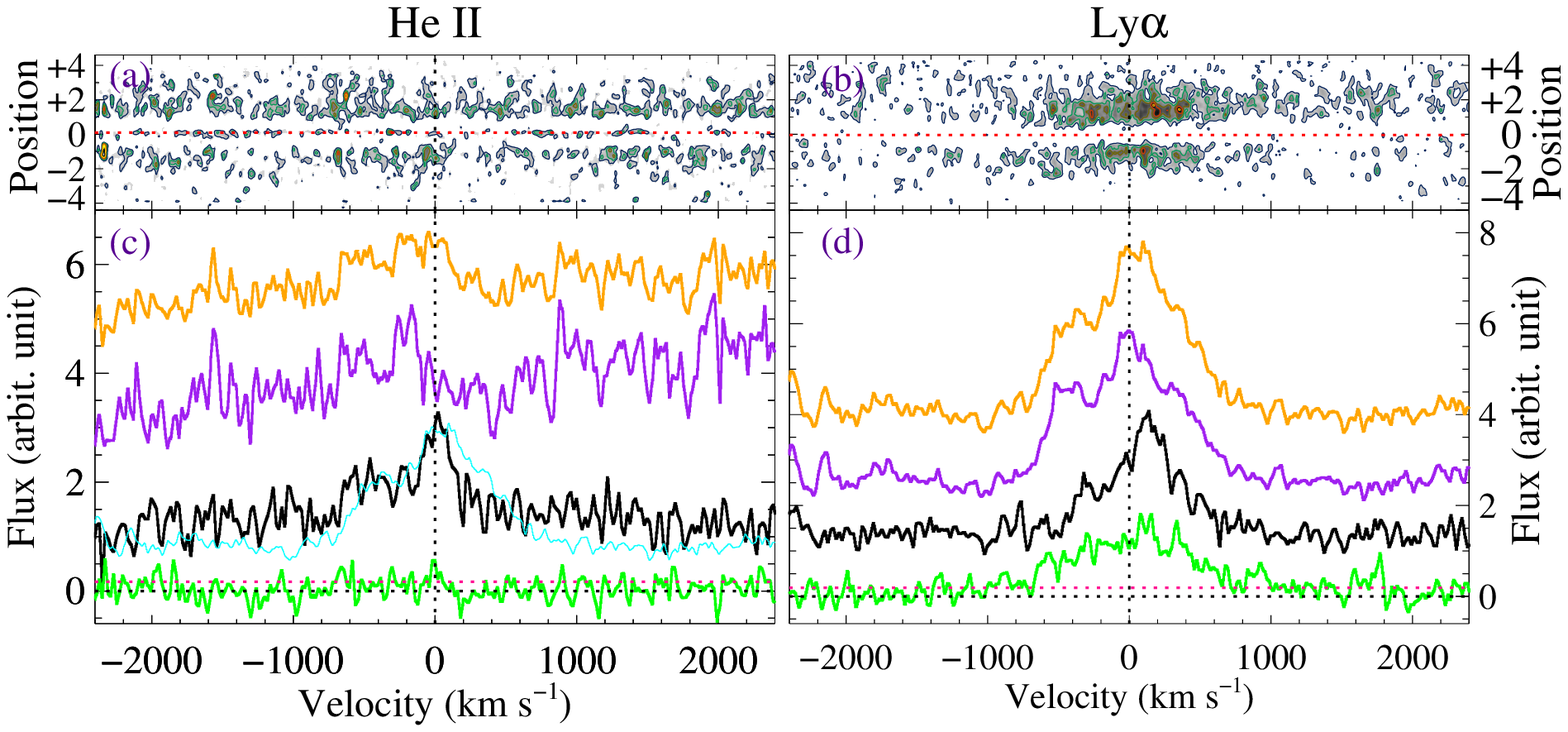} \\
\includegraphics[bb=40 398 575 548,clip=,width=0.70\hsize]{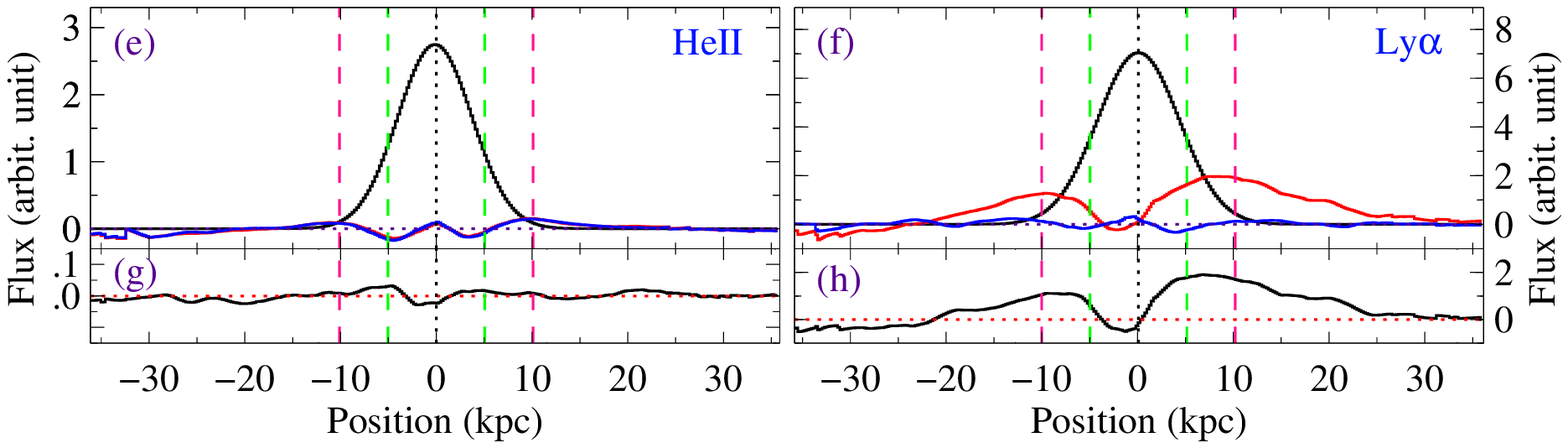}
\end{tabular}
\caption{{\sl Panels~(a) and (b)}: Stacked 2D spectra around the
He\,{\sc ii} and Ly$\alpha$ spectral regions after subtracting the
emission on the trace. The red dotted horizontal lines mark the
center of the trace. The y-axis is position in arcsecond. {\sl
Panels~(c) and (d)}: The green curves show the intensity profiles of
the He\,{\sc ii} and Ly$\alpha$ emission lines integrated over the
regions outside the quasar's trace. Actually, the green curves are
the 1D profiles of the 2D spectra shown in panels~(a) \& (b)
integrated over the y-axis. The black curves are the stacked 1D
spectra of the three quasars (i.e. J0112$-$0048, J1058$+$0315, and
J1154$-$0215) for which the He\,{\sc ii} emission line is detected
in the quasar spectrum on the trace. The purple curves are the
stacked 1D spectra of the three quasars (i.e. J0823$+$0529,
J0953$+$0349, and J1253$+$1007) with uncertain He\,{\sc ii}
detection on the quasar trace. The orange curves are the stacked 1D
spectra of all six quasars. For the sake of comparison, the stacked
Ly$\alpha$ emission line profile (scaled arbitrarily) is also
overplotted as a cyan curve in panel~(c). The horizontal pink dotted
lines mark the 1\,$\sigma$ uncertainty in the flux. The origin of
the velocity plots is at $z$~=~2.5742 which is the mean redshift of
all six quasars. Note that, for the sake of better illustration, the
black, purple, and orange profiles are arbitrarily shifted along the
y-axis. {\sl Panels~(e) and (f)}: the black curve shows the emission
in the quasar's PSF centered on the trace, integrated over the
spectral width of the He\,{\sc ii} (resp. Ly$\alpha$) emission line.
The red curve is the intensity profile of the same region but after
subtracting the quasar trace and the blue curve is the control
profile. The control profile is integrated across a wavelength range
of the same width as the red and black curves but centered
100\,\textup{\AA} redder (resp. bluer) than the He\,{\sc ii} (resp.
Ly$\alpha$) emission line. Note that the mean flux of the control
region outside the quasar trace was subtracted from the green curve
in panel~(c) to correct for the non-zero residual flux (see the
text). The green and pink vertical dashed lines show the FWHM and
full width of the quasar PSF, respectively. {\sl Panels~(g) and
(h)}: The black curves show the difference between the red and blue
curves in panels~(e) \& (f).}
 \label{heii_stack}
\end{figure*}
%*******************************************************

\subsection{He\,{\sc ii} emission line}

Since it is possible that the extended Ly$\alpha$ emission seen
around quasars might arise from the resonant scattering of
Ly$\alpha$ photons from the central AGN, observation of other lines
with non-resonant characteristic, like He\,{\sc ii}\,$\lambda$1640,
is needed to constrain the importance of this effect. If the
observed extension of the Ly$\alpha$ envelope results from resonant
scattering, then the corresponding He\,{\sc ii} emission should not
be extended. Similarly, detections of spatially extended He\,{\sc
ii} and Ly$\alpha$ emission on the same scale would imply that
resonant scattering of Ly$\alpha$ photons is not important.

The He\,{\sc ii}\,$\lambda$1640 emission line is detected in the spectra
of three of our quasars (i.e. J0112$-$0048, J1058$+$0315 and
J1154$-$0215). The spectra of the remaining quasars (i.e.
J0823$+$0529, J0953$+$0349, and J1253$+$1007) also show some weak
emission features near the expected position of the He\,{\sc ii}
emission line. However, the low SNR of the latter spectra makes
the corresponding detections highly uncertain.

We calculated the integrated flux in the expected extended He\,{\sc
ii} emission line region outside the quasar trace as the sum of the
flux in all He\,{\sc ii} pixels of the 2D spectra corresponding to
the detected Ly$\alpha$ extended region. The result is listed in the
seventh column of Table~\ref{flux_extent}. We calculated the
corresponding 3\,$\sigma$ detection limit as three times the square
root of the sum of the variance of the flux for the same pixels. The
result is reported in the eighth column of Table~\ref{flux_extent}.
By comparing the values in the seventh and eighth column of
Table~\ref{flux_extent} we can see that no extended He\,{\sc ii}
emission is detected above the 3\,$\sigma$ detection limit outside
the quasar trace in any of our objects.

We also investigate the possible spatial extension of the He\,{\sc
ii} emission by co-adding and comparing the 2D spectra around the
position of the He\,{\sc ii}\,$\lambda$\,1640 and Ly$\alpha$
emission lines (see Fig.\,\ref{heii_stack}). Although Ly$\alpha$
emission is very extended off the quasar trace, its corresponding
extended He\,{\sc ii} emission is apparently not detected in the
final stacked spectrum. To quantify this, we integrated the 2D
spectra along the spatial direction and compared the result to the
background noise level. The result is shown as green curves in
panels~(c) and (d) of Fig.\,\ref{heii_stack} where the 1\,$\sigma$
background noise level is marked as horizontal pink dotted lines.
Integrating the green curve over the velocity range
$-500$~$\le$~$v$~$\le$~500 results in $f_{\rm
HeII}$~=~(1.09~$\pm$~0.89)~$\times$~10$^{-17}$
erg~s$^{-1}$\,cm$^{-2}$ which is consistent with no detection.

We also integrated the stacked 2D spectra along the spectral
direction and the result is shown as red curves in panels~(e) and
(f) of Fig.~\ref{heii_stack}. For the He\,{\sc ii} (resp.
Ly$\alpha$) flux, the integration was performed over the velocity
range $-$700~$<$~$v$~$<$~+200~km\,s$^{-1}$ (resp.
$-$800~$<$~$v$~$<$~+600~km\,s$^{-1}$). The black curves show the
emission in the PSF centered on the trace. The blue curves are the
control profiles which are integrated across a wavelength range of
the same width as the red lines but $\sim$\,100~\textup{\AA} away
from the He\,{\sc ii} and Ly$\alpha$ emission lines center. The
difference between the red and blue curves is shown in panels~(g)
and (h) of Fig.\,\ref{heii_stack}. The profile in panel~(g) is
consistent with no He\,{\sc ii} emission detection outside the
quasar trace at the limit of our observations.

All this may imply that the extended Ly$\alpha$ emission seen around
these quasars has predominantly resonant-scattering origin, or the
ratio of the He\,{\sc ii}/Ly$\alpha$ emission is much smaller than
what is usually seen in quasars, radio galaxies and/or Lyman alpha
blobs (Lehnert \& Becker 1998; Arrigoni Battaia et al. 2015). If one
assumes that the expected He\,{\sc ii} to Ly$\alpha$ flux ratio is
$\sim$\,0.1 which is the typical value observed in Ly$\alpha$ blobs
and radio galaxies (Villar-Mart\'in et al. 2007; Prescott et al.
2013; Arrigoni Battaia et al. 2015), then one would expect to detect
He\,{\sc ii} emission at $\sim$~4.5\,$\sigma$ towards the two
quasars J0112$-$0048 (along the EW direction) and J0823+0529 (along
both PAs). The expected He\,{\sc ii} emission flux towards the
remaining quasars would still be below the 3\,$\sigma$ detection
limit. The fact that no He\,{\sc ii} emission is detected above the
3\,$\sigma$ detection limit towards J0112$-$0048 and J0823+0529
suggests that the He\,{\sc ii}/Ly$\alpha$ ratio towards these two
quasars is below 0.1, and thus argues for resonant scattering of
Ly$\alpha$ photons being a dominant process in these quasars.

%************************ Table 1 *********************
\begin{landscape}
 \begin{table}
   \begin{threeparttable}[b]
 \centering
\caption{Derived parameters of emission lines. First column: slit
position angles. Second column: spatial extent of the Ly$\alpha$
emission (i.e. the diameter) of the envelope in both arcsecond and
kpc. Third column: flux of the unresolved region which is seen on
the quasar trace. Fourth column: flux of the extended Ly$\alpha$
envelope which is seen beyond the quasar trace. Fifth column:
3\,$\sigma$ detection limit (DL) of the Ly$\alpha$ emission envelope
outside the quasar trace. Sixth column: 1\,$\sigma$ uncertainty in
the Ly$\alpha$ surface brightness. Seventh column: integrated flux
over the expected He\,{\sc ii} emission line region outside the
quasar trace. Eighth column: 3\,$\sigma$ detection limit of the
He\,{\sc ii} emission envelope outside the quasar trace. Ninth
column: 3\,$\sigma$ upper limit He\,{\sc ii}/Ly$\alpha$ ratio
determined from the values given in the eighth and fourth columns. }
 \setlength{\tabcolsep}{9.5pt}
\renewcommand{\arraystretch}{1.3}
\begin{tabular}{c c c c c c c c c}
\hline

 PA    & Extent & NEL flux(Ly$\alpha$) & Envlp. flux(Ly$\alpha$) & 3\,$\sigma$~DL$_{\rm Ly\alpha}$\tnote{a} & 1\,$\sigma$ uncertainty\tnote{b}  & Envlp. flux(HeII)\tnote{c} & 3\,$\sigma$~DL$_{\rm HeII}$\tnote{c} & Flux ratio\\
     & ("; kpc) & [erg~s$^{-1}$\,cm$^{-2}$] & [erg~s$^{-1}$\,cm$^{-2}$] & [erg~s$^{-1}$\,cm$^{-2}$] & [erg~s$^{-1}$\,cm$^{-2}$\,arcsec$^{-2}$]   & [erg~s$^{-1}$\,cm$^{-2}$]  & [erg~s$^{-1}$\,cm$^{-2}$]  &  (HeII/Ly$\alpha$)\\
\hline
\multicolumn{9}{c}{SDSS J0112$-$0048} \\
NS &  (5.1; 43) & 2.54($\pm$\,0.10)\,$\times$\,10$^{-16}$ & 1.04($\pm$\,0.12)\,$\times$\,10$^{-16}$ &  4.46\,$\times$\,10$^{-17}$ & 0.92\,$\times$\,10$^{-17}$  & 1.65\,$\times$\,10$^{-18}$ & 2.01\,$\times$\,10$^{-17}$ & $<$\,0.19 \\
EW &  (4.9; 42) & 2.46($\pm$\,0.37)\,$\times$\,10$^{-16}$ & 2.53($\pm$\,0.44)\,$\times$\,10$^{-16}$ &  1.36\,$\times$\,10$^{-16}$ & 3.55\,$\times$\,10$^{-17}$  & 6.42\,$\times$\,10$^{-18}$ & 1.69\,$\times$\,10$^{-17}$ & $<$\,0.07\\

\multicolumn{9}{c}{SDSS J0823$+$0529} \\
NS &  (5.5; 42) & 3.59($\pm$\,0.10)\,$\times$\,10$^{-16}$ & 3.97($\pm$\,0.10)\,$\times$\,10$^{-16}$ &  4.05\,$\times$\,10$^{-17}$ & 0.74\,$\times$\,10$^{-17}$  & 1.22\,$\times$\,10$^{-17}$ & 2.21\,$\times$\,10$^{-17}$ & $<$\,0.06 \\
EW &  (6.0; 46) & 3.36($\pm$\,0.10)\,$\times$\,10$^{-16}$ & 2.68($\pm$\,0.10)\,$\times$\,10$^{-16}$ &  4.15\,$\times$\,10$^{-17}$ & 0.92\,$\times$\,10$^{-17}$  & 1.24\,$\times$\,10$^{-17}$ & 2.04\,$\times$\,10$^{-17}$ & $<$\,0.08 \\

\multicolumn{9}{c}{SDSS J0953+0349} \\
NS &  (3.6; 30) & 1.27($\pm$\,0.12)\,$\times$\,10$^{-16}$ & 0.52($\pm$\,0.11)\,$\times$\,10$^{-16}$ &  4.08\,$\times$\,10$^{-17}$ & 0.93\,$\times$\,10$^{-17}$  & 8.47\,$\times$\,10$^{-18}$ & 3.73\,$\times$\,10$^{-17}$ & $<$\,0.71 \\
EW &  (3.1; 26) & 1.22($\pm$\,0.10)\,$\times$\,10$^{-16}$ & 0.54($\pm$\,0.10)\,$\times$\,10$^{-16}$ &  3.41\,$\times$\,10$^{-17}$ & 1.15\,$\times$\,10$^{-17}$  & 6.70\,$\times$\,10$^{-18}$ & 3.06\,$\times$\,10$^{-17}$ & $<$\,0.57 \\

\multicolumn{9}{c}{SDSS J1058+0315} \\
NS &  (3.7; 31) & 0.70($\pm$\,0.19)\,$\times$\,10$^{-16}$ & 0.65($\pm$\,0.20)\,$\times$\,10$^{-16}$ &  4.39\,$\times$\,10$^{-17}$ & 1.59\,$\times$\,10$^{-17}$  & 7.82\,$\times$\,10$^{-18}$ & 1.78\,$\times$\,10$^{-17}$ & $<$\,0.27 \\
EW &  (4.1; 35) & 1.11($\pm$\,0.18)\,$\times$\,10$^{-16}$ & 1.15($\pm$\,0.17)\,$\times$\,10$^{-16}$ &  5.42\,$\times$\,10$^{-17}$ & 1.80\,$\times$\,10$^{-17}$  & 7.35\,$\times$\,10$^{-18}$ & 2.18\,$\times$\,10$^{-17}$ & $<$\,0.19 \\

\multicolumn{9}{c}{SDSS J1154$-$0215} \\
NS &  (6.0; 51) & 1.19($\pm$\,0.10)\,$\times$\,10$^{-16}$ & 1.01($\pm$\,0.10)\,$\times$\,10$^{-16}$ &  3.91\,$\times$\,10$^{-17}$ & 0.53\,$\times$\,10$^{-17}$  & 6.24\,$\times$\,10$^{-18}$ & 2.11\,$\times$\,10$^{-17}$ & $<$\,0.21 \\
EW &  (5.4; 46) & 1.23($\pm$\,0.15)\,$\times$\,10$^{-16}$ & 1.82($\pm$\,0.15)\,$\times$\,10$^{-16}$ &  4.51\,$\times$\,10$^{-17}$ & 1.03\,$\times$\,10$^{-17}$  & 3.47\,$\times$\,10$^{-18}$ & 2.22\,$\times$\,10$^{-17}$ & $<$\,0.12 \\

\multicolumn{9}{c}{SDSS J1253+1007} \\
NS &  (6.4; 51) & 2.66($\pm$\,0.18)\,$\times$\,10$^{-16}$ & 2.26($\pm$\,0.21)\,$\times$\,10$^{-16}$ &  5.93\,$\times$\,10$^{-17}$ & 1.32\,$\times$\,10$^{-17}$  & 1.75\,$\times$\,10$^{-17}$ & 4.07\,$\times$\,10$^{-17}$ & $<$\,0.18 \\
EW &  (4.9; 39) & 2.62($\pm$\,0.15)\,$\times$\,10$^{-16}$ & 2.34($\pm$\,0.18)\,$\times$\,10$^{-16}$ &  6.23\,$\times$\,10$^{-17}$ & 1.77\,$\times$\,10$^{-17}$  & 2.37\,$\times$\,10$^{-17}$ & 4.45\,$\times$\,10$^{-17}$ & $<$\,0.19 \\
\hline

\end{tabular}
\label{flux_extent}
\begin{tablenotes}
\item [a] The 3$\,\sigma$ detection limit is calculated from the standard deviation (in the flux) of the spectral and spatial pixels over which the Ly$\alpha$ emission outside the quasar trace is detected.
\item [b] The 1\,$\sigma$ uncertainty in the flux is used to determine the spatial extent of the Ly$\alpha$ envelope.
\item [c] Since the He\,{\sc ii} emission envelopes are apparently not detected outside the quasar trace in
all our spectra, the integration is performed over the same spectral
and spatial extent as the corresponding Ly$\alpha$ emission
envelope.
%\item [d] This is the ratio of the values given in the seventh and fourth columns.
\end{tablenotes}
\end{threeparttable}
\renewcommand{\footnoterule}{}
\end{table}
\end{landscape}
%******************************************************

%************************ Table 1 *********************
%\begin{landscape}
 \begin{table*}
 \begin{threeparttable}[b]
 \centering
\caption{ The Ly$\alpha$ flux (erg~s$^{-1}$\,cm$^{-2}$), luminosity
(erg~s$^{-1}$) and surface brightness (SB;
erg~s$^{-1}$\,cm$^{-2}$~arcsec$^{-2}$) of the envelopes detected
around the quasars in the sample. Note that the flux and luminosity
of the extended envelopes are corrected for the Galactic extinction
and the slit-clipping effects (see Text). First row: the ID of the
quasars in the sample. Second row: The integrated Ly$\alpha$ flux in
the quasars' PSF centered on the trace (i.e. the NEL unresolved
luminosity). Third row: the integrated flux in the extended
envelopes. Fourth row: the integrated luminosity of the quasars NEL.
Fifth row: the luminosity of the extended envelopes. Sixth row: the
ionizing luminosity of the quasars, i.e. $L_{\rm <912}$. Seventh
row: PCA prediction of the Ly$\alpha$ luminosity of the quasars in
the BELR (note that this luminosity is measured in the rest-frame
wavelength interval 1200$-$1230\,\textup{\AA}). Eighth row: the mean
surface brightness of the envelopes. }
 \setlength{\tabcolsep}{12.4pt}
\renewcommand{\arraystretch}{1.3}
\begin{tabular}{c c c c c c c}
\hline
    ID        & J0112$-$0048   & J0823$+$0529 & J0953$+$0349 & J1058$+$0315 & J1154$-$0215 & J1253$+$1007 \\
\hline
  $f_{\rm NEL}$($\times$10$^{-16}$) & 2.82\,$\pm$\,0.22 & 7.57\,$\pm$\,0.20 & 1.43\,$\pm$\,0.10 & 0.83\,$\pm$\,0.16 &  1.35\,$\pm$\,0.10 & 2.80\,$\pm$\,0.12 \\
  $f_{\rm envlp}$($\times$10$^{-16}$) & 8.83\,$\pm$\,1.14 & 18.77\,$\pm$\,0.50 & 2.35\,$\pm$\,0.33 & 3.86\,$\pm$\,0.56 &  6.71\,$\pm$\,0.41 & 10.88\,$\pm$\,0.67 \\
  $L_{\rm NEL}$($\times$10$^{43}$) & 1.03\,$\pm$\,0.10 & 3.61\,$\pm$\,0.19 & 0.83\,$\pm$\,0.10 & 0.35\,$\pm$\,0.10 &  0.51\,$\pm$\,0.10 & 2.36\,$\pm$\,0.10 \\
  $L_{\rm envlp}$($\times$10$^{43}$) & 3.21\,$\pm$\,0.41 &17.88\,$\pm$\,0.49 & 1.36\,$\pm$\,0.19 & 1.65\,$\pm$\,0.24 &  2.54\,$\pm$\,0.15 & 9.18\,$\pm$\,0.56 \\
  $L_{\rm <912}$($\times$10$^{44}$)  &        2.12       &        3.98       &        6.51       &        2.31       &         1.71       &        8.05       \\
  $L_{\rm BLR}$($\times$10$^{44}$)   &        0.60       &        6.78       &        2.69       &        1.12       &         2.19       &        2.75       \\
   SB($\times$10$^{-17}$) & 4.00\,$\pm$\,0.50 & 5.95\,$\pm$\,0.16 & 1.91\,$\pm$\,0.27 & 3.07\,$\pm$\,0.45 &  2.84\,$\pm$\,0.17 & 4.71\,$\pm$\,0.29 \\
\hline
\end{tabular}
 \label{flux_luminosity}
%\begin{tablenotes}
%            \item [a] Doppler parameter (km\,s$^{-1}$).
%        \end{tablenotes}
        \end{threeparttable}
 \renewcommand{\footnoterule}{}
\end{table*}
%\end{landscape}

Foltz et al. (1988) and Heckman et al. (1991b) found that the quasar
He\,{\sc ii} emission flux is $\sim$\,30\,\% of that of the
Ly$\alpha$ envelope while in our case, it is $\sim$\,5$-$10\,\%. The
discrepancy could in part be due to the uncertain redshift of the
Ly$\alpha$ envelope in Heckman et al. (1991a) narrow band images,
which led to $\sim$\,30\,\% of uncertainty in their Ly$\alpha$
fluxes. The uncertain redshift of the envelope makes it difficult to
precisely place the Ly$\alpha$ emission within the bandpass of an
interference filter. Moreover, in Heckman et al. (1991a) sample, in
contrast to ours, the Ly$\alpha$ emission from the BLR of the quasar
is not extinguished by the presence of a strong eclipsing DLA.
Therefore, de-convolution of the strong broad Ly$\alpha$ emission
could lead to the underestimation of the Ly$\alpha$ flux of the
envelope by over-subtracting the emission close to the central
region. In our sample, this effect is minimum.

%************************ Z_vs_SB *********************
\begin{figure}
\centering
\begin{tabular}{c}
\includegraphics[bb=98 398 458 617,clip=,width=0.95\hsize]{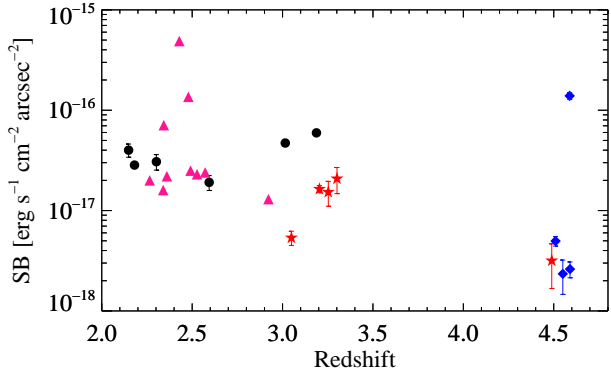}
\end{tabular}
\caption{Average surface brightness of the Ly$\alpha$ envelopes as a
function of redshift for our sample (black filled circles), together
with the CJ06 (red filled stars), NC12 (blue filled squares) and
Villar-Mart\'in et al. (2003, pink triangles) objects, all corrected
for the redshift dimming effect. The mean redshift of our quasars
(i.e. $z$~$\sim$~2.5) is taken as the reference redshift to correct
for this effect.}
 \label{Z_vs_SB}
\end{figure}
%*******************************************************

%************************ Z_vs_L *********************
\begin{figure}
\centering
\begin{tabular}{c}
\includegraphics[bb=112 398 458 613,clip=,width=0.95\hsize]{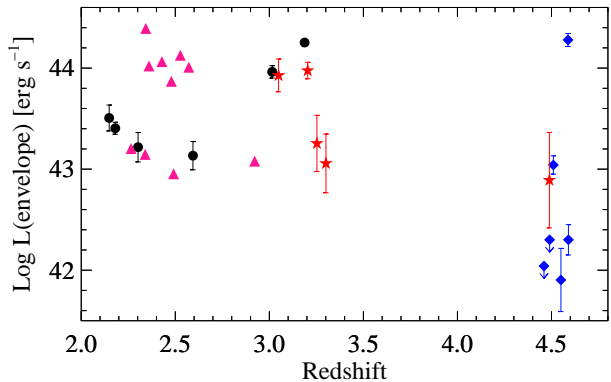}
\end{tabular}
\caption{Ly$\alpha$ luminosity of the envelopes versus quasar
redshift. The symbols represent the same samples as in
Fig.\ref{Z_vs_SB}.}
 \label{Z_vs_L}
\end{figure}
%*******************************************************

%************************ Tr_vs_env *********************
\begin{figure}
\centering
\begin{tabular}{c}
\includegraphics[bb=112 395 459 613,clip=,width=0.95\hsize]{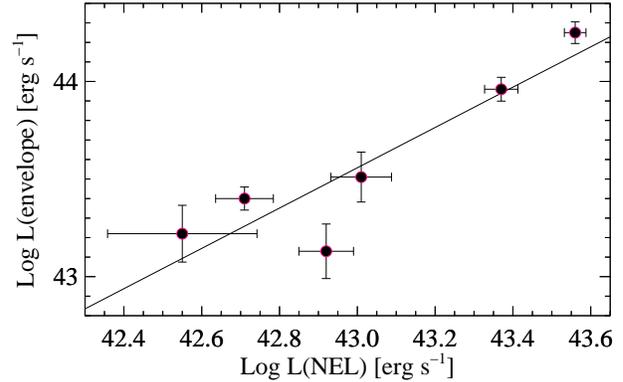}
\end{tabular}
\caption{Luminosity of the extended Ly$\alpha$ envelopes versus that
of the unresolved NEL for our sample. The black line is the fit to all points.}
 \label{Tr_vs_env}
\end{figure}
%*******************************************************

%************************ BLR_vs_env *********************
\begin{figure}
\centering
\begin{tabular}{c}
\includegraphics[bb=112 389 458 613,clip=,width=0.95\hsize]{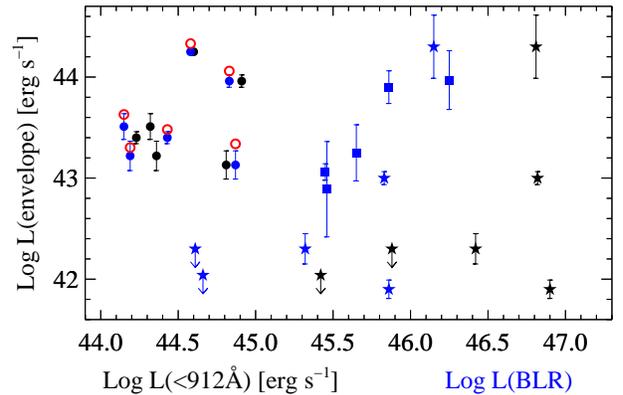}
\end{tabular}
\caption{Luminosity of the Ly$\alpha$ envelope versus that of the
Ly$\alpha$ BLR (blue symbols) and quasar ionizing radiation (black
symbols) for the NC12 (stars), CJ06 (squares) and our sample (filled
circles). The red circles show the total observed Ly$\alpha$
luminosity as a function of that of the expected Ly$\alpha$ BLR in
our sample.}
 \label{BLR_vs_env}
\end{figure}
%*******************************************************

%************************ L_vs_Ext *********************
\begin{figure}
\centering
\begin{tabular}{c}
\includegraphics[bb=102 395 458 614,clip=,width=0.95\hsize]{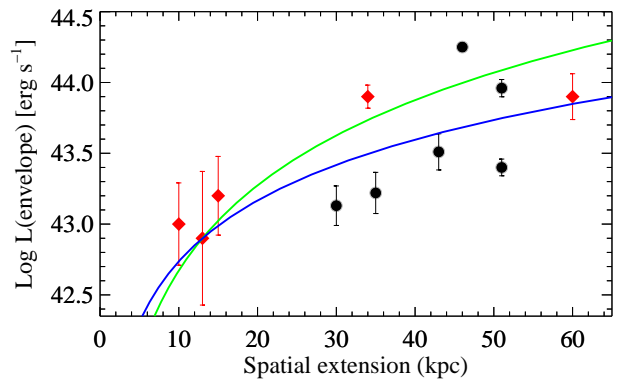}
\end{tabular}
\caption{Luminosity of the Ly$\alpha$ envelopes as a function of
their spatial extension for the CJ06 (red squares) and our sample
(black filled circles). The green and blue curves show power law
functions with index $\alpha$~=~0.5 and 0.7, respectively (see the text).}
 \label{L_vs_Ext}
\end{figure}
%*******************************************************

%************************ dvLya_dvCIV *********************
\begin{figure}
\centering
\begin{tabular}{c}
\includegraphics[bb=116 396 458 614,clip=,width=0.95\hsize]{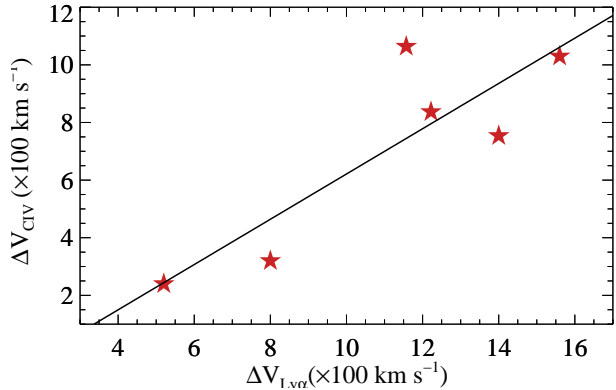}
\end{tabular}
\caption{Relation between the velocity extent of the Ly$\alpha$
emission and that of the C\,{\sc iv} absorption profile in our
sample. The black line is a fit to all data points.}
 \label{dv_dv}
\end{figure}
%*******************************************************

\subsection{Redshift dependence}

NC12 suggested that quasars at higher redshifts tend to have fainter
envelopes. In Fig.~\ref{Z_vs_SB}, we plot the average surface
brightness of the Ly$\alpha$ envelopes as a function of redshift for
our sample (black filled circles), together with the CJ06 (red
filled stars), NC12 (blue filled squares) and Villar-Mart\'in et al.
(2003, pink triangles) objects, all corrected for redshift dimming
effect. The envelopes in our sample are $\gtrsim$\,10 times brighter
than those at redshift $z\sim 4$. The surface brightness seems to
steadily decrease with redshift.

Figure~\ref{Z_vs_L} presents the luminosity of the Ly$\alpha$
envelopes versus redshift. The symbols are the same as in
Fig.~\ref{Z_vs_SB}. The luminosity of the Ly$\alpha$ envelopes in
our sample are almost the same as those in Villar-Mart\'in et al.
(2003) and CJ06. Our brightest envelope has the same total
luminosity as the brightest object in NC12 while their faintest
object is $\sim$\,17 times less luminous than our faintest one.

NC12 argued that their lower surface brightness may be due to the
deeper sensitivity of their observation which led to the more
extended envelopes compared to those of CJ06. The mean size of the
envelopes in the NC12, CJ06 and our sample are 45, 26.4 and 41~kpc,
respectively. The mean detection limit of the NC12 observations is
$\sim$~2.5~$\times$~10$^{-17}$~erg~s$^{-1}$\,cm$^{-2}$ (after taking
into account the redshift dimming effect) while in our case it is
$\sim$~4.7~$\times$~10$^{-17}$~erg~s$^{-1}$\,cm$^{-2}$. This
indicates that the large extent and higher surface brightness of our
envelopes cannot be attributed to the depth of the observation. Our
envelopes, despite having less depth, still are almost as extended
as the NC12's envelopes. This further strengthens the idea put forth
by NC12 that higher redshift envelopes may intrinsically be fainter
than their lower redshift descendants.

We also checked for any correlation between the surface brightness
of the envelope and the velocity width of the Ly$\alpha$ emission
line in our sample, and found a mild correlation between them. The
correlation coefficient is 0.64 implying a 17\,\% probability of
finding this trend by chance occurrence. If we take into account
the NC12 and CJ06 samples, then the correlation disappears,
suggesting that the trend found in our sample may not be real.
However, the number of quasars is not large enough to be
statistically significant. Observation of a large sample of quasars
with extended Ly$\alpha$ envelope may help us settle this question.

\subsection{Quasar narrow Ly$\alpha$ emission}

If the narrow Ly$\alpha$ emission on the spectrum trace is
associated with the unresolved inner part of the extended Ly$\alpha$
envelope as suggested by Heckman et al. (1991b) then one would
expect a correlation between the Ly$\alpha$ luminosity on the trace
and in the extended region. Since the narrow Ly$\alpha$ emission
line in a quasar spectrum is usually strongly blended and
contaminated by the emission from the BLR, no attempt has ever been
made to correlate its luminosity with that of the extended envelope.
For the first time, in this work we can directly examine this
correlation for the objects in our sample. Figure~\ref{Tr_vs_env}
presents the luminosity of the extended Ly$\alpha$ envelopes as a
function of the luminosity of the Ly$\alpha$ emission seen on the
quasar trace for the objects in our sample. A strong correlation
with a coefficient of 0.90 is found between the two luminosities.
This correlation implies that much of the narrow spatially
unresolved Ly$\alpha$ emission seen on the quasar trace is probably
associated with the inner part of the extended Ly$\alpha$ envelope.
This correlation can be interpreted as a consequence of the direct
influence of the quasars on their Ly$\alpha$ envelopes.

\subsection{Quasar vs. the envelope luminosities}

We investigate the possible correlation between the broad Ly$\alpha$
line luminosity of the quasars and that of the extended Ly$\alpha$
envelopes. Since the broad Ly$\alpha$ emission line is fully
extinguished in our spectra, its luminosity is calculated from the
principle component analysis (PCA) reconstruction of this line
(P\^aris et al. 2011, 2014). To be consistent with previous work, we
integrate the 1D quasar spectra in the rest-frame wavelength
interval of 1200$-$1230\,$\textup{\AA}$ and then measure the
luminosity. The result is shown in Fig.\,\ref{BLR_vs_env} as blue
symbols. The correlation coefficient for the data points in the CJ06
(blue squares), NC12 (blue stars), our quasars (blue filled circles)
and for the combination of the three samples are 0.92, 0.61,
0.24 and 0.11, respectively. The lack of a correlation in the
combined sample mainly stems from the absence of a correlation in our sample.
We added the luminosity of the unresolved NEL to that of the extended envelope
in our sample (the result is shown as red open circles in
Fig.\,\ref{BLR_vs_env}), and still found no correlation with
the Ly$\alpha$ BLR luminosity.

We also investigated whether the extended Ly$\alpha$ envelope
luminosities are correlated with the ionizing radiation flux from
the quasars. We used the same technique described in paper I to
estimate the ionizing flux of the quasars in our sample. The result
is shown as black symbols in Fig.\,\ref{BLR_vs_env}. We found
correlation coefficients of 0.20, 0.32 and $-$0.41 for the objects
in the NC12, our sample and the total sample, respectively. The lack
of a strong correlation between the quasar luminosity and that of
its envelope may suggest that processes other than direct
photo-ionization are at play such as variations of covering factors
and/or dust extinction from one line-of-sight to the other.

%************************ AlIII/SiII vs HI *********************
\begin{figure}
\centering
\begin{tabular}{c}
\includegraphics[bb=112 393 458 649,clip=,width=0.95\hsize]{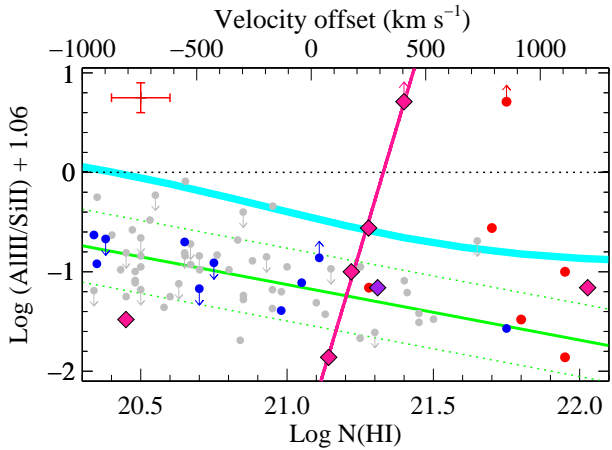}
\end{tabular}
\caption{The Al\,{\sc iii}/Si\,{\sc ii} ratio as a function of
$N$(H\,{\sc i}) for the Berg et al. (2015, grey points), EP10 (blue
points), and our eclipsing DLAs (red points). The red error bar on
the upper left side of the plot shows the uncertainty in the column
densities of our data points.  The green solid line is the fit to
the Berg et al. (2015) DLAs. The two green dotted lines show the
standard deviation of the grey points. The thick cyan line shows the
result of a grid of {\sc cloudy} models constructed for the
ionization parameter log\,U~=~0.0, a nominal hydrogen density
log\,n$_{\rm HI}$~=~1.0, and hydrogen column density in the range
20.30~$\le$~log$N$(H\,{\sc i})~$\le$~22.10. The pink diamonds show
the Al\,{\sc iii}/Si\,{\sc ii} ratios in our sample as a function of
the velocity offset between the DLAs and their background quasars
(upper x-axis scale where positive values indicate a redshift i.e.
the DLA falling towards the quasars). The pink line shows the fit to
the data points with positive velocity offsets, after excluding the
point at log~$N$(H~{\sc i})$<$21.5 corresponding to J1253$+$1007.
The purple diamond shows the value for the DLA towards J1253$+$1007,
after correcting the quasar redshift for systemic shift (see the
text). }
 \label{Al3Si2}
\end{figure}
%*******************************************************

\subsection{Spatial and velocity extent of the nebulae}

The mean sizes (i.e. diameter) of the envelopes in the NC12, CJ06
and our sample are 45, 26 and 41~kpc, respectively. The depth of the
observation in our sample is very similar to that of the CJ06 while
the NC12 observations are the deepest. This suggests that our
envelopes could be more extended if our observations were as deep as
the NC12 ones.

The Ly$\alpha$ envelopes around our quasars exhibit two-sided
emissions along both PAs. This is in contrast to the CJ06 sample of
RQQs in which only one-sided emission is detected. Although
one-sided and two-sided envelopes are both reported around RQQs
(e.g. M$\o$ller et al. 2000 and Weidinger et al. 2005), RLQs are
usually observed to have asymmetric two-sided nebulae (Heckman et
al. 1991a). Our envelopes, therefore, seem to morphologically
resemble those found around RLQs. Figure~\ref{L_vs_Ext} presents the
luminosity of the Ly$\alpha$ envelopes as a function of their
spatial extension for the CJ06 (red diamonds) and our sample (black
filled circles). A strong correlation with a high probability of
$\sim$~99.6\% is found between the Ly$\alpha$ luminosity of the
extended envelope and its spatial extension for the full sample.
This strong correlation confirms the result by CJ06 that more
luminous envelopes are also spatially more extended.

Kaspi et al. (2000; 2005) found a relation between the luminosity of
the AGN and the size of the BLR, $R_{\rm BLR}$~$\sim$~L$^{\alpha}$
(also see Bentz et al. 2013). Their best value of $\alpha$ is in the
range 0.5~$<$~$\alpha$~$<$~0.7 which is consistent with simple
photo-ionization expectations. Interestingly, similar relation seems
to be present between the size of the extended Ly$\alpha$ envelope
and its luminosity. In Fig.~\ref{L_vs_Ext}, the green and blue
curves show the power law functions for $\alpha$~=~0.5 and 0.7,
respectively. This could imply that the Ly$\alpha$ envelopes may be
strongly affected by the ionizing radiation emanating from the
central AGN. The scatter seen in our observed data points around
this mean power law relation may result from the non-uniform
conditions in the emission regions. Part of this scatter could also
be due to the presence of dust in these regions.

In paper~I, we suggested that the high ionization gas seen towards
J0823+0529 may arise from some warm absorber driven by the accretion
disk luminosity. The outflowing material is then stopped by
collision with the ISM of the host galaxy, producing shock waves.
The shocked material then gets cooled and mixed with the ISM. The
outflowing material could also affect the kinematics of the ISM gas.
If true, then one would expect a correlation between the kinematics
of the ISM gas and that of the outflowing material. In
Fig.~\ref{dv_dv}, we plot the relation between the total velocity
extension of the Ly$\alpha$ emission and that of the C\,{\sc iv}
absorption profile reported in Table~\ref{redshifts}. A correlation
coefficient of 0.86 is found in this sample. This correlation may
hint at the possibility that the Ly$\alpha$ emission envelopes
around the quasars in our sample may be affected by the outflowing
material originated from the innermost region of the AGN. However,
observation of more quasars with extended Ly$\alpha$ emission are
needed to confirm whether this correlation is real.

\subsection{Physical properties of the eclipsing DLAs}

The first systematic study of the physical properties of PDLAs was
done by Ellison et al. (2010, hereafter EP10; also see Ellison et
al. 2002). They studied in detail, a sample of 16 PDLAs with a
velocity separation $\delta$V~$<$~3000~km\,s$^{-1}$ from the
quasars. They compared the properties of their PDLAs with those of a
control sample of intervening DLAs to investigate the differences
between the two populations. Their control sample contains
intervening DLAs with $\delta$V~$>$~10\,000~km\,s$^{-1}$ from the
Berg et al. (2015b) catalog. In this section, we compare our DLAs
with those of EP10 and the control sample.

The velocity offset between the quasars and the eclipsing DLAs in
our sample is in the range
$-$809~$\lesssim$~$\delta$V~$\lesssim$~1207~km\,s$^{-1}$ (see
Table~\ref{redshifts} column 5). 5 out of 6 DLAs have positive
velocity offset (i.e. DLAs falling towards the quasars) while in the
EP10 sample it is 4 out of 16. It is interesting to note that the
only object with a negative velocity offset in our sample (i.e.
J1058$+$0315) has also the smallest Ly$\alpha$ velocity extent
($\Delta$V$_{Ly\alpha}$~$\sim$~520~km\,s$^{-1}$) and the simplest
C\,{\sc iv} absorption feature with only one single component.
Moreover, the maximum velocity offset of $\sim$~882~km\,s$^{-1}$
between the DLA and the extended Ly$\alpha$ emission in our sample
is also seen along this line of sight.

\subsubsection{Ionization properties}

In our eclipsing DLAs we detect high ionization species which in
many cases extend hundreds of km\,s$^{-1}$ to the blue of the main
low-ion component. C\,{\sc iv} and Si\,{\sc iv} absorption are
detected in all of our eclipsing DLAs. The O\,{\sc vi} region in
most of our spectra suffers from low SNR and contamination with
Ly$\alpha$ forest absorption. Nonetheless, the O\,{\sc vi} doublet
is detected in the relatively high SNR spectrum of the quasar
J1253$+$1007 in the velocity range
$-$850~$\lesssim$~$\delta$V~$\lesssim$~$-$100~km\,s$^{-1}$ (see
Fig.~\ref{J1253_1D}). Although the O\,{\sc vi}\,$\lambda$\,1037
absorption seems to be slightly blended with some Ly$\alpha$ forest
absorption, the similarity between the O\,{\sc vi}\,$\lambda$\,1031
and the N\,{\sc v} doublet absorption confirms the detection. At
zero velocity, we have one detection (towards J0823$+$0529) and one
strict upper limit (towards J1253$+$1007) of N\,{\sc v}, with no
confident N\,{\sc v} detection towards the remaining objects. Offset
N\,{\sc v} and Si\,{\sc iv} absorption are seen to the blue at
$v$~$\lesssim$~$-$100~km\,s$^{-1}$ towards J0953$+$0349,
J1154$-$0215, and J1253$+$1007. Moreover, offset C\,{\sc iv}
absorption is also seen at $v$~$\lesssim$~$-$100~km\,s$^{-1}$
towards all our objects except J1058$+$0315. EP10 found no low-ions
associated with their offset high ions while in our case, some
strong low-ion transitions (e.g. O\,{\sc i}, C\,{\sc ii}, C\,{\sc
ii}$^{*}$) are seen towards J0953$+$0349 at
$v$~$\sim$~$-$660~km\,s$^{-1}$.

In J0953$+$0349, the velocity separation between the quasar and the
two low-ion components are $-$484 and 175~km\,s$^{-1}$ (see
panel~(a) in Fig.~\ref{J0953_1D}). This is reminiscent of the PDLA
studied by Rix et al. (2007) towards the quasar Q2343$-$BX415. They
found two low-ion components, one redshifted by
$\sim$~400~km\,s$^{-1}$ and the other blueshifted by
$\sim$~$170$~km\,s$^{-1}$ relative to the quasar systemic redshift
(see their figure~5). They concluded that the presence of prominent
high ionization lines in the redshifted component along with the
partial coverage of the continuum source (Ofengeim et al. 2015;
Klimenko et al. 2015) are indicative of gas falling onto the central
AGN; and the blueshifted component may be part of the outflowing
material of the quasar host galaxy.

The column density ratio $N$(Al\,{\sc iii})/$N$(Al\,{\sc ii}) can
be used to study the ionization properties of DLAs. Vladilo et al.
(2001) showed that, in the presence of a soft stellar-type ionizing
radiation, this ratio decreases when $N$(H\,{\sc i}) increases.
Figure~\ref{Al3Si2} shows Al\,{\sc iii}/Si\,{\sc ii} ratios (after
correcting for the solar value of Al/Si) as a function of
$N$(H\,{\sc i}) for the Berg et al. (2015, grey points), EP10 (blue
points), and our DLAs (red points). In this figure, Si\,{\sc ii} is
used as a proxy for Al\,{\sc ii} which is mostly saturated in DLAs.
EP10 found no apparent difference between the intervening DLAs and
their PDLAs. In Fig.~\ref{Al3Si2}, the green solid line, which
clearly shows the anti-correlation with $N$(H\,{\sc i}), is the fit
to the Berg et al. (2015) DLAs. The two green dotted lines mark the
1\,$\sigma$ uncertainty in the Al\,{\sc iii}/Si\,{\sc ii} ratios,
calculated from the standard deviation of the grey data points in
Fig.~\ref{Al3Si2}. Interestingly, three of our eclipsing DLAs seem
to follow this anti-correlation, implying that the DLA gas is
photoionized by a soft radiation and/or if it is close to the AGN,
the ionization parameter is very low (Howk \& Sembach 1999). The
remaining three eclipsing DLAs indicate rather elevated Al\,{\sc
iii}/Si\,{\sc ii} ratios, with one DLA (towards J1154$-$0215) having
the highest Al\,{\sc iii}/Si\,{\sc ii} ratio ($\gtrsim$\,+0.71) ever
reported for a DLA or PDLA.

In Fig.~\ref{Al3Si2}, the thick cyan line shows the result of a grid
of {\sc Cloudy} models constructed for the ionization parameter
log\,U~=~0.0 and a nominal hydrogen density log\,n$_{\rm HI}$~=~1.0.
The combination of the mean AGN spectrum of Mathews \& Ferland
(1987), the cosmic microwave background (CMB) radiation at
$z$~=~2.5742 (mean redshift of all our eclipsing DLAs), and the
Haardt-Madau extragalactic spectrum (Haardt \& Madau 1996) at the
same redshift is taken as the incident radiation hitting the cloud.
Note that the mean metallicity $<$[Si/H]$>$~=~$-$1.2 is assumed
although the result is not very sensitive to the choice of the
metallicity and redshift. As shown in Fig.~\ref{Al3Si2}, two of our
eclipsing DLAs seem to have Al\,{\sc iii}/Si\,{\sc ii} ratios
consistent with this {\sc Cloudy} model.

As mentioned above, the Al\,{\sc iii}/Si\,{\sc ii} ratio in the
eclipsing DLA towards J1154$-$0215 is $\gtrsim$\,+0.71 which is
$\sim$~6 times more than the highest value in the Berg et al. (2015)
DLA sample (see Fig.~\ref{Al3Si2}). We note that this value can even
be considered as a {\it strict} lower limit because the Si\,{\sc ii}
(resp. Al\,{\sc iii}) column density can be considered as a {\it
strict} upper limit (resp. lower limit; see below and
Fig.~\ref{J1154_1D}). To get the Si\,{\sc ii} column density, a
Voigt profile fit was conducted on the Si\,{\sc ii}\,$\lambda$\,1808
and $\lambda$\,1526 absorption lines. Since the former transition is
weak and the latter one is slightly blended to the red, the derived
$N$(Si\,{\sc ii}) is taken as an upper limit. On the other hand, the
Al\,{\sc iii} absorption profiles are most probably saturated as the
two components seem to have almost the same optical depth. The high
value of Al\,{\sc iii}/Si\,{\sc ii} ratio towards J1154$-$0215 could
be explained if one assumes a two- or multi-phase medium.

In Fig.~\ref{Al3Si2}, the pink diamonds show the Al\,{\sc
iii}/Si\,{\sc ii} ratios in our sample as a function of the velocity
offset between the DLAs and their quasars. In this figure, the upper
x-axis represents the velocity offsets in km\,s$^{-1}$, and positive
values indicate a redshift (i.e. the DLAs falling towards the
quasar). Taken at face value, there appears to be a strong
correlation (with the coefficient of 0.99) between the velocity
offsets and the Al\,{\sc iii}/Si\,{\sc ii} ratios among the DLAs
with positive velocity offset. The only exception to this is the DLA
towards J1253$+$1007 which exhibits a high positive velocity offset
of $\sim$~1207~km\,s$^{-1}$, and apparently does not follow this
trend. We note that the redshift of this quasar was obtained by
fitting its C\,{\sc iv} emission line (see Fig.~\ref{J1253_1D}). So,
if we apply the known systematic shift of C\,{\sc iv} emission line
(as reported in Shen et al. 2007) to correct the redshift, we get a
velocity offset of $\sim$~290~km\,s$^{-1}$ (instead of
$\sim$~1207~km\,s$^{-1}$) between the DLA and the quasar. This new
value, which is now more consistent with the correlation mentioned
above, is shown as a purple square in Fig.~\ref{Al3Si2}. Note that
the pink line shows the fit to the data points of DLAs with positive
velocity offset, after excluding the one towards J1253$+$1007. It
could be possible that eclipsing DLAs with positive velocity offset
are chemically distinct from those with negative velocity offset.
The DLA towards J1058+0315 with negative velocity offset of
$-$809~km\,s$^{-1}$ is probably an intervening gas.

This correlation strongly suggests that DLAs with higher velocity
towards the quasars have also higher ionization and therefore may be closer
to the AGN which is consistent with the fact that they could be
part of an infall flow onto the host galaxy.

\subsubsection{The size of the eclipsing DLAs}

Since extended Ly$\alpha$ emission is detected in the trough of our
eclipsing DLAs, this indicates that the size of the absorbing cloud
should be smaller than that of the Ly$\alpha$ emitting central
unresolved region. In paper~I, we studied in detail the eclipsing
DLA seen towards the quasar J0823+0529, which also exhibits some
indications of partial coverage of the BELR. We used the {\sc
cloudy} photo-ionization code to model the absorber including the
Si\,{\sc ii}$^{*}$ absorption and derived hydrogen density in the
range 180~$<$~$n_{\rm H}$~$<$~710~cm$^{-3}$, resulting in the cloud
size in the range 2.3~$<$~$l$(H\,{\sc i})~$<$~9.1~pc.

Si\,{\sc ii}$^{*}$ absorption is also detected in the eclipsing DLA
towards quasar J1253+1007 (see Fig.~\ref{J1253_1D}). Following the
same approach as in paper~I, we get $n_{\rm H}$~$\sim$~13~cm$^{-3}$
and $l$(H\,{\sc i})$~\sim$~50~pc. Unfortunately, in the other 4
objects in our sample, the Si\,{\sc ii}$^{*}$ absorption is not
detected.

Assuming, as in e.g. Prochaska and Hennawi (2009) that the optically
thick eclipsing DLAs in our sample have $n_{\rm
H}$~$>$~10~cm$^{-3}$, we can derive an indicative upper limit on
their size (i.e. $l$~$\sim$~$N_{\rm HI}$/$n_{\rm H}$). The result is
given in the last column of Table~\ref{redshifts} where the values
are in pc. As shown in this table, all our eclipsing DLAs appear to
have sizes smaller than 300~pc. Since NELRs in quasars have
dimensions of several hundreds of parsecs, our upper limits on the
size of the eclipsing DLAs are consistent with the argument that the
clouds are smaller than the central unresolved emission region.

\section{Summary and conclusions}

We present spectroscopic observations of six high redshift ($z_{\rm
em}$\,$>$~2) quasars, which have been selected because of the
presence of a strong proximate ($z_{\rm abs}$~$\sim$~$z_{\rm em}$)
coronagraphic DLA which covers only the broad line region of the
quasar, thus revealing emission from the host galaxy. We detect
spatially extended Ly$\alpha$ emission envelopes around the six
quasars, with projected spatial extent in the range 26~$\le$~$d_{\rm
Ly\alpha}$~$\le$~51~kpc. No correlation is found between the quasar
ionizing luminosity and the Ly$\alpha$ luminosity of their extended
envelopes. This could be related to the limited covering factor of
the extended gas and/or due to the AGN being obscured in other
directions than towards the observer. Indeed, we find a strong
correlation between the luminosity of the envelope and its spatial
extent, which suggests that the envelopes are probably ionized by
the AGN. The metallicity of the coronagraphic DLAs is low and varies
in the range $-$1.75\,$<$\,[Si/H]\,$<$\,$-$0.63. Highly ionized gas
is observed associated with most of these DLAs, probably indicating
ionization by the central AGN. One of these DLAs has the highest
Al\,{\sc iii}/Si\,{\sc ii} ratio ever reported for any intervening
and/or proximate DLA. Most of these DLAs are redshifted with respect
to the quasar, implying that they might represent infalling gas
probably accreted onto the quasar host galaxies through filaments.

\subsection{Nature of the emitting gas}

The main mechanisms proposed to power the extended Ly$\alpha$
emission envelopes around high redshift quasars are photo-ionization
by the AGN, and cold accretion of neutral hydrogen gas into the
quasar dark matter halo. If ionizing photons from the AGN are
responsible for powering the extended Ly$\alpha$ emission, a
correlation between the quasar and its envelope luminosity is
expected. On the other hand, in the cold accretion scenario, no such
correlation is expected.

CJ06 found no correlation between the quasar ionizing luminosity and
the luminosity of the Ly$\alpha$ envelope. They argued that the lack
of correlation cannot be due to extinction by dust or
absorption by neutral hydrogen in the quasar environments which
could potentially affect the quasar luminosity at
912~\textup{\AA}. Similarly, NC12 found no correlation between the
two luminosities for their full sample, apparently confirming the
CJ06 result, but at higher redshifts and for a larger luminosity range.
NC12 argued that removing the object BR\,2237$-$0607
from their sample would restore the correlation, implying that the
quasar might be the main powering source of the
Ly$\alpha$ envelope. But the lack of a correlation for their full
sample seems to favor the cold accretion scenario.

In our sample, the quasar ionizing luminosity does not seem to
correlate with the extended Ly$\alpha$ envelope luminosity. This is
consistent with what is seen in CJ06 and NC12 full samples. However,
in section 4.5, we found a strong correlation between the luminosity
of the Ly$\alpha$ envelope and its spatial extension. We suggested
that this correlation could imply that the quasar ionizing flux
might be powering the Ly$\alpha$ envelopes. This seems to be at odds
with the apparent lack of a correlation between the quasar ionizing
luminosity and the Ly$\alpha$ envelope luminosity. It is possible
that the level of extinction by dust significantly varies from one
object to the other in our sample, resulting in highly uncertain
measurement of the intrinsic quasar luminosities at
912~\textup{\AA}. Moreover, in contrast to the CJ06 and NC12
samples, the presence of the strong eclipsing DLAs at the quasar
emission redshift in our sample would also mean higher absorption of
the quasar flux by the surrounding neutral hydrogen. These effects
could in principle remove the possible correlation that might have
existed between the luminosity of the quasar at 912~\textup{\AA} and
that of the Ly$\alpha$ envelope in our sample. To circumvent this
problem, one can use the infrared spectra of the quasars to better
calibrate the quasar template spectrum and consequently get a better
estimate of the quasar luminosity at 912~\textup{\AA}. Better
estimate of the quasars ionizing luminosities might restore the
correlation.

\subsection{Nature of the eclipsing DLAs}

An important question regarding the origin of the proximate DLAs is
whether they are due to outflowing material intrinsic to the quasar
or some (infalling) material unrelated to the AGN environment but
coincident in redshift with the background quasar. If the DLA is
part of some low density ($n_{\rm H}$~$<$~10~cm$^{-3}$) infalling
material close to the central AGN, the quasar's ionizing photons
could ionize and photoevaporate it (Prochaska \& Hennawi 2009).
Although infalling gas is usually expected to have lower
metallicity, associated outflowing absorbers tend to have solar or
super solar metallicities (Petitjean et al. 1994). This would in
principle allow one to distinguish whether the gas is associated or
intervening.

EP10 found that their PDLAs with higher H\,{\sc i} column densities
have metallicities $\sim$~3 times higher than what is seen for the
intevening DLAs, but still below those expected for an associated
absorber. They argued that their PDLAs are probably neither
associated nor intervening. They suggested that PDLAs probably
probe overdense regions in massive galaxies with higher
metallicities in dense clusters.

The metallicities of our eclipsing DLAs are low and similar to what
is seen in intervening DLAs. This would imply that these absorbers
might not be outflowing material driven by the central AGN. The fact
that most of our eclipsing DLAs have redshifts larger than that of
their background quasars further strengthens the idea that these
DLAs could be associated with some less enriched infalling material
accreting onto the quasar host galaxies, probably through filaments
as was concluded for the quasar J0823+0529 (Fathivavsari et al.
2015).

\section*{Acknowledgments}
We would like to thank the anonymous referee for his/her
constructive comments, which helped us to improve the paper. We also
thank George Becker for advices on MagE data reduction. We also
thank Hadi Rahmani for useful discussion. HFV was supported by the
Agence Nationale pour la Recherche under program
ANR-10-BLAN-0510-01-02. SL has been supported by FONDECYT grant
number 1140838 and partially by PFB-06 CATA.

\appendix
\section{Figures A1 to A8}

%************************ J0953_1D *********************
\begin{figure*}
\centering
\begin{tabular}{c}
\includegraphics[bb=44 363 559 702,clip=,width=0.65\hsize]{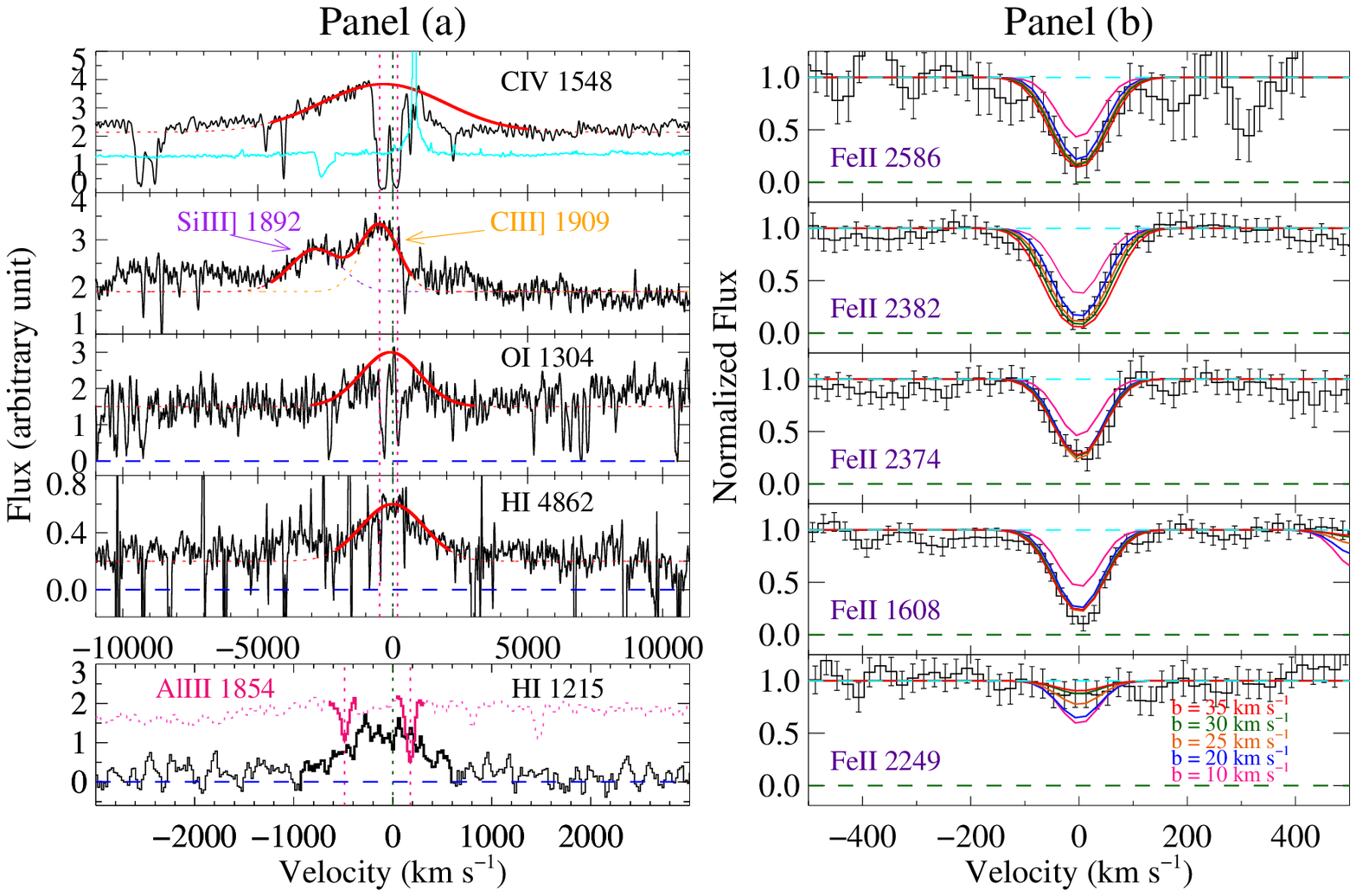}\\
\includegraphics[bb=63 351 562 687,clip=,width=0.65\hsize]{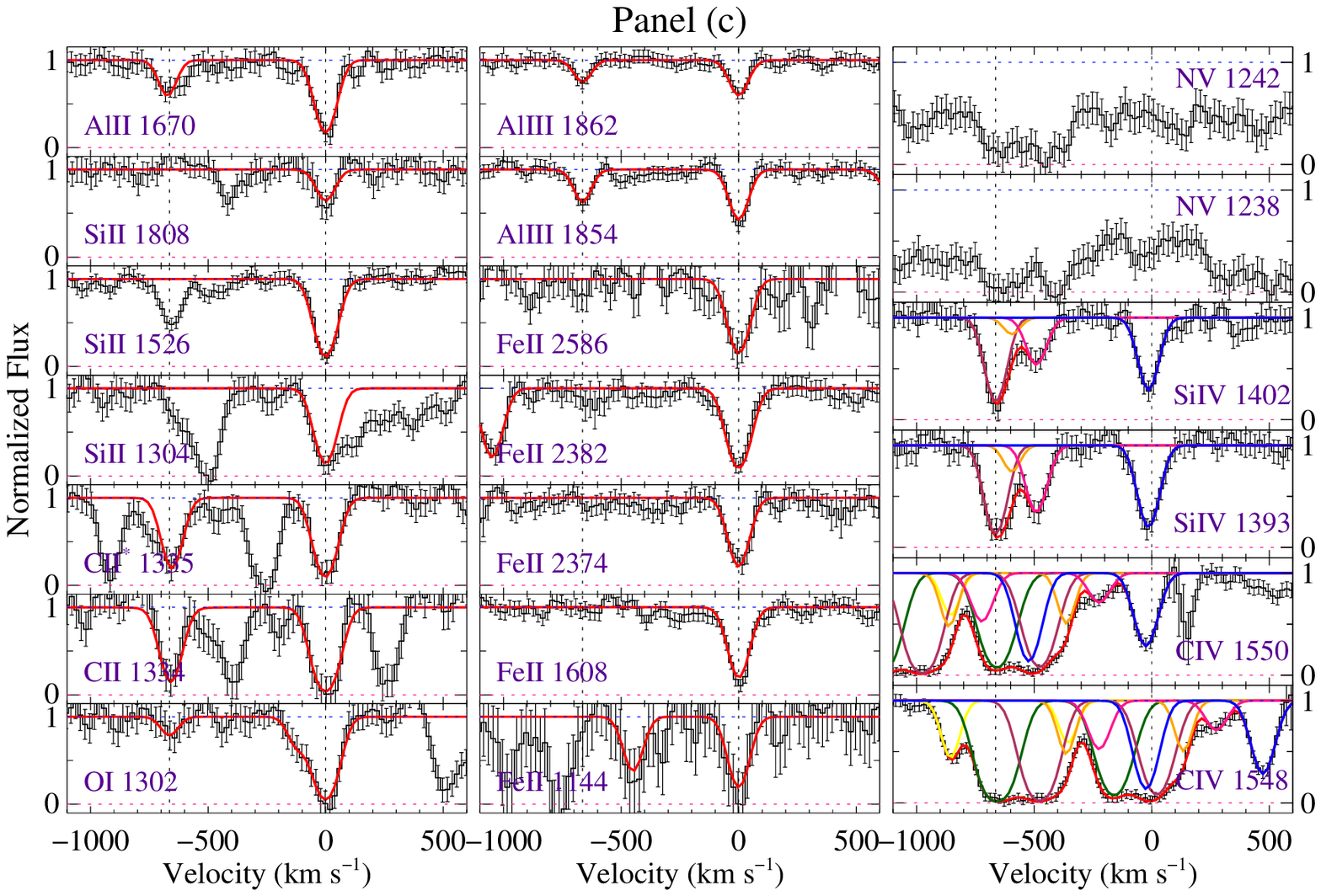}\\
\includegraphics[bb=58 435 533 594,clip=,width=0.65\hsize]{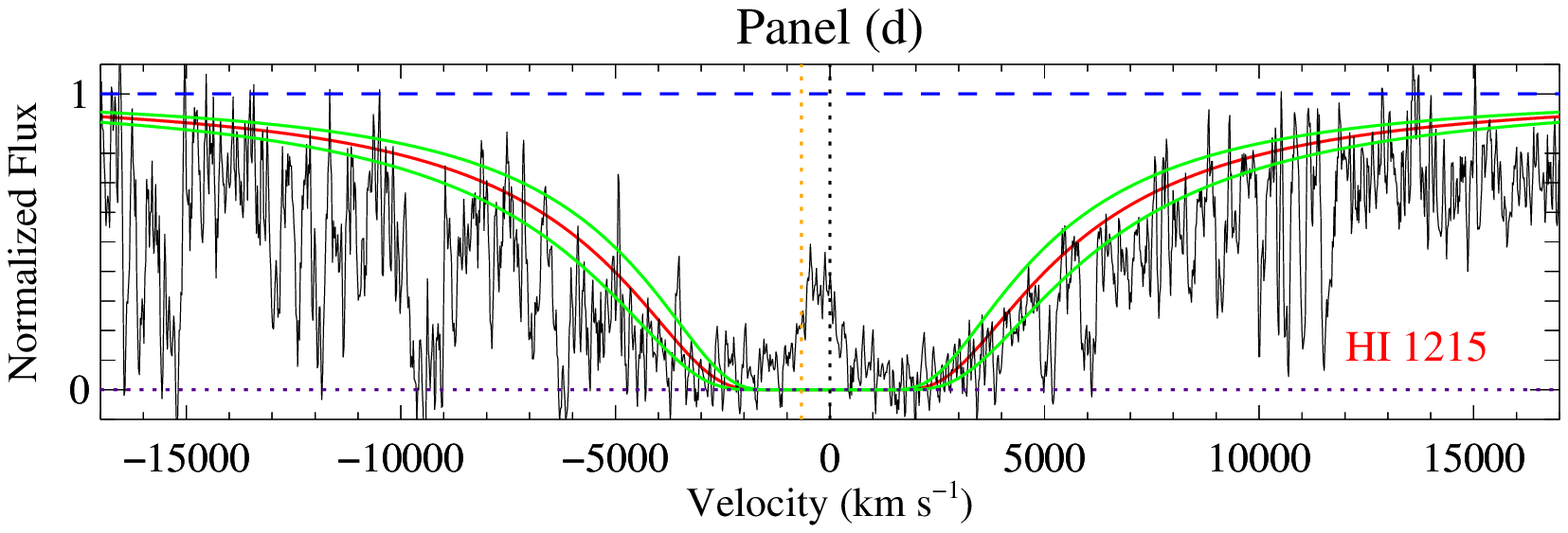}\\
\end{tabular}
\caption{Same as Fig.\,1 but for the quasar J0953$+$0349 with
$z_{\rm em}$\,=\,2.5940 and $z_{\rm abs}$\,=\,2.5961. The H\,{\sc
i}\,$\lambda$4862 spectrum in Panel~(a) is from XSHOOTER
observation. }
 \label{J0953_1D}
\end{figure*}
%*******************************************************

%************************ J0953_2D *********************
\begin{figure*}
\centering
\begin{tabular}{c}
\includegraphics[bb=71 390 546 675,clip=,width=0.65\hsize]{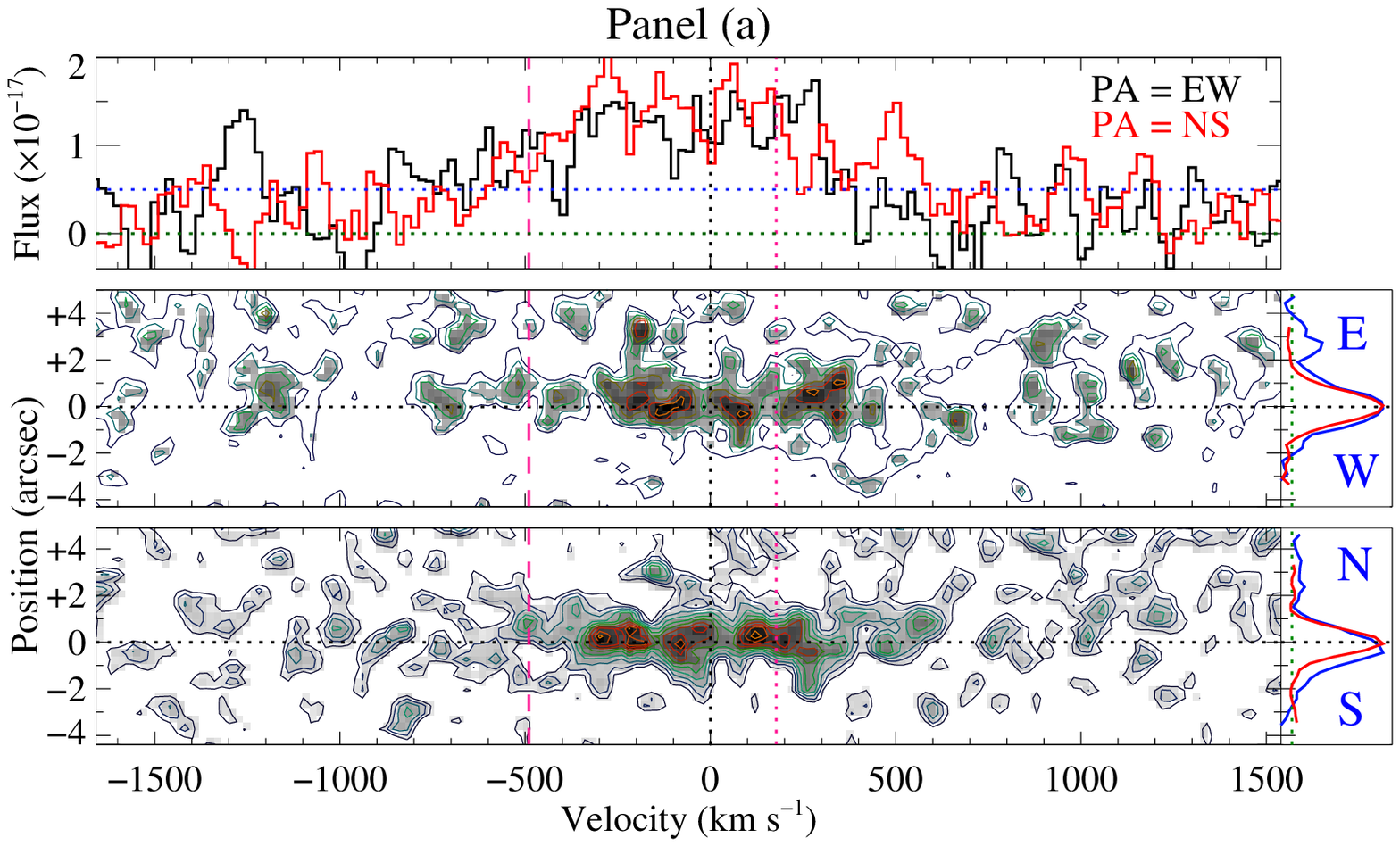}\\
\includegraphics[bb=43 399 566 629,clip=,width=0.65\hsize]{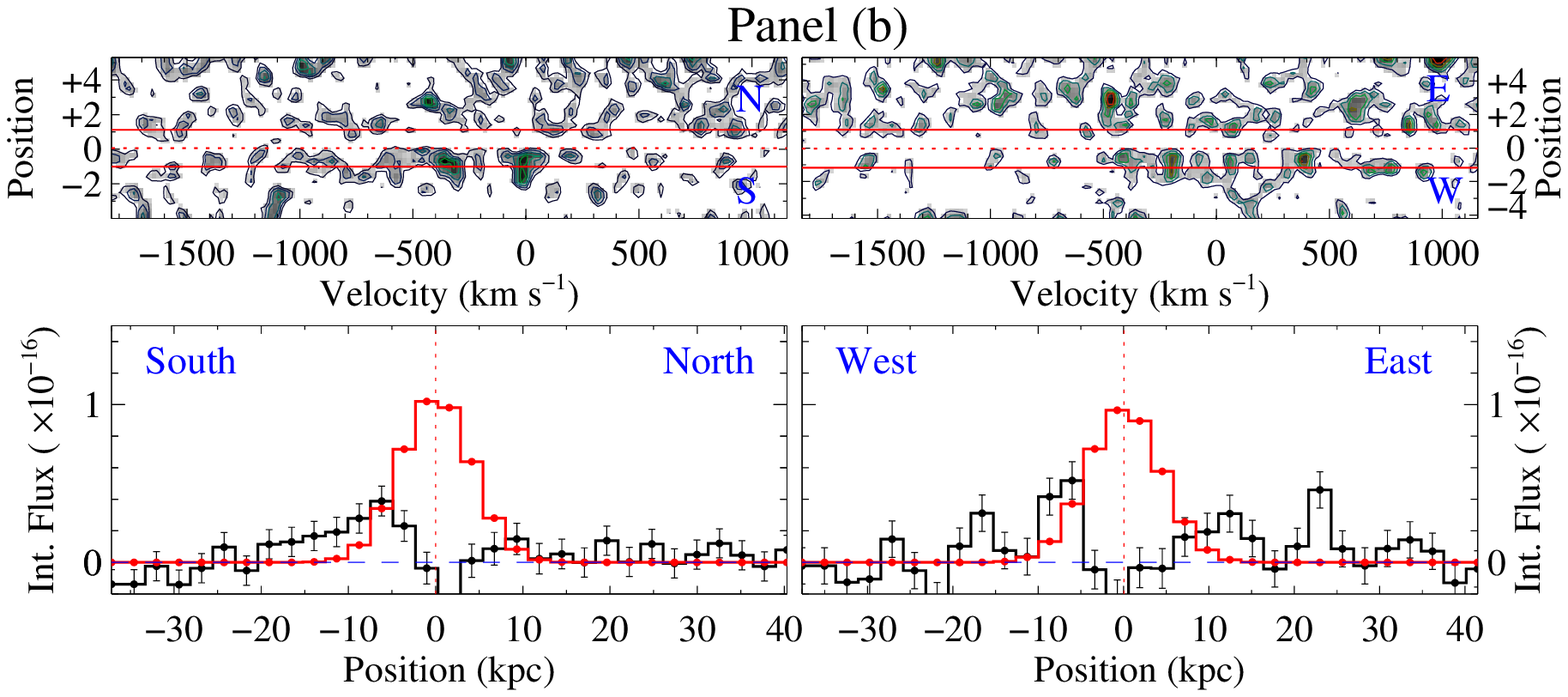}\\
\includegraphics[bb=49 363 566 575,clip=,width=0.65\hsize]{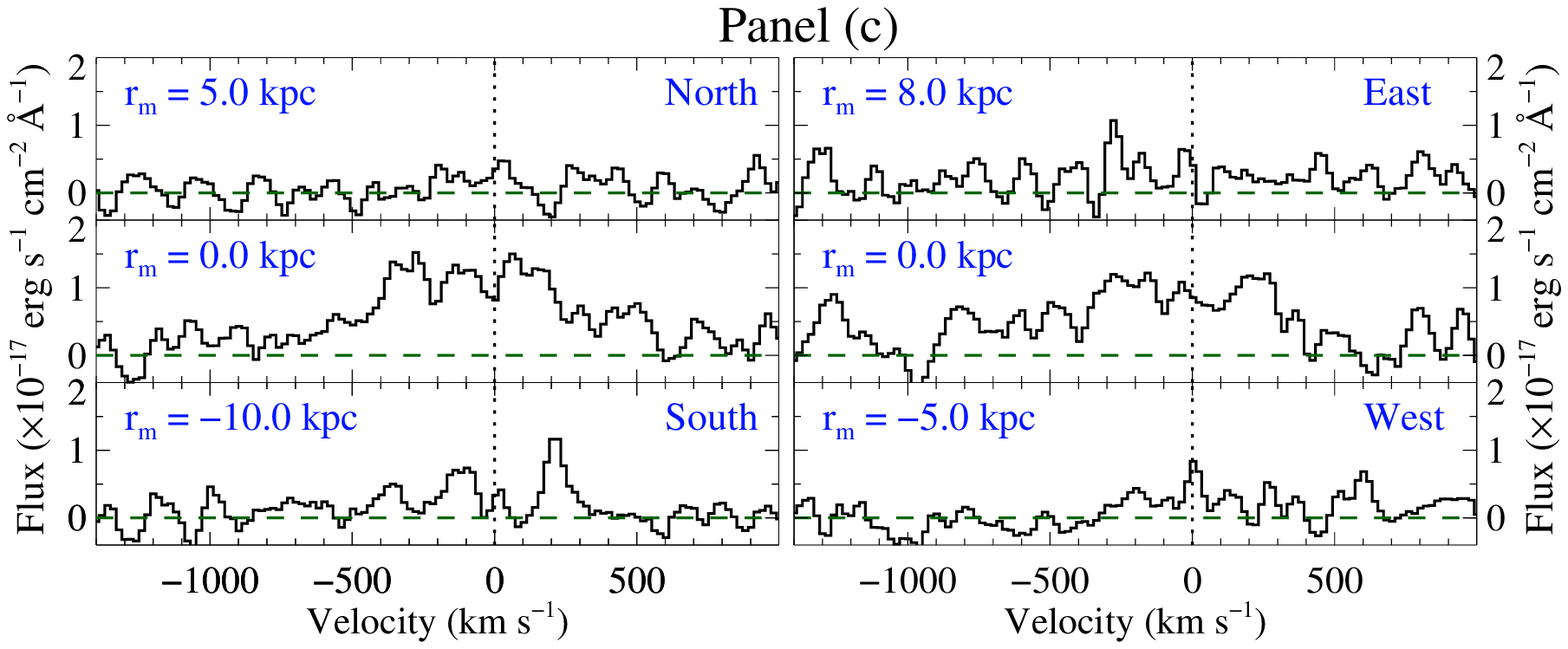}\\
\end{tabular}
\caption{Same as Fig.\,2 but for the quasar J0953$+$0349. The zero
velocity is at $z_{\rm em}$\,=\,2.5940. In panel~(a), the pink
dotted lines mark the position of the DLA at $z_{\rm
abs}$\,=\,2.5961 and the pink dashed lines show the position of the
absorber at $z_{\rm abs}$\,=\,2.5880. In panels\,(a) and (b) the
outermost contour along PA=NS (resp. PA=EW) corresponds to a flux
density of 9.10\,$\times$\,10$^{-20}$ (resp.
1.36\,$\times$\,10$^{-19}$)
erg~s$^{-1}$\,cm$^{-2}$\textup{\AA}$^{-1}$ and each contour is
separated by 9.10\,$\times$\,10$^{-20}$ (resp.
1.36\,$\times$\,10$^{-19}$)
erg~s$^{-1}$\,cm$^{-2}$\textup{\AA}$^{-1}$ from its neighboring
contour.}
 \label{J0953_2D}
\end{figure*}
%*******************************************************

%************************ J1058_1D *********************
\begin{figure*}
\centering
\begin{tabular}{c}
\includegraphics[bb=52 348 559 597,clip=,width=0.65\hsize]{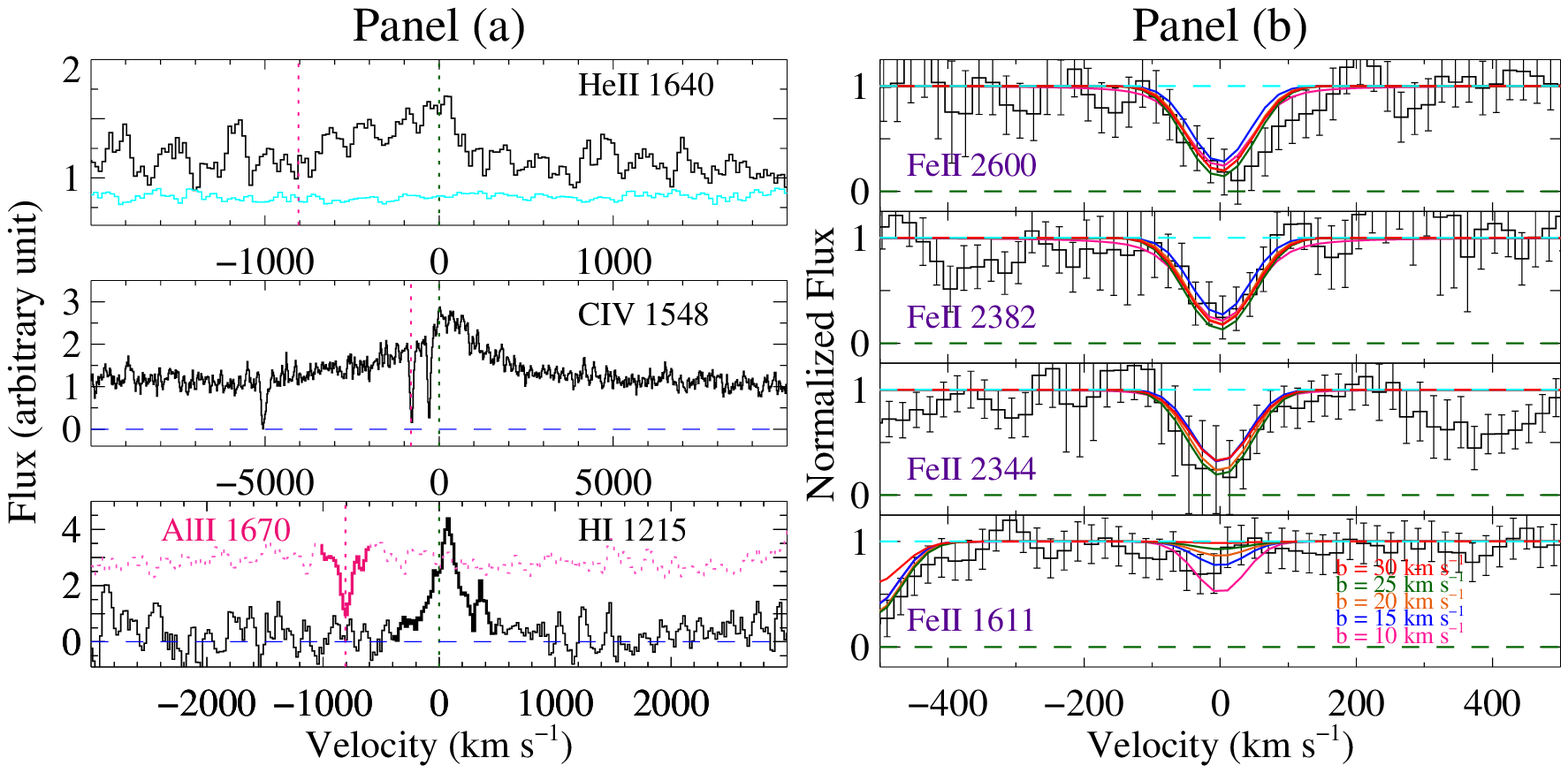}\\
\includegraphics[bb=60 348 554 648,clip=,width=0.65\hsize]{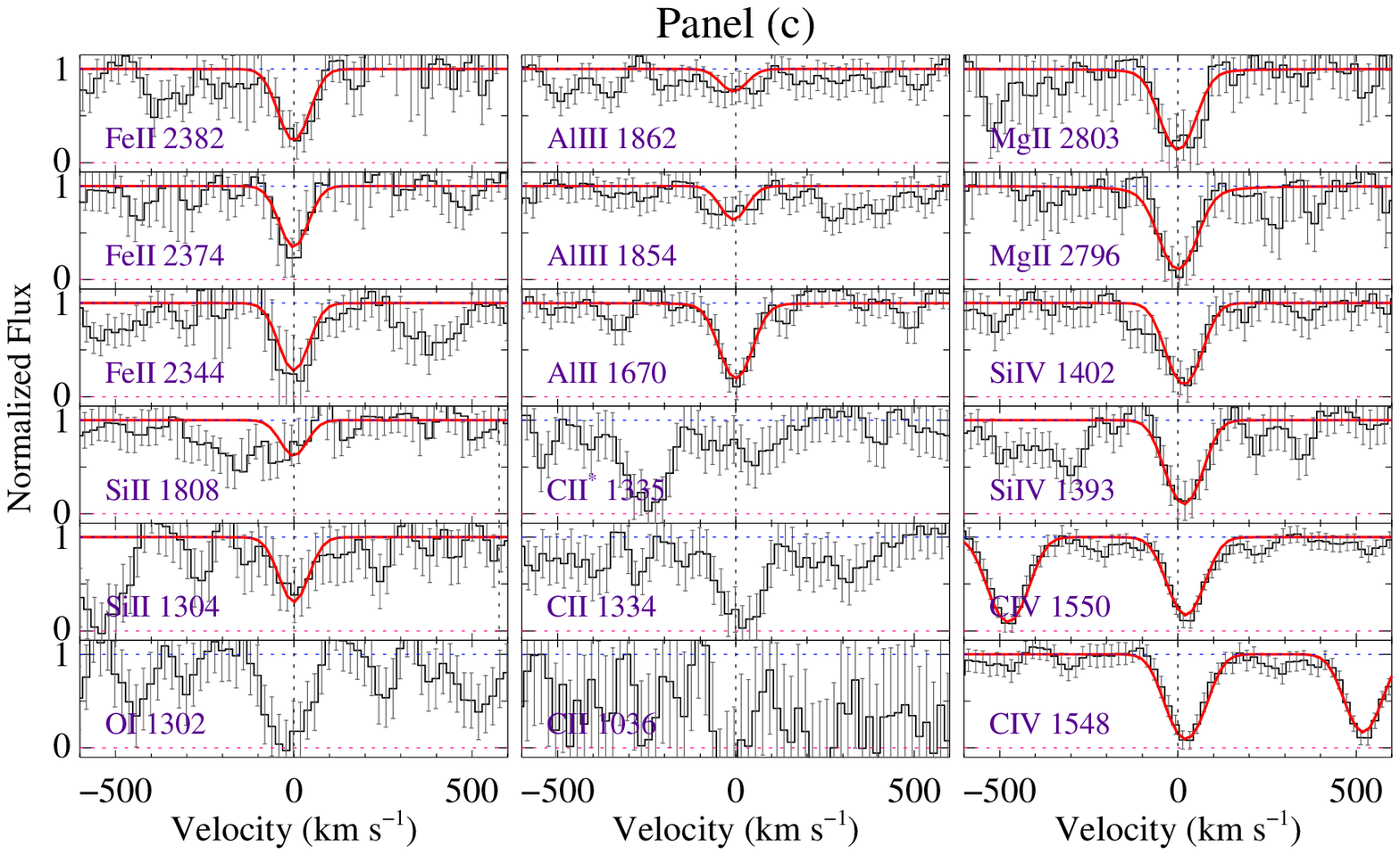}\\
\includegraphics[bb=58 435 554 599,clip=,width=0.65\hsize]{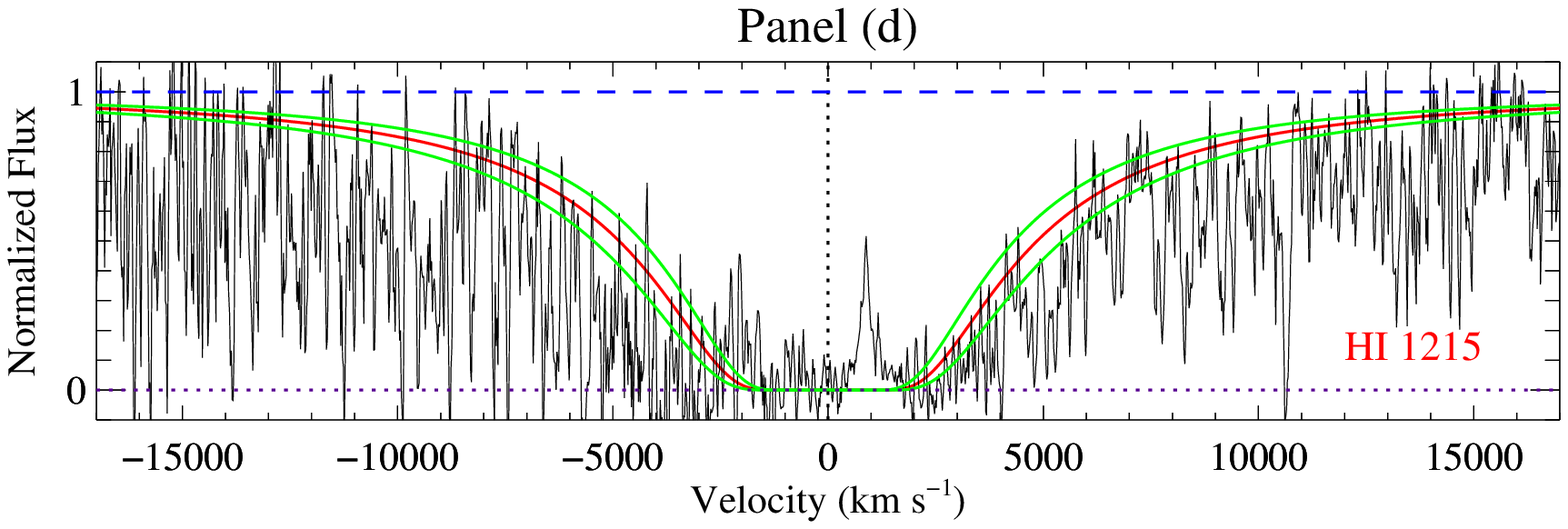}\\
\end{tabular}
\caption{Same as Fig.\,1 but for the quasar J1058$+$0315 with
$z_{\rm em}$\,=\,2.3021 and $z_{\rm abs}$\,=\,2.2932.}
 \label{J1058_1D}
\end{figure*}
%*******************************************************

%************************ J1058_2D *********************
\begin{figure*}
\centering
\begin{tabular}{c}
\includegraphics[bb=73 390 546 675,clip=,width=0.65\hsize]{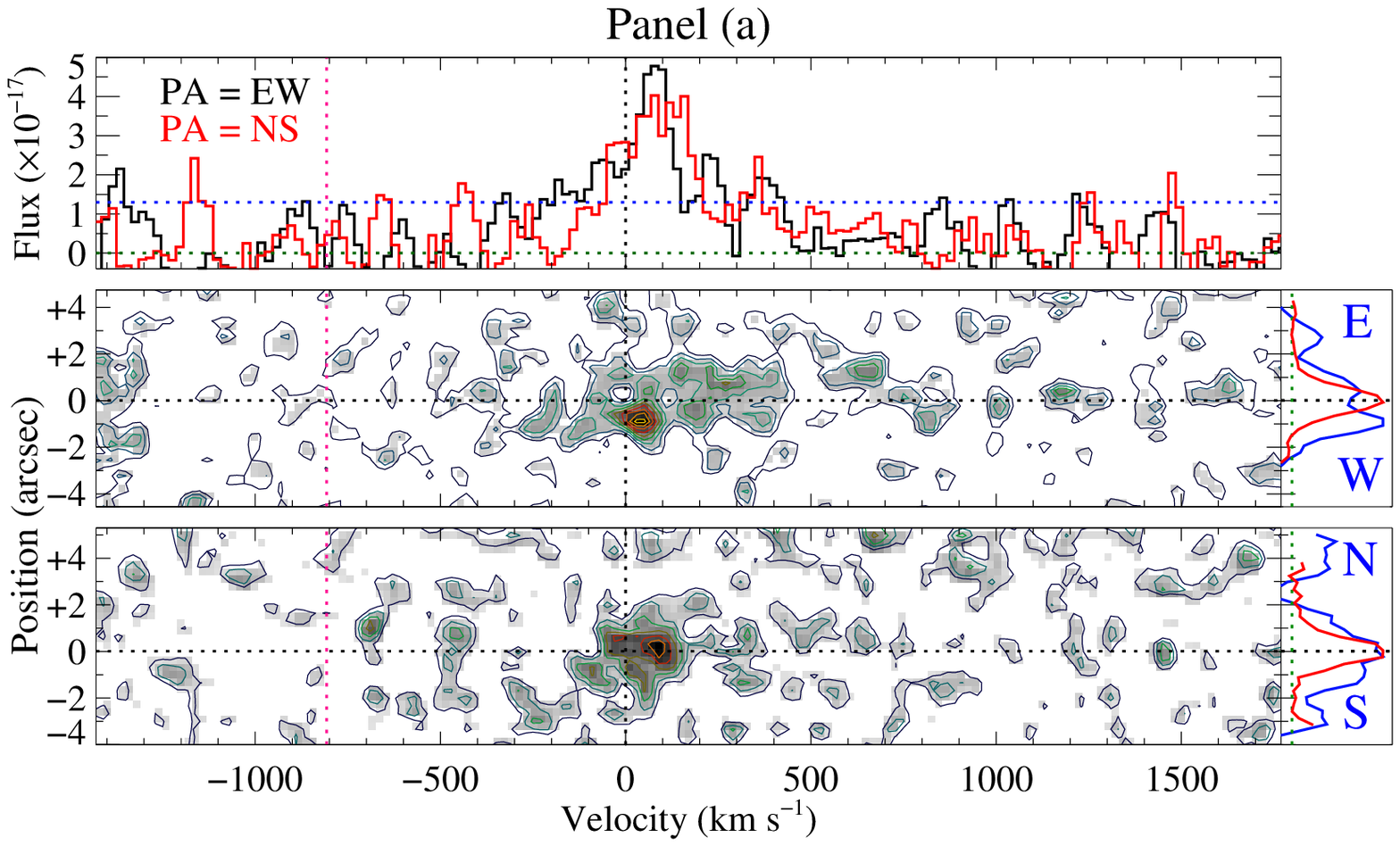}\\
\includegraphics[bb=43 399 566 629,clip=,width=0.65\hsize]{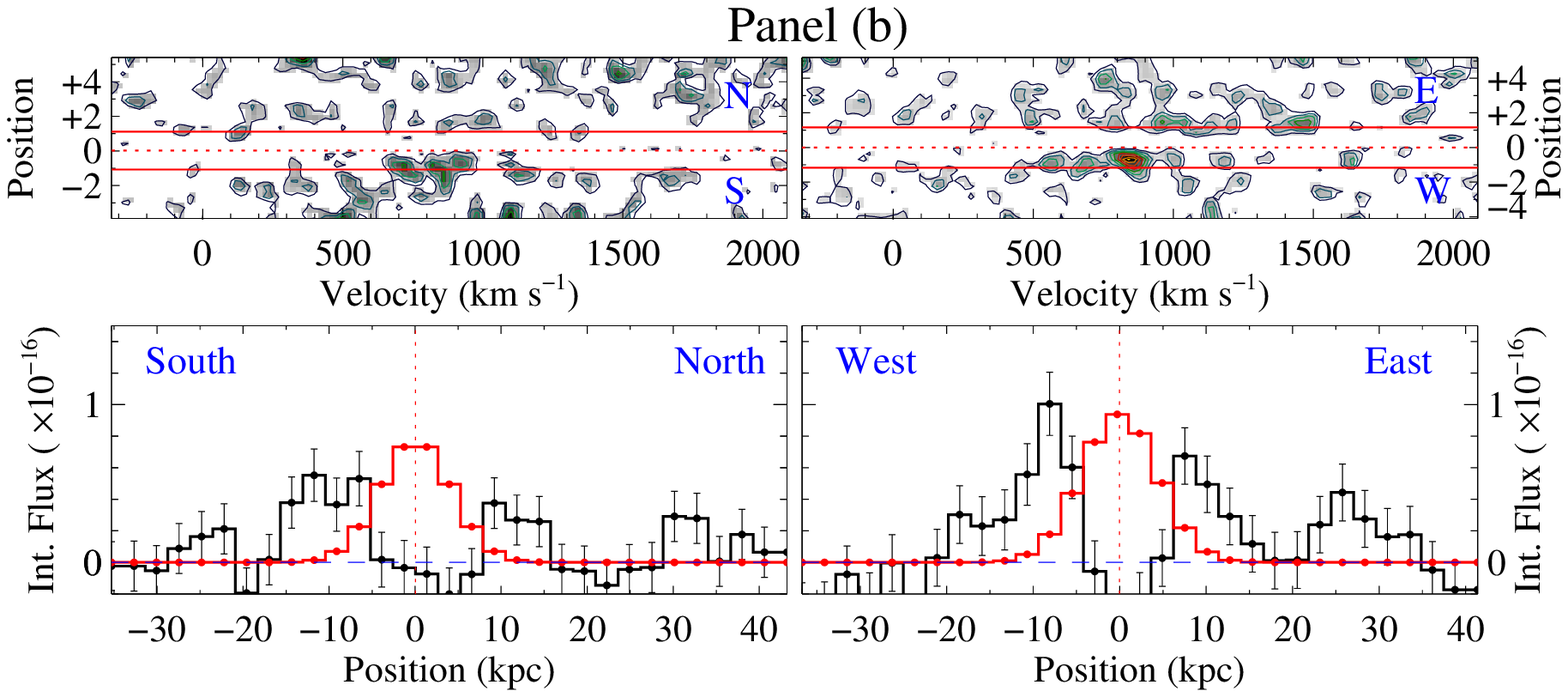}\\
\includegraphics[bb=49 363 566 575,clip=,width=0.65\hsize]{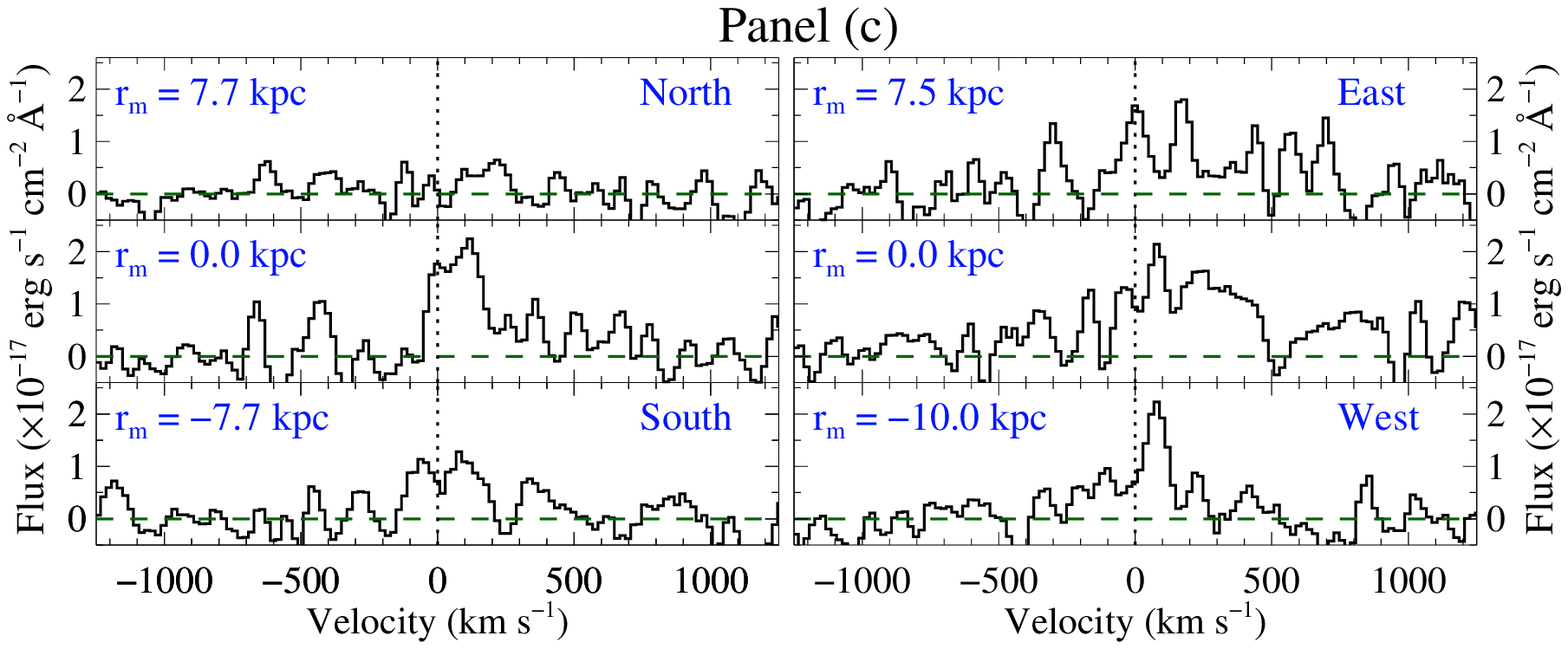}\\
\end{tabular}
\caption{Same as Fig.\,2 but for the quasar J1058$+$0315. The zero
velocity is at $z_{\rm em}$\,=\,2.3021. In panel~(a), the pink
dotted lines mark the position of the DLA with $z_{\rm
abs}$\,=\,2.2932.  In panels\,(a) and (b) the outermost contour
along PA=NS (resp. PA=EW) corresponds to a flux density of
2.94\,$\times$\,10$^{-19}$ (resp. 3.36\,$\times$\,10$^{-19}$)
erg~s$^{-1}$\,cm$^{-2}$\textup{\AA}$^{-1}$ and each contour is
separated by 2.94\,$\times$\,10$^{-19}$ (resp.
3.36\,$\times$\,10$^{-19}$)
erg~s$^{-1}$\,cm$^{-2}$\textup{\AA}$^{-1}$ from its neighboring
contour.}
 \label{J1058_2D}
\end{figure*}
%*******************************************************

%************************ J1154_1D *********************
\begin{figure*}
\centering
\begin{tabular}{c}
\includegraphics[bb=42 348 559 619,clip=,width=0.65\hsize]{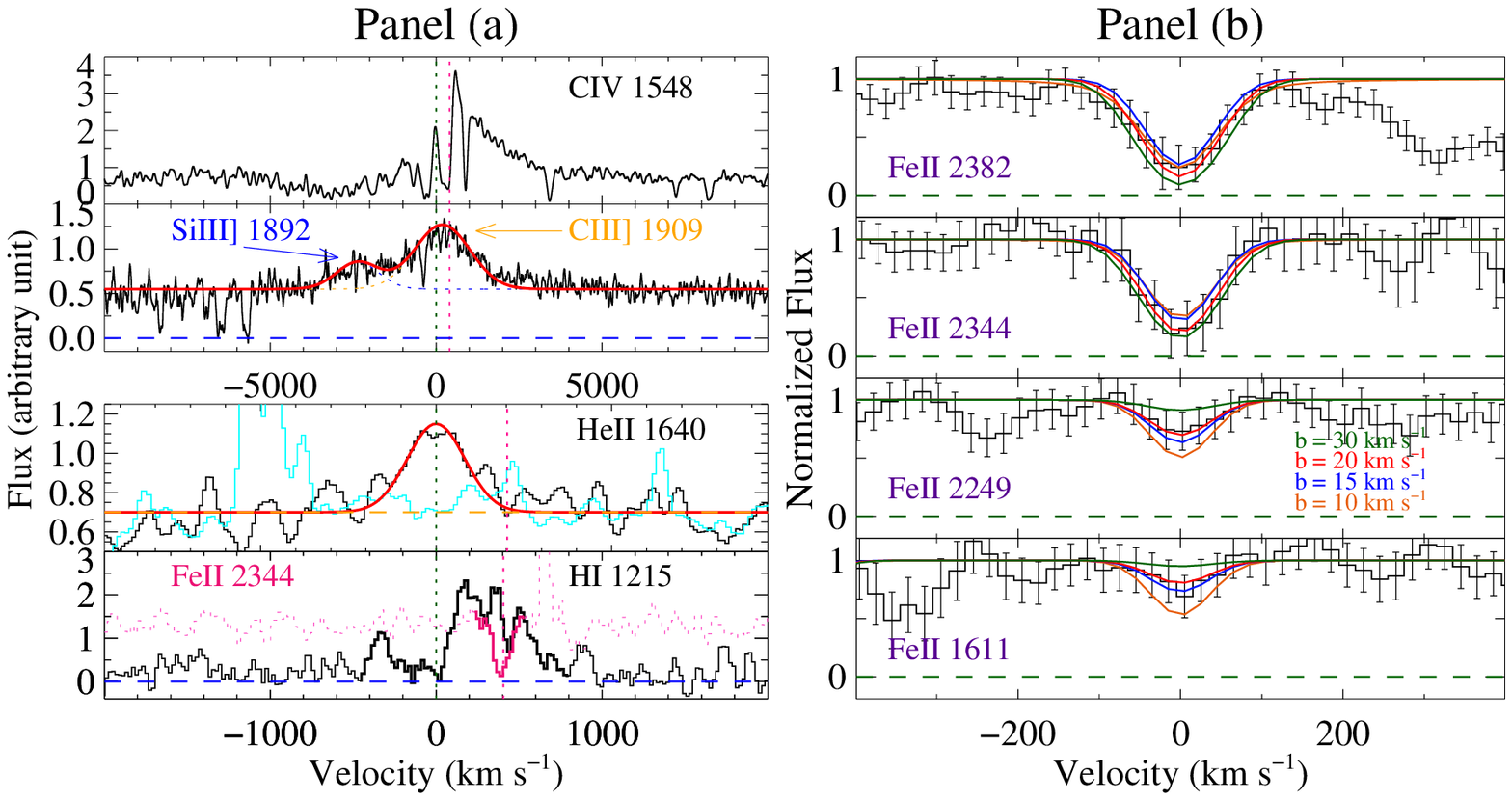}\\
\includegraphics[bb=63 351 562 687,clip=,width=0.65\hsize]{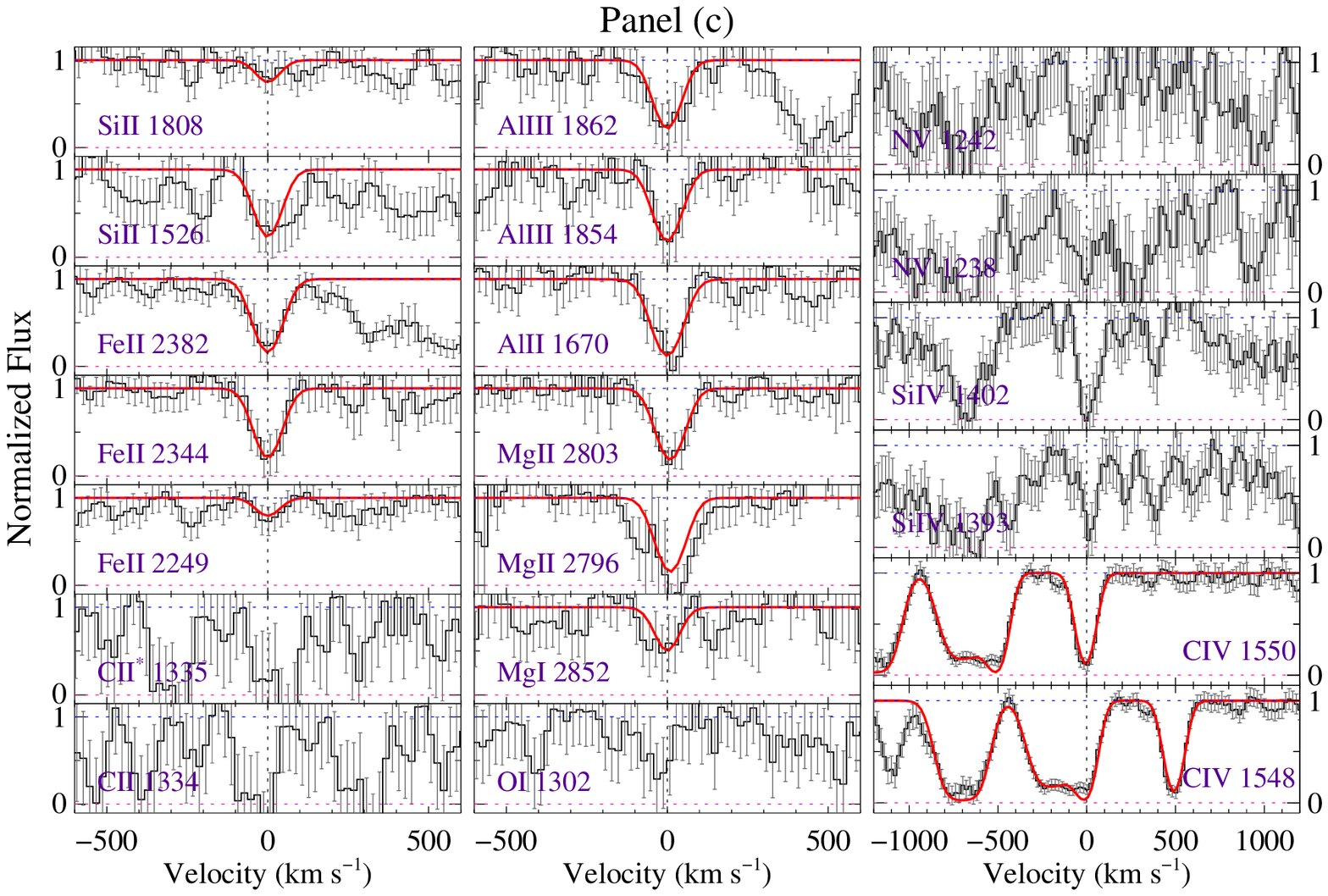}\\
\includegraphics[bb=58 435 554 599,clip=,width=0.65\hsize]{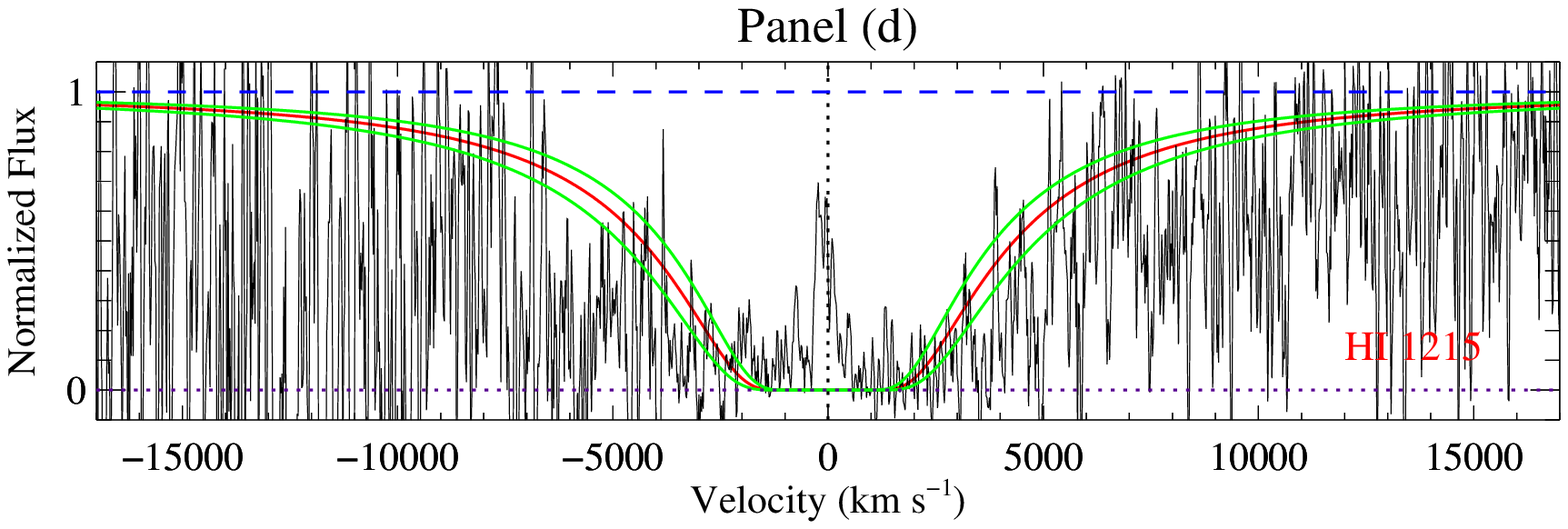}\\
\end{tabular}
\caption{Same as Fig.\,1 but for the quasar J1154$-$0215 with
$z_{\rm em}$\,=\,2.1810 and $z_{\rm abs}$\,=\,2.1853.}
 \label{J1154_1D}
\end{figure*}
%*******************************************************

%************************ J1154_2D *********************
\begin{figure*}
\centering
\begin{tabular}{c}
\includegraphics[bb=73 390 546 675,clip=,width=0.65\hsize]{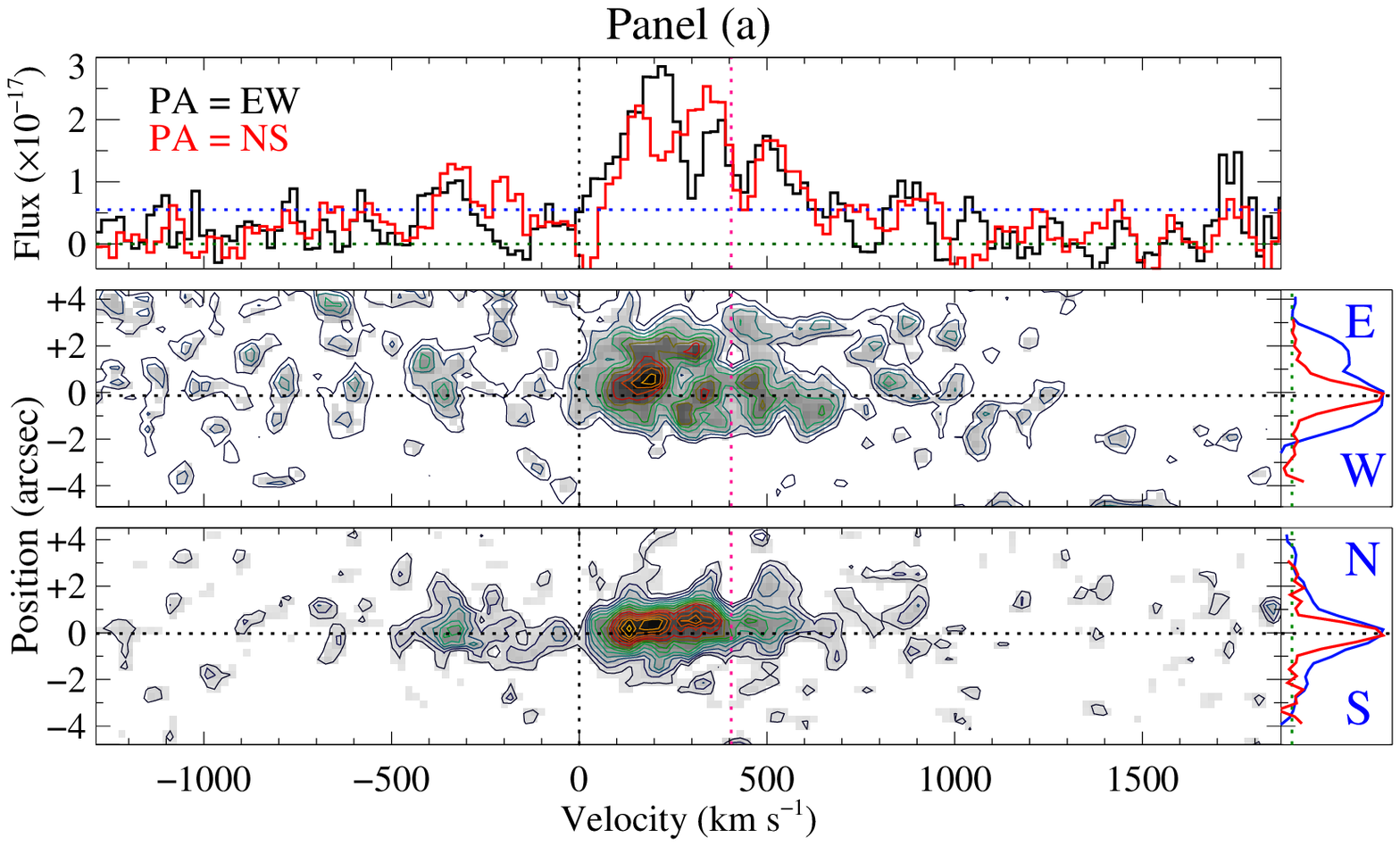}\\
\includegraphics[bb=43 399 566 629,clip=,width=0.65\hsize]{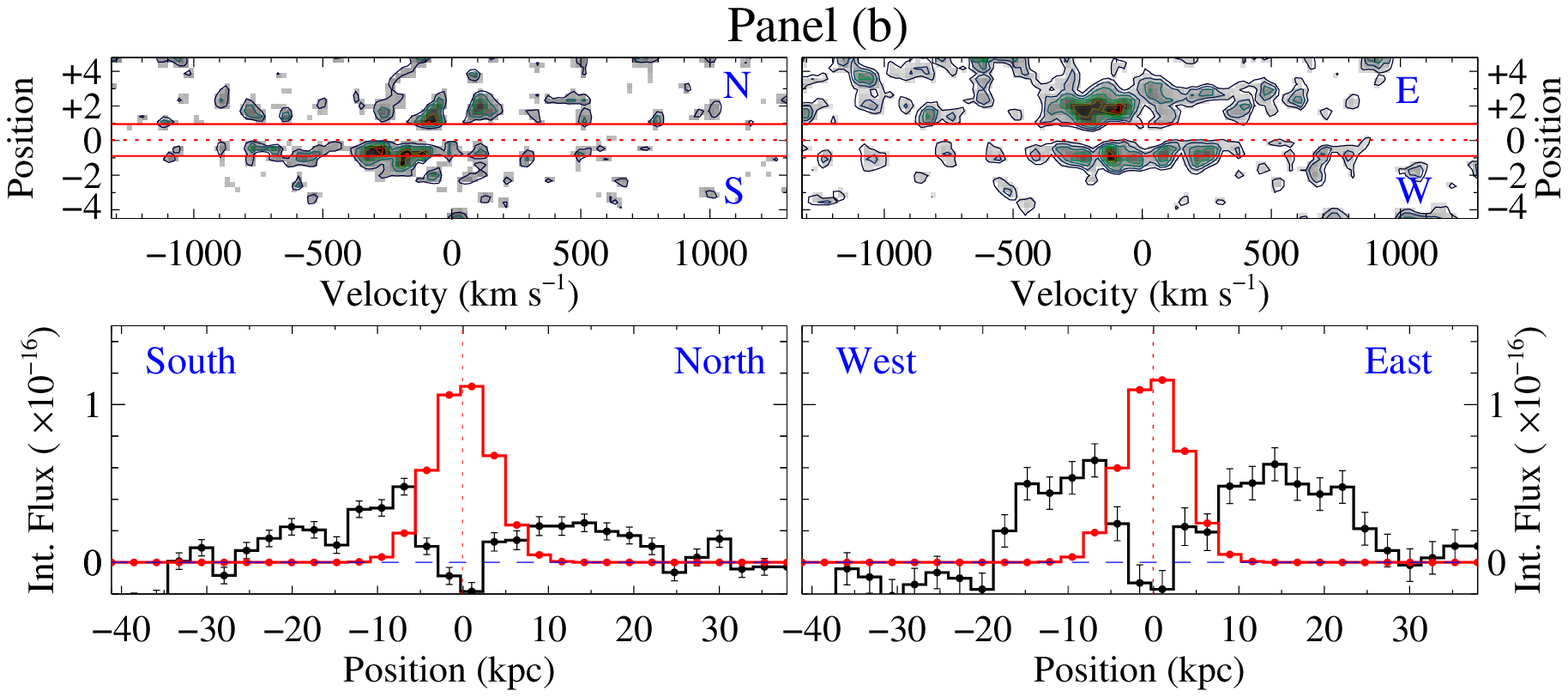}\\
\includegraphics[bb=49 363 566 575,clip=,width=0.65\hsize]{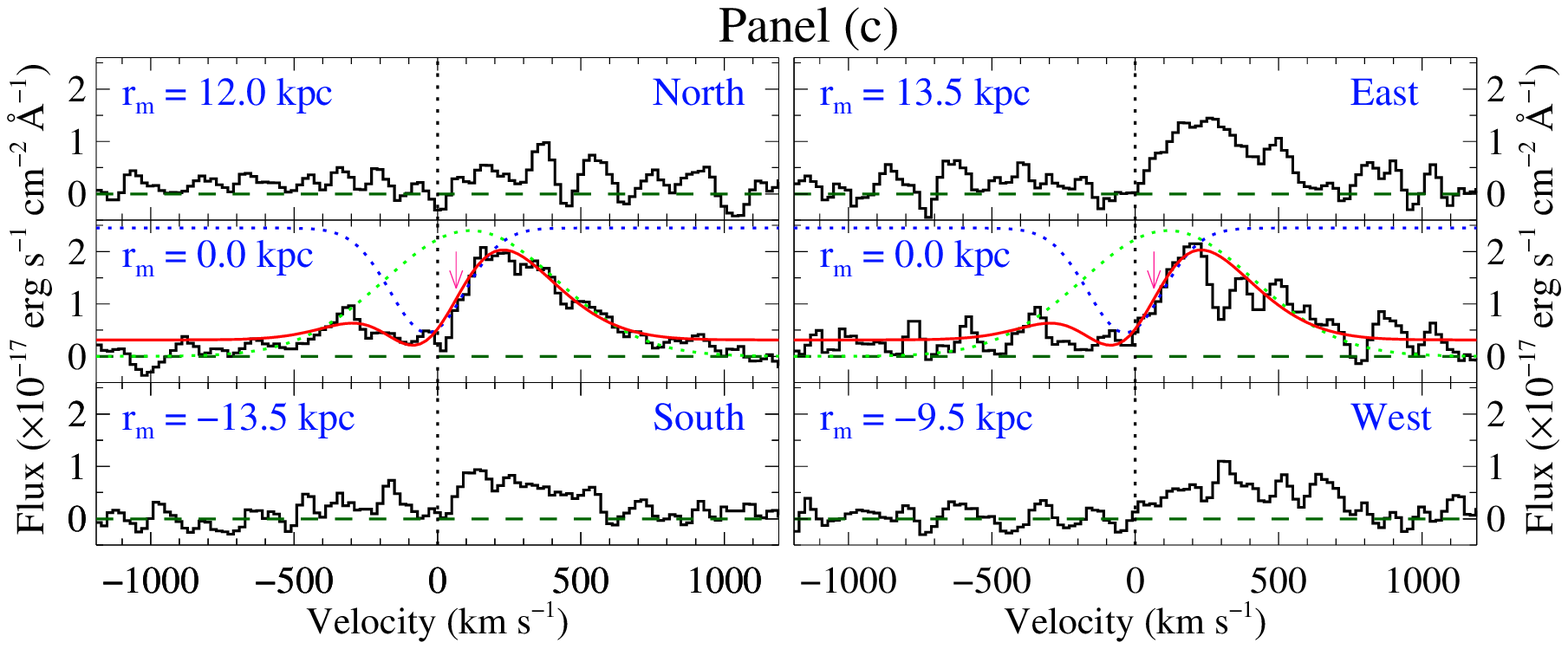}\\
\end{tabular}
\caption{Same as Fig.\,2 but for the quasar J1154$-$0215. The zero
velocity is at $z_{\rm em}$\,=\,2.1810. In panel~(a), the pink
dotted lines mark the position of the DLA with $z_{\rm
abs}$\,=\,2.1853. In panels\,(a) and (b) the outermost contour along
PA=NS (resp. PA=EW) corresponds to a flux density of
1.60\,$\times$\,10$^{-19}$ (resp. 1.29\,$\times$\,10$^{-19}$)
erg~s$^{-1}$\,cm$^{-2}$\textup{\AA}$^{-1}$ and each contour is
separated by 8.00\,$\times$\,10$^{-20}$ (resp.
1.29\,$\times$\,10$^{-19}$)
erg~s$^{-1}$\,cm$^{-2}$\textup{\AA}$^{-1}$ from its neighboring
contour. See the text for the description of the red solid line, and
the blue and green dotted lines shown in panel~(c).}
 \label{J1154_2D}
\end{figure*}
%*******************************************************

%************************ J1253_1D *********************
\begin{figure*}
\centering
\begin{tabular}{c}
\includegraphics[bb=52 363 559 630,clip=,width=0.65\hsize]{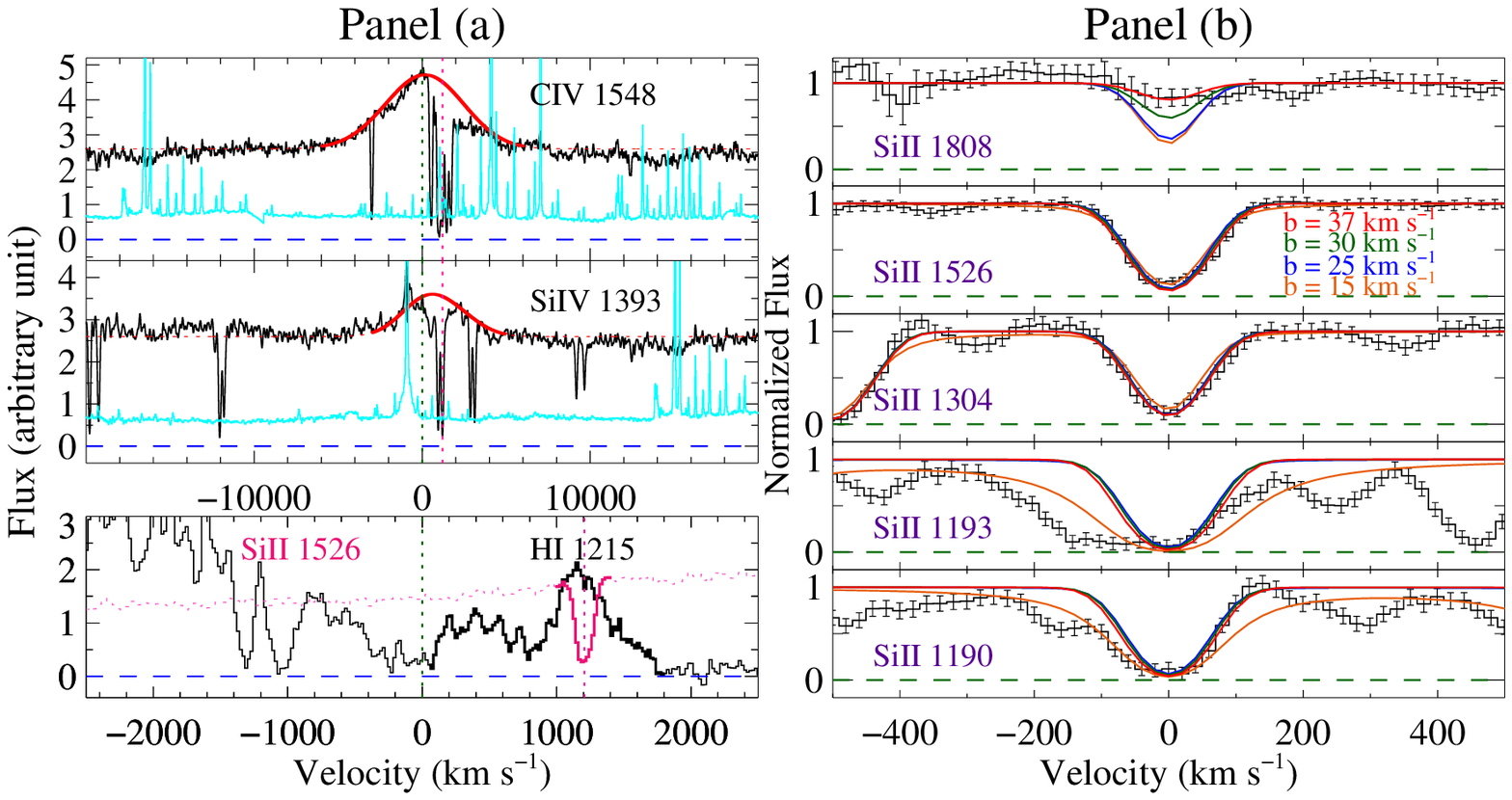}\\
\includegraphics[bb=62 348 523 730,clip=,width=0.65\hsize]{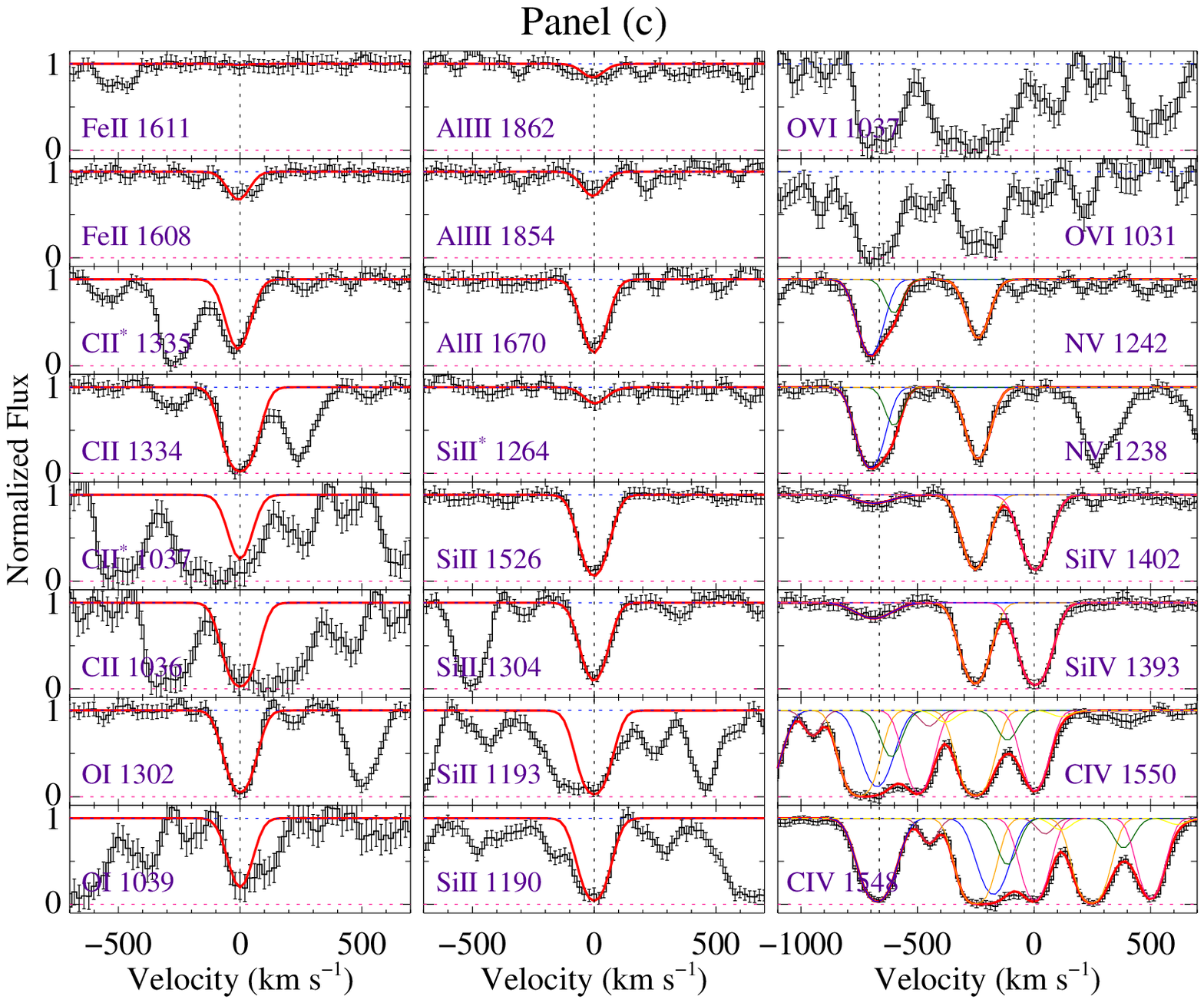}\\
\includegraphics[bb=63 399 554 630,clip=,width=0.65\hsize]{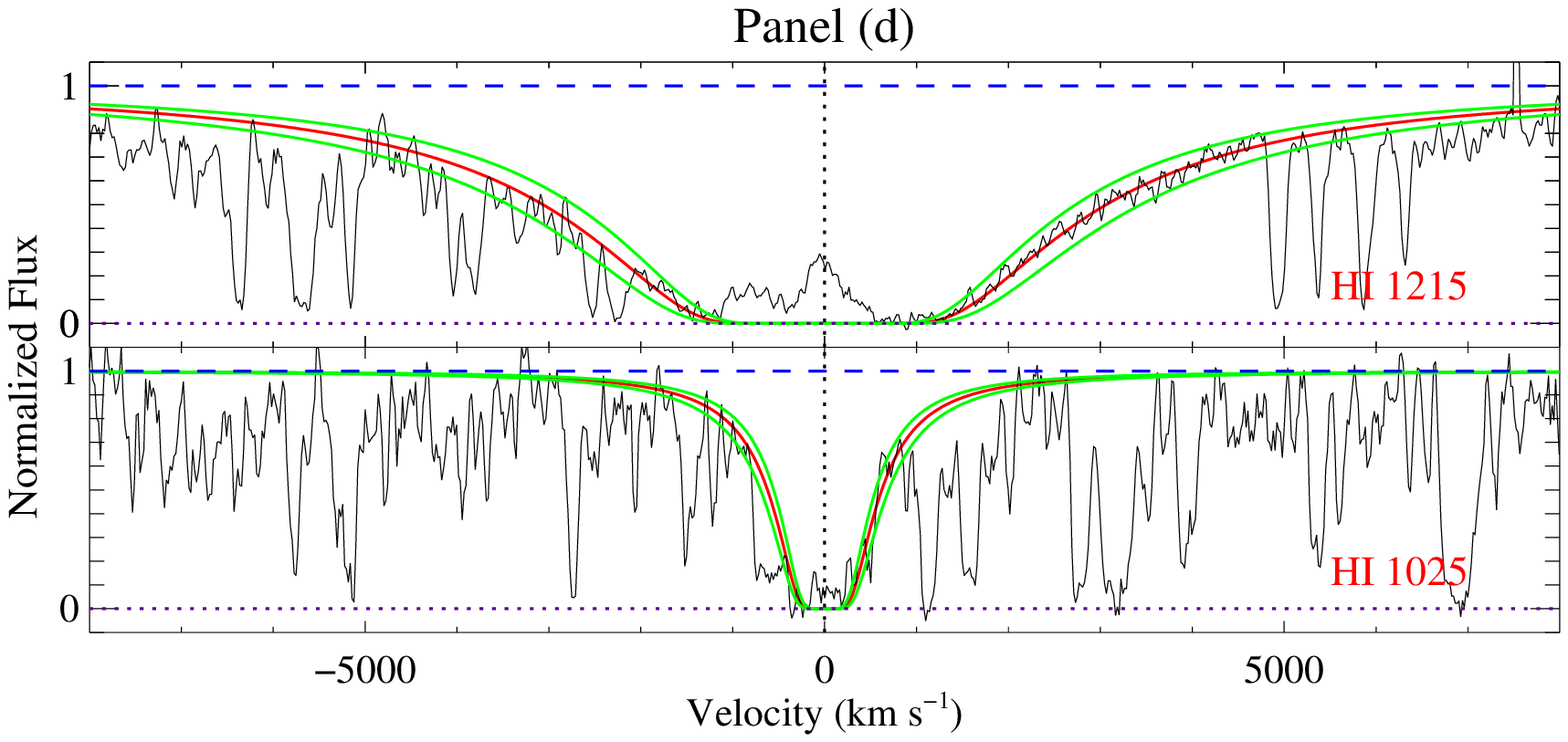}\\
\end{tabular}
\caption{Same as Fig.\,1 but for the quasar J1253$+$1007 with
$z_{\rm em}$\,=\,3.0150 and $z_{\rm abs}$\,=\,3.0312.}
 \label{J1253_1D}
\end{figure*}
%*******************************************************

%************************ J1253_2D *********************
\begin{figure*}
\centering
\begin{tabular}{c}
\includegraphics[bb=73 390 546 675,clip=,width=0.65\hsize]{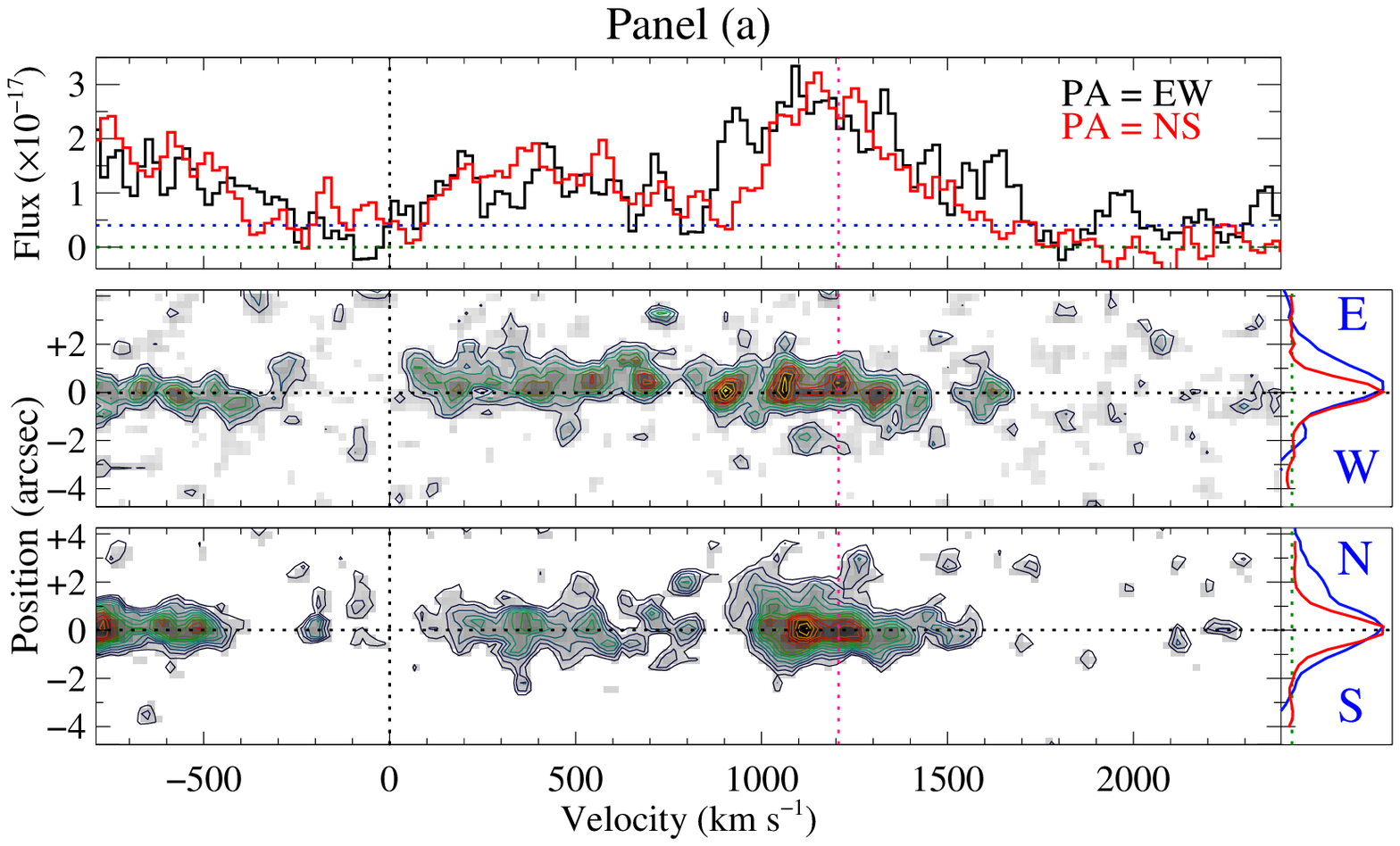}\\
\includegraphics[bb=43 399 566 629,clip=,width=0.65\hsize]{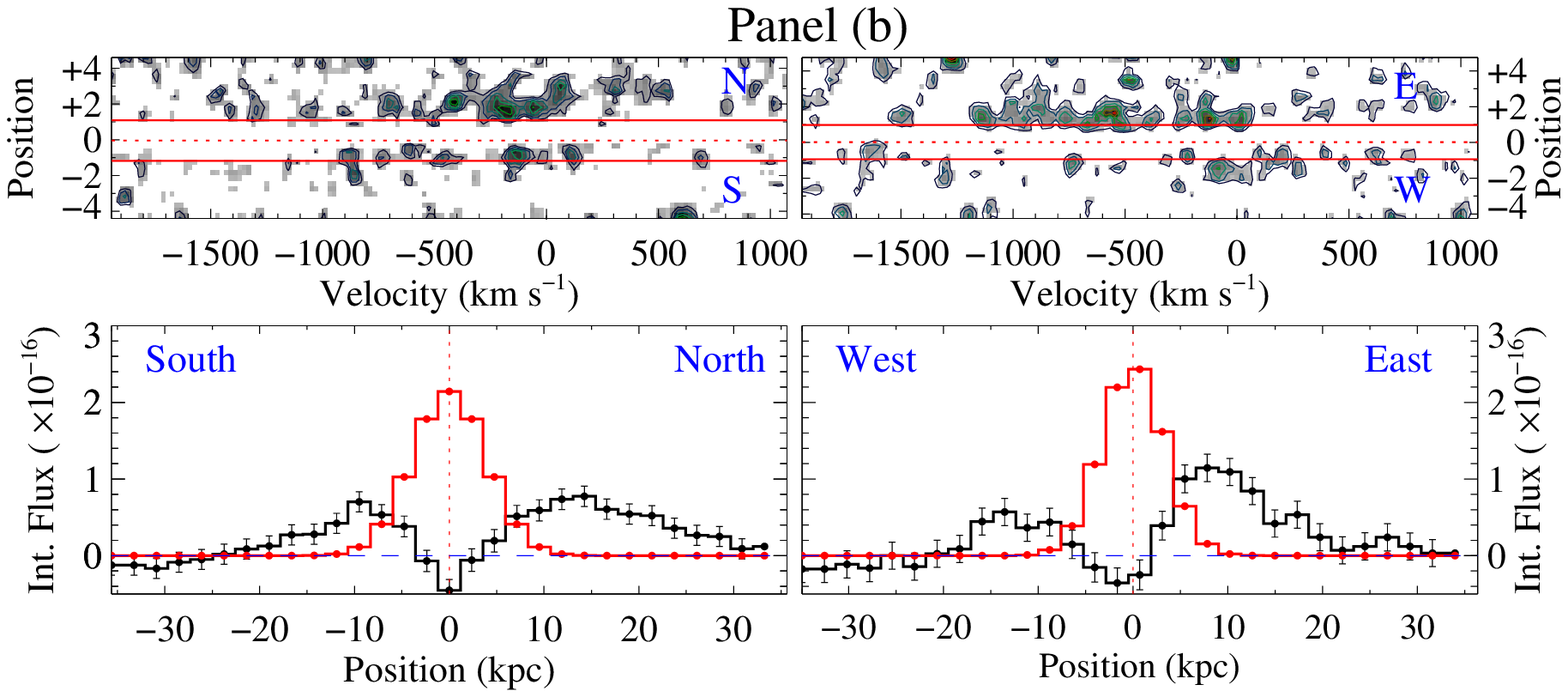}\\
\includegraphics[bb=49 363 566 575,clip=,width=0.65\hsize]{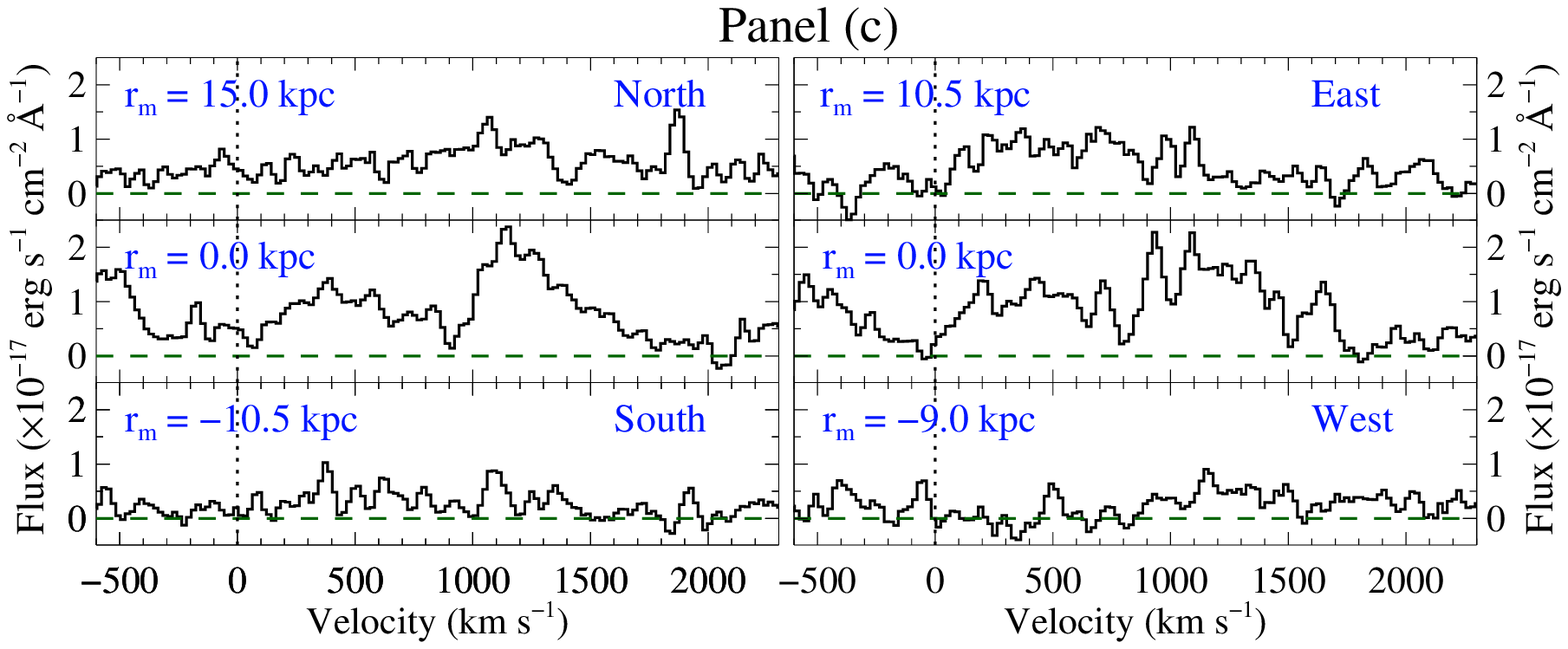}\\
\end{tabular}
\caption{Same as Fig.\,2 but for the quasar J1253$+$1007. The zero
velocity is at $z_{\rm em}$\,=\,3.0150. In panel~(a), the pink
dotted lines mark the position of the DLA with $z_{\rm
abs}$\,=\,3.0312. In panels\,(a) and (b) the outermost contour along
PA=NS (resp. PA=EW) corresponds to a flux density of
2.49\,$\times$\,10$^{-19}$ (resp. 2.26\,$\times$\,10$^{-19}$)
erg~s$^{-1}$\,cm$^{-2}$\textup{\AA}$^{-1}$ and each contour is
separated by 8.30\,$\times$\,10$^{-20}$ (resp.
1.13\,$\times$\,10$^{-19}$)
erg~s$^{-1}$\,cm$^{-2}$\textup{\AA}$^{-1}$ from its neighboring
contour. }
 \label{J1253_2D}
\end{figure*}
%*******************************************************

\end{document}